\documentclass[preprint,twocolumn]{aastex63}
\usepackage{float, bm, graphicx, amsmath, morefloats}
\usepackage[caption=false]{subfig}
\usepackage{lipsum, babel}
\PassOptionsToPackage{hyphens}{url}

\submitjournal{ApJ}

\turnoffeditone
\turnoffedittwo

\shorttitle{AT2020xnd}
\shortauthors{Ho et al.}

\newcommand{\chandra}{{\em Chandra}}

\newcommand{\erg}{\mbox{$\rm erg$}}
\newcommand{\jy}{\mbox{$\rm Jy$}}

\newcommand{\gauss}{\mbox{$\rm G$}}

\newcommand{\cm}{\mbox{$\rm cm$}}
\newcommand{\km}{\mbox{$\rm km$}}
\newcommand{\mpc}{\mbox{$\rm Mpc$}}
\newcommand{\pmpc}{\mbox{$\rm Mpc^{-1}$}}

\newcommand{\pcmsq}{\mbox{$\rm cm^{-2}$}}

\newcommand{\degsq}{\mbox{$\rm deg^{2}$}}

\newcommand{\pcmcub}{\mbox{$\rm cm^{-3}$}}

\newcommand{\pgpccub}{\mbox{$\rm Gpc^{-3}$}}

\newcommand{\msol}{\mbox{$\rm M_\odot$}}

\newcommand{\days}{\mbox{$\rm d$}}

\newcommand{\psec}{\mbox{$\rm s^{-1}$}}
\newcommand{\pyr}{\mbox{$\rm yr^{-1}$}}

\newcommand{\ghz}{\mbox{$\rm GHz$}}

\newcommand{\phz}{\mbox{$\rm Hz^{-1}$}}

\begin{document}

\title{
Luminous Millimeter, Radio, and X-ray Emission from
ZTF20acigmel (AT2020xnd)
} 

\correspondingauthor{Anna Y. Q. Ho}
\email{annayqho@berkeley.edu}

\author[0000-0002-9017-3567]{Anna Y. Q.~Ho}
\affiliation{Department of Astronomy, University of California, Berkeley, 501 Campbell Hall, Berkeley, CA, 94720, USA}
\affiliation{Miller Institute for Basic Research in Science, 468 Donner Lab, Berkeley, CA 94720, USA}

\author[0000-0001-8405-2649]{Ben Margalit}
\altaffiliation{NASA Einstein Fellow}
\affiliation{Astronomy Department and Theoretical Astrophysics Center, University of California, Berkeley, Berkeley, CA 94720, USA}

\author{Michael Bremer}
\affiliation{Institut de Radio Astronomie Millimétrique (IRAM), 300 rue de la Piscine, 38406 Saint Martin d’Hères, France}

\author[0000-0001-8472-1996]{Daniel A.~Perley}
\affiliation{Astrophysics Research Institute, Liverpool John Moores University, IC2, Liverpool Science Park, 146 Brownlow Hill, Liverpool L3 5RF, UK}

\author[0000-0001-6747-8509]{Yuhan Yao}
\affiliation{Cahill Center for Astrophysics, 
California Institute of Technology, MC 249-17, 
1200 E California Boulevard, Pasadena, CA, 91125, USA}

\author{Dougal Dobie}
\affiliation{Centre for Astrophysics and Supercomputing, Swinburne University of Technology, Hawthorn, Victoria, Australia}
\affiliation{ARC Centre of Excellence for Gravitational Wave Discovery (OzGrav), Hawthorn, VIC 3122, Australia}

\author[0000-0001-6295-2881]{David L.\ Kaplan}
\affiliation{Department of Physics, University of Wisconsin-Milwaukee, P.O. Box 413, Milwaukee, WI 53201, USA}

\author[0000-0003-4609-2791]{Andrew O'Brien}
\affiliation{Department of Physics, University of Wisconsin-Milwaukee, P.O. Box 413, Milwaukee, WI 53201, USA}

\author{Glen Petitpas}
\affiliation{Harvard-Smithsonian Center for Astrophysics, 60 Garden Street, Cambridge, MA 02138, USA}

\author{Andrew Zic}
\affiliation{ATNF, CSIRO Space and Astronomy, PO Box 76, Epping, New South Wales 1710, Australia}
\affiliation{Department of Physics and Astronomy, and Research Centre in Astronomy, Astrophysics and Astrophotonics, Macquarie University,
NSW 2109, Australia}

\begin{abstract}

Observations of the extragalactic ($z=0.0141$) transient AT2018cow established a new class of energetic explosions shocking a dense medium, which produce luminous emission at millimeter and sub-millimeter wavelengths.
Here we present detailed millimeter- through centimeter-wave observations of a similar transient, ZTF20acigmel (AT2020xnd) at $z=0.2433$.
Using observations from the NOrthern Extended Millimeter Array and the Very Large Array,
we model the unusual millimeter and radio emission from AT2020xnd under several different assumptions, and ultimately favor synchrotron radiation from a thermal electron population (relativistic Maxwellian).
The thermal-electron model implies a fast but sub-relativistic ($v\approx0.3c$) shock and a high ambient density ($n_e\approx4\times10^{3}\,\pcmcub$) at $\Delta t\approx40\,\days$.
The X-ray luminosity of $L_X\approx10^{43}\,\erg\,\psec$ exceeds simple predictions from the radio and UVOIR luminosity and likely has a separate physical origin, such as a central engine.
Using the fact that month-long luminous ($L_\nu\approx 2\times10^{30}\,\erg\,\psec\,\phz$ at 100\,\ghz) millimeter emission appears to be a generic feature of transients with fast ($t_{1/2}\approx3\,\days$) and luminous ($M_\mathrm{peak}\approx-21\,$mag) optical light curves,
we estimate the rate at which transients like AT2018cow and AT2020xnd will be detected by future wide-field millimeter transient surveys like CMB-S4,
and conclude that energetic explosions in dense environments may represent a significant population of extragalactic transients in the 100\,GHz sky.

\end{abstract}

\section{Introduction}
\label{sec:introduction}

In a cosmic explosion, high-velocity material shocks the ambient medium, accelerating electrons to relativistic speeds and producing synchrotron radiation.
Centimeter-wavelength observations have been widely used to model the forward-shock properties from a variety of energetic phenomena, including supernovae (SNe; e.g. \citealt{Chevalier1998,Kulkarni1998,Bietenholz2021}), gamma-ray bursts (GRBs; \citealt{Chandra2012}), and tidal disruption events (TDEs; \citealt{Alexander2020}).
Observations at millimeter (mm) wavelengths have been less common for both technical and astrophysical reasons:
previous generations of mm telescopes had low sensitivity,
and mm emission from cosmic explosions tends to be shorter-lived than emission at cm wavelengths.

The landscape has changed due to the enhanced sensitivity of mm telescopes and the routine discovery of young explosions by high-cadence optical surveys.
Rapid mm follow-up observations of GRBs and supernovae (SNe) is enabling
modeling of the reverse shock \citep{Laskar2018_GRB161219B} and the innermost circumstellar medium in massive stars \citep{Maeda2021}.
Surprisingly, the nearby ($z=0.014$) fast optical transient AT2018cow \citep{Prentice2018,Perley2019cow} had luminous millimeter emission that persisted for weeks \citep{Ho2019cow},
significantly exceeding expectations from the model used to describe the late-time ($\Delta t\gtrsim 80\,\days$) low-frequency ($\nu \lesssim 40\,\ghz$) data \citep{Margutti2019}.
\citet{Margutti2019} suggested that the unusual millimeter emission could arise from a distinct component like a reverse shock.
\citet{Ho2019cow} suggested that the millimeter emission was produced while the shock was in a dense confined region,
and that it abruptly diminished when the shock passed into lower-density material.

Here we present millimeter, radio, and X-ray observations of ZTF20acigmel (AT2020xnd), which appears to be a distant ($z=0.2433$) analog to AT2018cow.
The optical light curves and spectra of AT2020xnd were published in \citet{Perley2021}. In short,
AT2020xnd was discovered on 2020 October 12 by the Zwicky Transient Facility \citep{Graham2019,Bellm2019_ztf} and flagged by filters designed to find optical transients that are faster evolving than ordinary supernovae \citep{Ho2020d,Perley2021}.
More precisely, the optical light curve of AT2020xnd had a duration above half-maximum of $t_{1/2}=3$--5\,d \citep{Perley2021}, similar to the $t_{1/2}\sim3\,$d duration of the optical light curves of AT2018cow \citep{Perley2019cow,Margutti2019} and ZTF18abvkwla (AT2018lug; \citealt{Ho2020b}), and much faster than the $t_{1/2}\gtrsim10\,$d of ordinary SNe \citep{Perley2020,Ho2021}.
Our observations represent only the second millimeter observations of an AT2018cow analog.
As was the case for AT2018cow, we find that the early-time millimeter-wavelength data are difficult to reconcile with the late-time centimeter-wavelength data.

The paper is organized as follows. In \S\ref{sec:observations} we describe observations from the NOrthern Extended Millimeter Array (NOEMA),
the Australia Telescope Compact Array (ATCA; \citealt{Frater1992}),
the Submillimeter Array (SMA; \citealt{Ho2004}),
the Very Large Array (VLA; \citealt{Perley2011}), and the Chandra X-ray Observatory (\chandra).
We model the forward shock in
\S\ref{sec:analysis} and explore several possible origins for the millimeter-wavelength emission. We conclude that the most likely explanation is synchrotron radiation from a thermal electron-energy distribution (relativistic Maxwellian).
In \S\ref{sec:xray} we discuss the origin of the X-ray emission.
In \S\ref{sec:rates} we
estimate the detection rates of events like AT2018cow and AT2020xnd in current and upcoming millimeter and radio time-domain surveys.

Throughout this paper we use MJD 59132.0 (2020 Oct 10.0) as the reference epoch $t_0$, following \citet{Perley2021}.
We assume a flat $\Lambda$CDM cosmology with $H_0=67.7\,\km\,\psec\,\pmpc$ and $\Omega_M=0.307$ \citep{Planck2016},
implying a luminosity distance to the source of 1261\,Mpc and an angular-diameter distance of 816\,Mpc.
Additional sub-mm and radio observations were obtained by an independent observing team and are presented and interpreted in \citet{Bright2022}.

\section{Observations}
\label{sec:observations}

In this section we present the millimeter, radio, and X-ray observations of AT2020xnd. We compare the observational properties to established classes of core-collapse SNe, as well as to the other `AT2018cow-like' events: AT2018cow itself \citep{Ho2019cow,Margutti2019,Nayana2021}, CSS161010 \citep{Coppejans2020}, and AT2018lug \citep{Ho2020b}.

Following the identification of AT2020xnd as a fast and luminous transient \citep{Perley2020ATel},
we triggered follow-up observations with the VLA.
Our first VLA observation began on 2020 October 22.99 UTC, at X-band (8--12\,\ghz).
We detected faint but significant ($24\pm6\,\mu$Jy) radio emission consistent with the position of the optical transient \citep{Ho2020ATel}.
The position of the radio source in our brightest X-band observation (on December 20), measured with a Gaussian fit, is $\alpha$(J2000)$=22^{\mathrm{h}}20^{\mathrm{m}}02\fs04$,
$\delta$(J2000)$=-02^{\mathrm{d}}50^{\mathrm{m}}25\fs4$.
The observation was taken in A configuration and the statistical uncertainty on the position is $0\farcs008$.
The uncertainty on the position is dominated by a systematic uncertainty of $0\farcs02$, calculated as $10\%$ of the full-width half-maximum of the synthesized beam\footnote{\url{https://science.nrao.edu/facilities/vla/docs/manuals/oss/performance/positional-accuracy}} at X-band in A-configuration.

\begin{figure*}[!htb]
    \centering
    \includegraphics{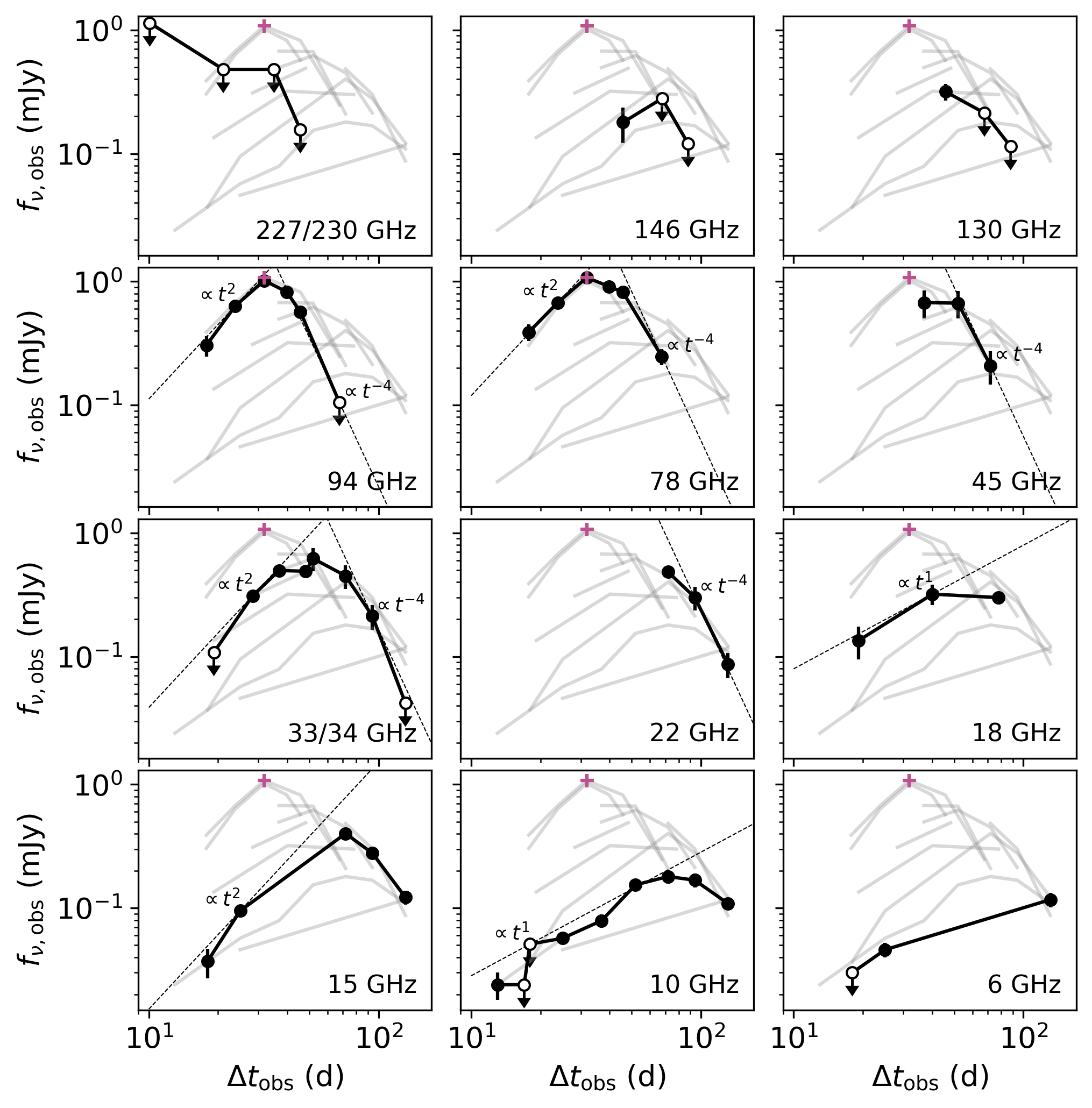}
    \caption{Millimeter and radio light curves of AT2020xnd from NOEMA, the ATCA, and the VLA.
    We include all frequencies that have two or more observations.
    The red cross marks the peak of the 79\,GHz NOEMA light curve, which was 1.1\,mJy at 32\,d.
    The full set of light curves are shown as grey lines in the background, and each panel highlights an individual observing band in black.
    Open circles represent 3-$\sigma$ upper limits.
    No cosmological correction has been applied, and time is in the observer-frame.
    }
    \label{fig:radio-lc}
\end{figure*}

The detection of radio emission similar in luminosity to that of AT2018cow motivated us to trigger other facilities.
A full description of our radio and millimeter observations and data reduction can be found in Appendix~\ref{sec:appendix-radio-data},
and the light curves are shown in Figure~\ref{fig:radio-lc}.
We obtained Director's Discretionary Time with the ATCA at 34\,GHz, to see whether (like AT2018cow) the emission was optically thick at these frequencies.
We triggered our SMA ToO program\footnote{Program 2020A-S037; P.I. Ho} and obtained Director's Discretionary Time with the NOrthern Extended Millimeter Array (NOEMA) to observe at 3\,mm, 2\,mm, and 1.3\,mm\footnote{Program D20AF and D20AG; P.I. Ho}.
We obtained several more epochs of VLA data\footnote{Program VLA/20A-374 and Program VLA/20B-205; PI Ho} from 2020 October--2021 May, spanning C-band (4--8\,\ghz) to Q-band (40--50\,\ghz).

As shown in Figure~\ref{fig:radio-lc},
the light curve at most frequencies rises as 
$f_\nu \propto t^{2}$ before the peak and fades as $F_\nu \propto t^{-4}$.
The rise at our lowest frequencies appears shallower, $f_\nu \propto t^{1}$.
A rise of $f_\nu \propto t^{2}$ was also observed at optically thick frequencies in AT2018cow \citep{Ho2019cow,Margutti2019} and interpreted as a constant-velocity shock.
Steeply declining radio light curves have been observed in all AT2018cow-like events at frequencies $\lesssim 10\,\ghz$ \citep{Coppejans2020,Ho2020b}.
From Figure~\ref{fig:radio-lc} it is clear that the steep decline is chromatic, beginning at later times at lower frequencies.
We discuss the origin of the chromatic steep decline in \S\ref{sec:analysis}.

\begin{figure}
    \centering
    \includegraphics{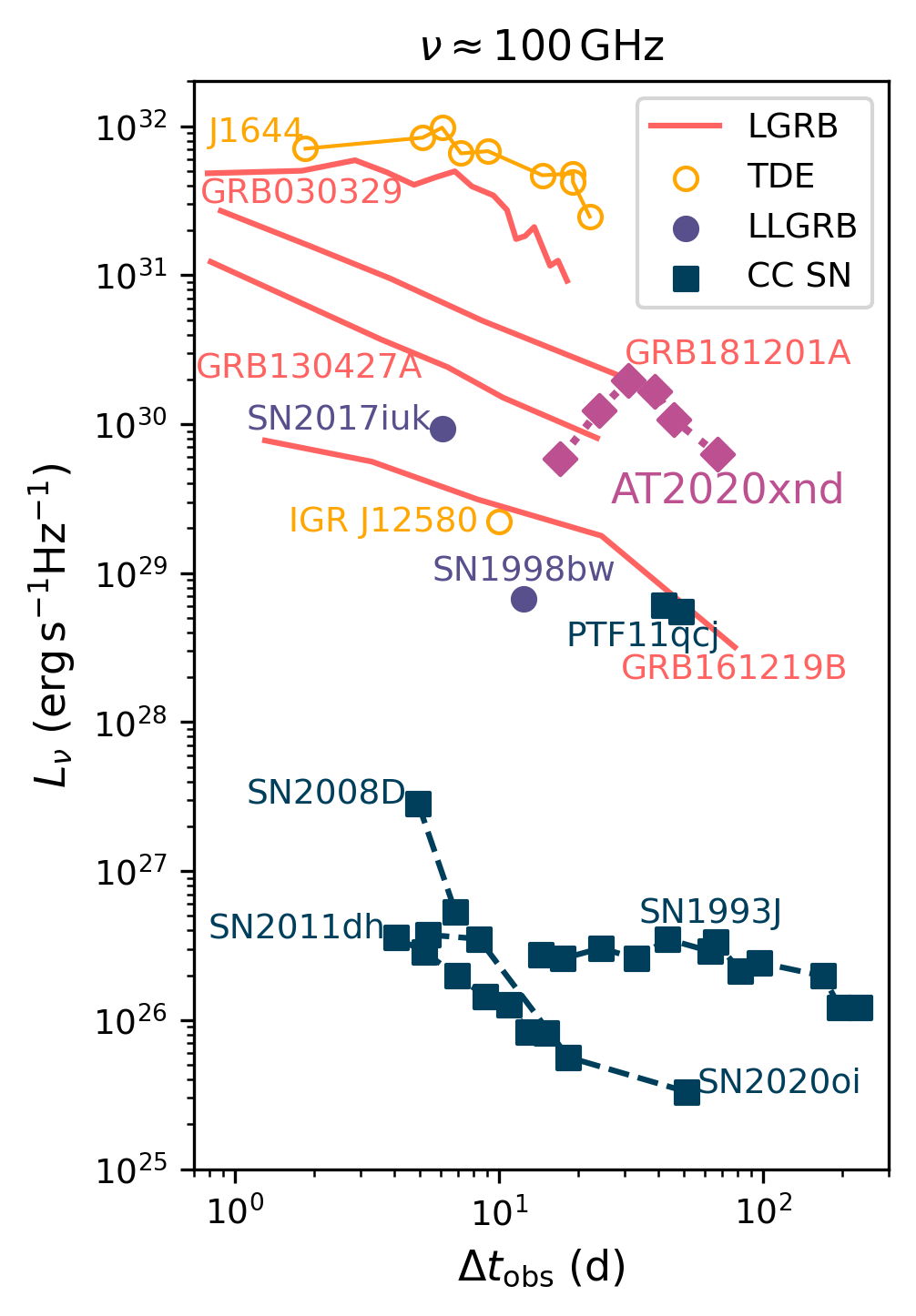}
    \caption{The NOEMA 94\,GHz light curve of AT2020xnd compared to light curves of millimeter-bright cosmic explosions at similar frequencies: long-duration gamma-ray bursts (LGRBs), tidal disruption events (TDEs), low-luminosity GRBs (LLGRBs), and core-collapse (CC) SNe. Data obtained from \citet{Kulkarni1998,Sheth2003,Weiler2007,Soderberg2010,Zauderer2011,Horesh2013,Corsi2014,Perley2014_130427A,Yuan2016,Perley2017-alma,Laskar2018_GRB161219B,Laskar2019,Maeda2021}. All observations are in the observer-frame.
    }
    \label{fig:mm-lc-comparison}
\end{figure}

After AT2018cow itself,
our NOEMA observations represent only the second detection of an AT2018cow-like transient at millimeter wavelengths.
The peak flux density of $1.08\pm0.05\,$mJy at 79\,\ghz\ (100\,GHz in the rest frame) corresponds to a spectral luminosity of $L_{79\,\mathrm{GHz}} = (2.05\pm0.09) \times 10^{30}\,\erg\,\psec\,\phz$.
As shown in Figure~\ref{fig:mm-lc-comparison},
the only transients in the literature with a higher luminosity at similar frequencies are relativistic explosions: long-duration GRBs (e.g., $10^{31}\,\erg,\,\psec\,\phz$ for GRB\,130427A; \citealt{Perley2014_130427A})
and tidal disruption events ($7\times10^{31}\,\erg\,\psec\,\phz$ for J1644+57; \citealt{Zauderer2011}).
However, the light curve of AT2020xnd rises to peak over a month instead of a few days.
In \S\ref{sec:rates} we use the 100\,\ghz\ light curve to estimate the detection rate for events like AT2020xnd in millimeter transient surveys.

The 10\,\ghz\ light curve peaks at $f_\nu=0.180\pm0.023\,$mJy,
or $L_\mathrm{10\,GHz} = (3.4\pm0.4)\times10^{29}\,\erg\,\psec\,\phz$.
The time to peak of $t_\mathrm{pk}\approx60$\,d is common for cm-wavelength emission from core-collapse supernovae \citep{Bietenholz2021}, but the luminosity is significantly greater.
The luminosity and timescale is similar to what was observed for AT2018cow \citep{Ho2019cow,Margutti2019}, CSS161010 \citep{Coppejans2020}, and AT2018lug \citep{Ho2020b}.

\begin{figure*}
    \centering
    \includegraphics{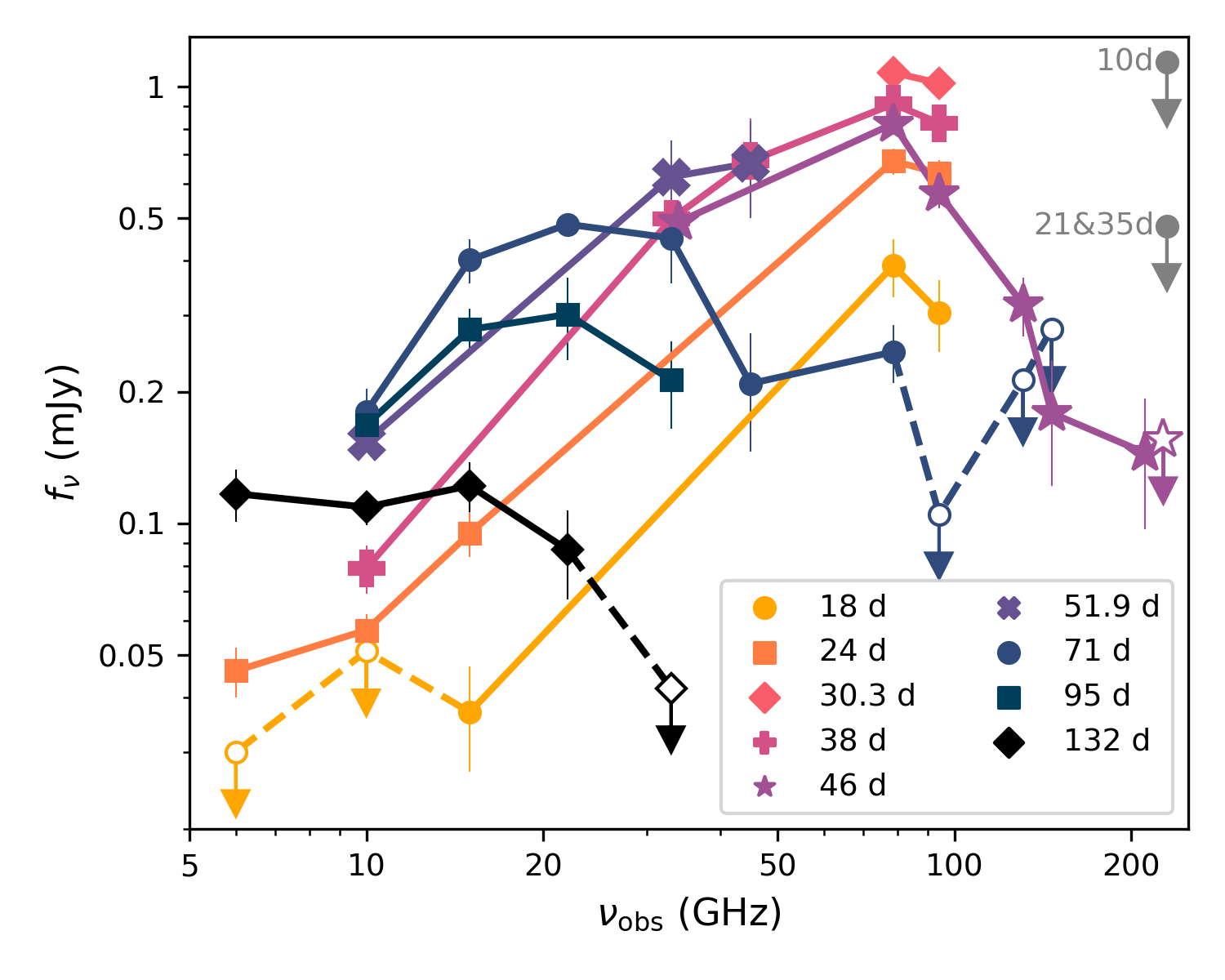}
    \caption{The evolution of the millimeter and radio spectral energy distribution of AT2020xnd. Observations are considered co-eval if they are within $\Delta t/10\,$d of each other. SMA 230\,GHz upper limits are shown in grey. At other epochs, upper limits are indicated with empty symbols and connected with dashed lines. Observations are in the observer-frame.}
    \label{fig:radio-sed}
\end{figure*}

\begin{figure}
    \centering
    \includegraphics{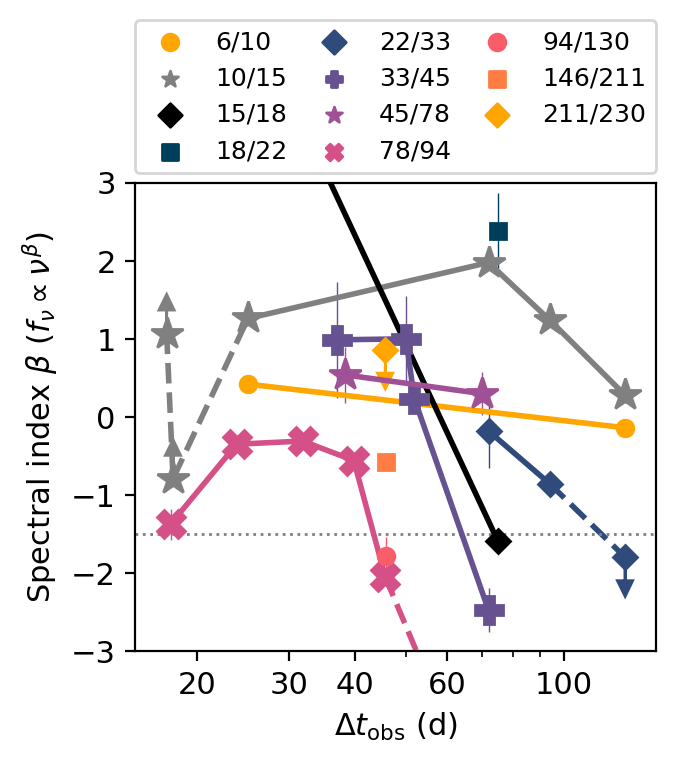}
    \caption{The evolution of the spectral index $\beta$ over time, where $f_\nu \propto \nu^{\beta}$. The spectral index is measured between adjacent frequency bands at every co-eval epoch $\Delta t$, defined as epochs where the observations take place within $\Delta t/10\,$d of each other. Epoch and frequencies are reported in the observer frame. The horizontal dotted line indicates $\beta=-1.5$, which might be expected from an electron energy distribution of $p=3$ in the fast-cooling regime. For clarity we do not show one 130/146\,GHz point ($\beta\approx-5$), one 15/18\,GHz point ($\beta\approx7$), and one 78/94 point ($\beta\approx-5$).}
    \label{fig:spindex}
\end{figure}

In Figure~\ref{fig:radio-sed} and Figure~\ref{fig:spindex}
we show the radio--mm spectral energy distribution (SED) as a function of time.
We regard data obtained within $\Delta t/10\,$d as co-eval, where $\Delta t$ is the time since $t_0$ (defined in \S\ref{sec:introduction}).
Based on these observations, we are motivated to consider the evolution of AT2020xnd in two stages (\S\ref{sec:analysis}).
Before $\Delta t=40\,$d, the spectral index from 79\,\ghz\ to 94\,\ghz\ is relatively flat and does not change with time,
even while the overall flux density changes.
At 46\,d we observe a steep spectral index across the NOEMA bands: a fit to the five high-frequency points gives $\beta=-2.00\pm0.23$ where $f_\nu \propto \nu^{\beta}$.
After $\Delta t=70$\,d, the SED cascades down in flux and frequency: the bulk of the radiation emerges at successively lower frequencies, with the peak luminosity also decreasing.

\begin{figure*}
    \centering
    \includegraphics[width=0.96\textwidth]{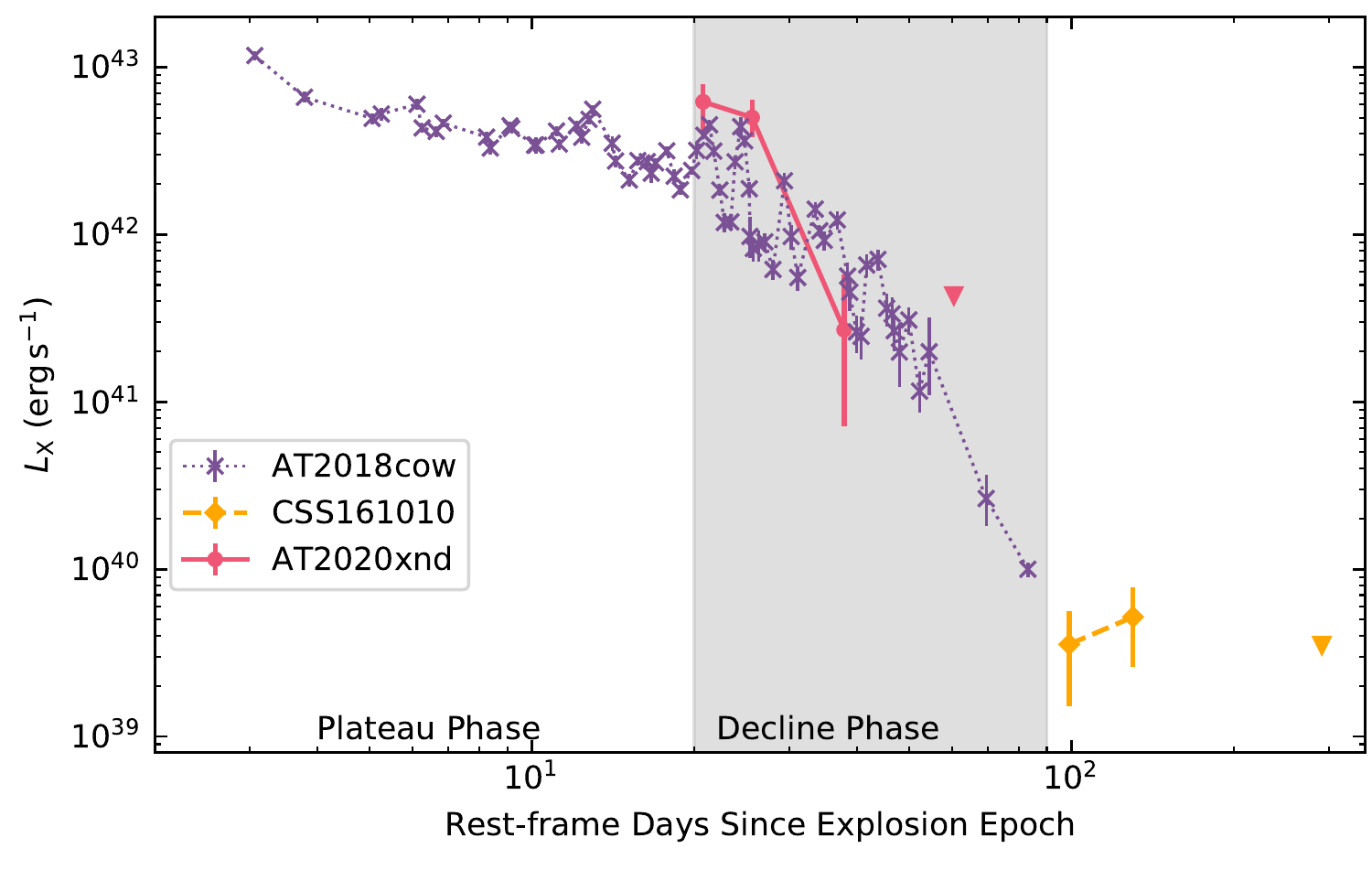}
    \caption{The 0.3--10\,keV X-ray light curve of AT2020xnd from \chandra\ compared to the 0.3--10\,keV X-ray light curves of AT2018cow \citep{RiveraSandoval2018,Kuin2019,Margutti2019,Ho2019cow} and CSS161010 \citep{Coppejans2020}. The grey shaded region marks the ``decline phase'' delineated in \citet{Ho2019cow}.
    The luminosity of AT2020xnd is similar to that of AT2018cow at the same phase, and we see tentative evidence of the same steep decline, although the data are significantly more sparse. Note that the AT2020xnd observations have no cosmological correction applied. \label{fig:xray-lc}}
\end{figure*}

In addition to radio and millimeter observations,
AT2020xnd was observed over $\Delta t=20$--150\,d with \chandra\ \citep{Matthews2020}.
We retrieved the observations from the \emph{Chandra} data archive and analyzed them with the procedure described in Appendix~\ref{sec:appendix-xray}.
The light curve is shown in Figure~\ref{fig:xray-lc}.
The peak luminosity of $7\times10^{42}\,\erg\,\psec$ (observer-frame) is almost identical to that of AT2018cow at the same epoch \citep{RiveraSandoval2018,Kuin2019,Margutti2019,Ho2019cow}.
CSS161010 was also detected in X-rays, but only at $\Delta t>100\,$d \citep{Coppejans2020}.

\section{Analysis}
\label{sec:analysis}

In \S\ref{sec:observations} we presented millimeter, radio, and X-ray observations of AT2020xnd.
In this section we use the data to derive basic properties of the forward shock.
We consider the origin of the X-rays separately (\S\ref{sec:xray}).
As discussed in \S\ref{sec:observations},
the evolution of AT2020xnd appears to proceed in two stages:
an early stage ($\Delta t<40\,$d) when the spectral index from 79\,\ghz\ to 94\,\ghz\ is relatively flat and unchanging,
and a later stage ($\Delta t>70\,$d) when the SED clearly cascades down both in flux and frequency.
We begin by considering the later stage, because the behavior is similar to what has been seen in previous events.

\subsection{Late Stage ($\Delta t>70\,\days$)}
\label{sec:model-late-stage}

To model the late-time centimeter-wavelength data,
we follow the standard approach for non-relativistic SNe \citep{Chevalier1998,Kulkarni1998,Soderberg2005-03L}.
We assume that the SEDs arise from synchrotron self-absorption of non-thermal electrons shock-accelerated into a power-law energy distribution of index $p=3$ down to a minimum Lorentz factor $\gamma_m$.
The same framework has been applied to AT2018cow and analogs to find shock speeds ranging from $v=0.1c$ (AT2018cow; \citealt{Ho2019cow,Margutti2019}) to $v=0.6c$ (CSS161010; \citealt{Coppejans2020}).
We perform a basic cosmological correction to the flux density measurements by dividing the observed values by a factor of $(1+z)$.

The assumptions behind this framework are summarized in Appendix~\ref{sec:appendix-model}.
We note that the standard equations in the literature \citep{Chevalier1998} assume that the synchrotron self-absorption (SSA) frequency is below the cooling frequency, $\nu_a < \nu_c$.
This is not necessarily valid:
for AT2018cow $\nu_a > \nu_c$ at early times \citep{Ho2019cow}, as a consequence of a large amount of energy being injected into a small volume of material, a regime selectively probed by high-frequency observations.
In Appendix~\ref{sec:appendix-model} we provide the corrected equations for the regime of $\nu_a > \nu_c$.

We model the SED as a broken power law. Following \citet{Granot2002_shape_spectral_breaks} we have:

\begin{equation}
    f_\nu = f_p \left[ \left( \frac{\nu}{\nu_{p}} \right)^{-s\beta_1} + \left( \frac{\nu}{\nu_{p}} \right)^{-s\beta_2} \right]^{-1/s}
\end{equation}

\noindent where $f_p$ and $\nu_p$ are the peak flux and peak frequency respectively, $\beta_1$ and $\beta_2$ are the spectral indices on either side of the break, and $s$ is a smoothing parameter.
We further assume that the peak flux and peak frequency evolve as power laws in time, with $f_p \propto t^{\alpha_1}$ and $\nu_p \propto t^{\alpha_2}$.
We begin by assuming that the peak is governed by SSA with an optically thick spectral index $\beta_1=5/2$ \citep{RL} and optically thin spectral index $\beta_2=-1$,
where $\beta_2=-(p-1)/2$ in the slow-cooling regime $\nu<\nu_c$.
We assume $p=3$.

For the fit, we must consider the effects of scintillation.
Radio point sources can exhibit significant variability in their centimeter-wavelength light curves due to inhomogeneities in the interstellar medium \citep{Rickett1990,Narayan1992,Walker1998}.
The light curves and SEDs of AT2020xnd are fairly smooth (\S\ref{sec:observations}), with the possible exception of the early-time 10\,GHz light curve (Figure~\ref{fig:radio-lc}) and the 6\,GHz flux density values in the SEDs (Figure~\ref{fig:radio-sed}).

The NE2001 model \citep{Cordes2002} predicts that the transition frequency in the direction of AT2020xnd is 9\,GHz,
and that the maximum source size subject to scintillation is 3--4\,$\mu$as \citep{Walker1998}.
Later in this section, we find $R\approx3\times10^{16}\,\cm$ at these epochs, which corresponds to $\theta\approx2\,\mu$as.
So, we conclude that observations with $\nu_\mathrm{obs}\lesssim 9\,\ghz$ could be affected by scintillation; at the transition frequency variations could be of order unity.
We therefore leave out the 6\,\ghz\ data points in our fitting.

\begin{figure*}[!htb]
    \centering
    \includegraphics{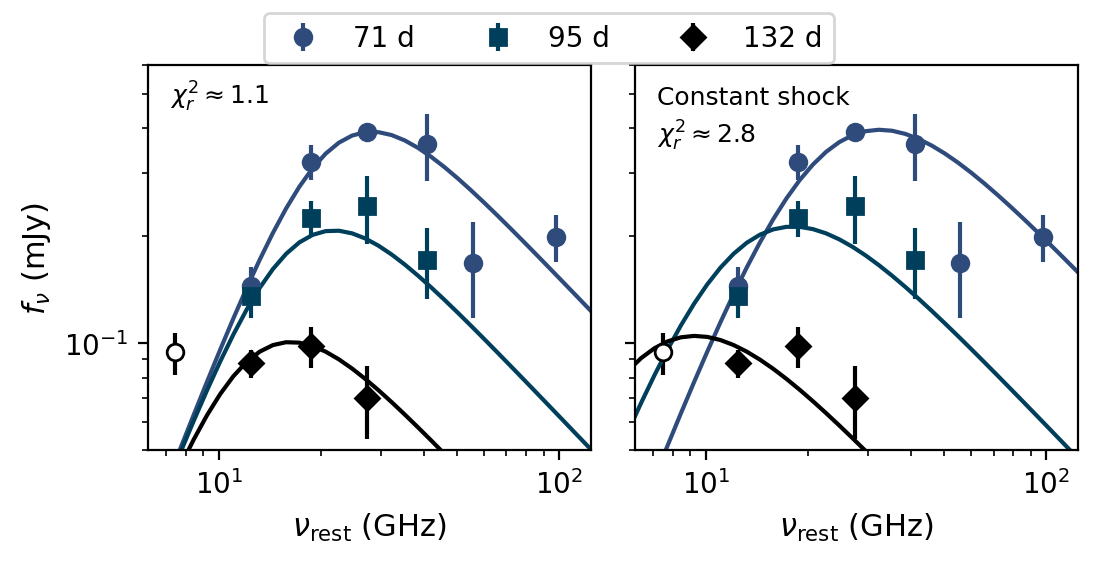}
    \caption{Broken power-law fit to the late-time data.
    Each color/symbol combination corresponds to a different observer-frame epoch, the same as in Figure~\ref{fig:radio-sed}.
    The single point below 9\,GHz is excluded from the fit due to possible scintillation, indicated with an unfilled circle. We assume that the peak flux and peak frequency also evolve as a power law in time, that the optically thick spectral index is $\beta=5/2$, and that the optically thin spectral index is $\beta=-1$, where $f_\nu \propto \nu^{\beta}.$ For the fit shown in the right-hand panel we also assume that the shock speed is constant, $R\propto t$. The flux density values have a basic cosmological correction applied, and frequency values are reported in the rest-frame.}
    \label{fig:vla-powlaw}
\end{figure*}

The resulting fit is shown in the left panel of Figure~\ref{fig:vla-powlaw}.
Using \texttt{curve\_fit} in \texttt{scipy},
we find that $f_p=0.68\pm0.08\,$mJy and $\nu_p=22\pm1\,\ghz$ at 58\,d in the rest-frame,
$\alpha_1=-2.2\pm0.1$, $\alpha_2=-0.88\pm0.20$, and $s=1.0\pm0.2$.
The reduced $\chi^2=1.1$ with $N=8$ degrees of freedom.
The corresponding forward-shock properties (using Equations C5 and C6 in Appendix~\ref{sec:appendix-model}) are $R\approx2\times10^{16}\,\cm$ and $B\approx0.9\,\gauss$,
with $R\propto t^{-0.2}$ and $B\propto t^{-0.7}$.
The magnetic field strength is close to what was observed for SN\,2003L \citep{Soderberg2005-03L} and SN\,2003bg \citep{Soderberg2006c-03bg}.

The constant or even decreasing radius we inferred above is not consistent with our assumption of an outwardly propagating shock.
So, we fit the same data fixing the shock speed to be constant
(a near-constant shock speed was observed in AT2018cow; \citealt{Margutti2019,Ho2019cow,Nayana2021}).
The results are shown in the right panel of Figure~\ref{fig:vla-powlaw}.
We find $f_p=0.83\pm0.11$, $\nu_p=23\pm1$, $\alpha_1=-2.1\pm0.1$, and $s=0.78\pm0.13$.
The reduced $\chi^2=2.8$ with $N=9$ degrees of freedom.
The magnetic field strength goes as $B \propto t^{-1.8}$.
This solution is also not physical:
the corresponding density profile is 
$n_e \propto B^2 \propto t^{-3.6} \propto R^{-3.6}$,
and the standard model does not apply to such a steep density profile ($k\geq3$ where $\rho \propto r^{-k}$).

\begin{figure}[htb!]
    \centering
    \includegraphics[width=\columnwidth]{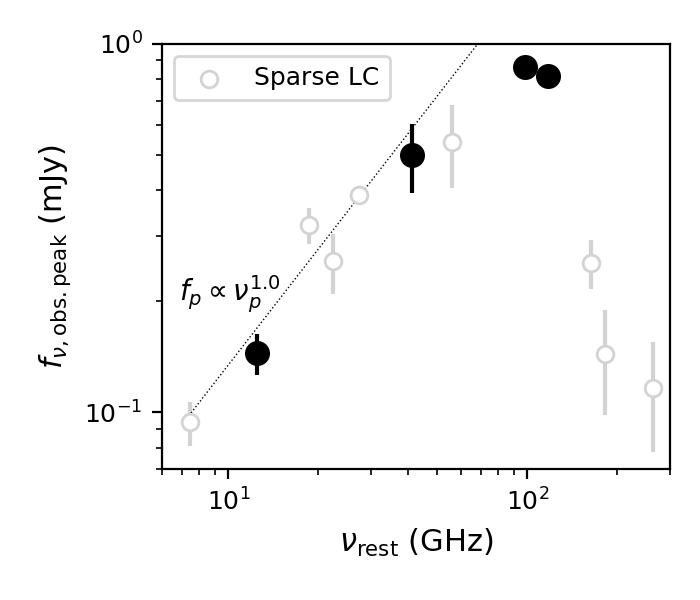}
    \caption{The peak observed flux density of the light curve at different frequencies. Filled points are from frequencies with a well-sampled light-curve peak (94\,\ghz, 79\,\ghz, 33\,\ghz, and 10\,\ghz). Empty points are from light curves that do not have a well-sampled peak.}
    \label{fig:fpeak-evol}
\end{figure}

Allowing the shock to be mildly decelerating (e.g., $R\propto t^{0.8}$, the value observed in CSS161010; \citealt{Coppejans2020}) results in a shallower density profile.
We can estimate the density profile for different rates of shock deceleration using the peak flux density of the light curve at each observing frequency, shown in Figure~\ref{fig:fpeak-evol}.
Including only points below 90\,\ghz, we find $f_p \propto \nu_p^{1.0\pm0.1}$.
Combining Equations C14 and C15, for this value of $d\ln(f_p)/d\ln(\nu_p)$ we find

\begin{equation}
    k=\frac{-20+54\alpha_r}{10\alpha_r}
\end{equation}

\noindent where $\alpha_r$ is defined as $R \propto t^{-\alpha_r}$. So, a constant-velocity shock $\alpha_r=1$ corresponds to $k=3.4$,
while a mildly decelerating shock $\alpha_r=0.8$ corresponds to $k=2.9$. For a wind profile $k=2$ we would require $\alpha_r=0.6$.
In summary, we cannot robustly constrain the hydrodynamics of the shock using our late-time VLA data alone.
However, under the reasonable physical assumption of a mildly decelerating shock the data could be explained by a medium with a steep density profile.

We can use our single-epoch estimates of $R$ and $B$ at 71\,\days\ in the observer frame (58\,\days\ in the rest frame) to estimate the mean velocity of the shock $v$, the total thermalized energy $U$, and the ambient density $n_e$.
Following the standard approach to modeling radio SNe \citep{Kulkarni1998,Soderberg2006b-06aj,Soderberg2005-03L,Horesh2013,Chevalier2006,Soderberg2010} we assume equipartition, $\epsilon_e=\epsilon_B=1/3$.
We find that the mean velocity $v\approx0.15c$: like the other AT2018cow analogs, fast but subrelativistic.
From Equation~12\footnote{We do not include a filling factor here.} in \citet{Ho2019cow} we have

\begin{equation}
    U = \frac{1}{\epsilon_B} \frac{4\pi}{3} R^3 \frac{B^2}{8\pi} \approx 2 \times 10^{48}\,\erg.
\end{equation}

\begin{figure*}[!htb]
    \centering
    \includegraphics{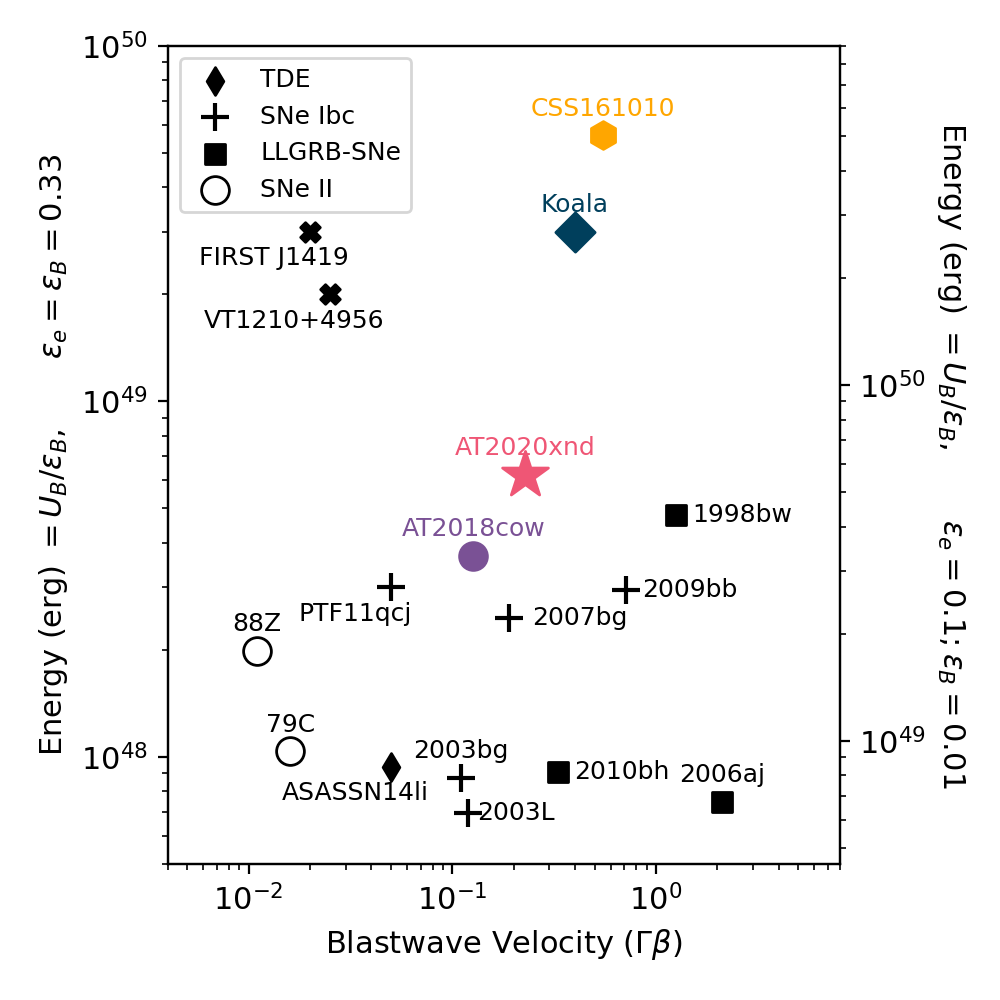}
    \caption{AT2020xnd (star) in velocity-energy space compared to other classes of radio-luminous transients: TDEs (filled diamonds; \citealt{Alexander2016}), Ibc supernovae (crosses; \citealt{Corsi2014,Soderberg2005-03L,Soderberg2006c-03bg,Salas2013,Soderberg2010}), SNe associated with LLGRBs (filled squares; \citealt{Kulkarni1998,Soderberg2006a-radio,Margutti2013}), Type II supernovae (open circles; \citealt{vanDyk1993,Weiler1986,Weiler1991}), and two luminous radio transients identified in radio survey data (filled `X'; \citealt{Law2018,Mooley2022,Dong2021}). For reference, GRBs lie above the plot at $10^{50}\,\erg < U < 10^{52}\,\erg$, and the relativistic TDE \emph{Swift} J1644+57 \citep{Zauderer2011,Berger2012,Eftekhari2018} lies at $10^{51}\,\erg$ in this framework.
    Transients similar to AT2018cow are shown as colored points: CSS161010 \citep{Coppejans2020}, AT2018lug (the ``Koala''; \citealt{Ho2020b}), and AT2018cow \citep{Ho2019cow,Margutti2019}.
    For more details see Appendix C in \citet{Ho2019cow}.}
    \label{fig:vel-e}
\end{figure*}

\noindent This is very similar to the value of $U$ found for AT2018cow at $\Delta t=22\,$d.
As shown in Figure~\ref{fig:vel-e}, AT2018cow and its analogs have very high measured energies compared to other subrelativistic cosmic explosions, with the exception of FIRST\,J1419 \citep{Law2018,Mooley2022} and VT1210+5946 \citep{Dong2021}.

To estimate the ambient density, we assume that the number densities of protons and electrons are equal ($n_e=n_p$) and that the medium is composed of fully ionized hydrogen, so that $n_e = \rho/(\mu_p m_p)$ where $\mu_p=1$. We therefore have

\begin{equation}
    n_e = \frac{B^2}{16 \pi \epsilon_B m_p v^2} \approx 4 \times 10^{3}\,\pcmcub.
\end{equation}

\noindent At a similar epoch (70\,d) an ambient density of $n_e=50\,\pcmcub$ was measured for CSS161010.

\begin{figure*}[!htb]
    \centering
    \includegraphics{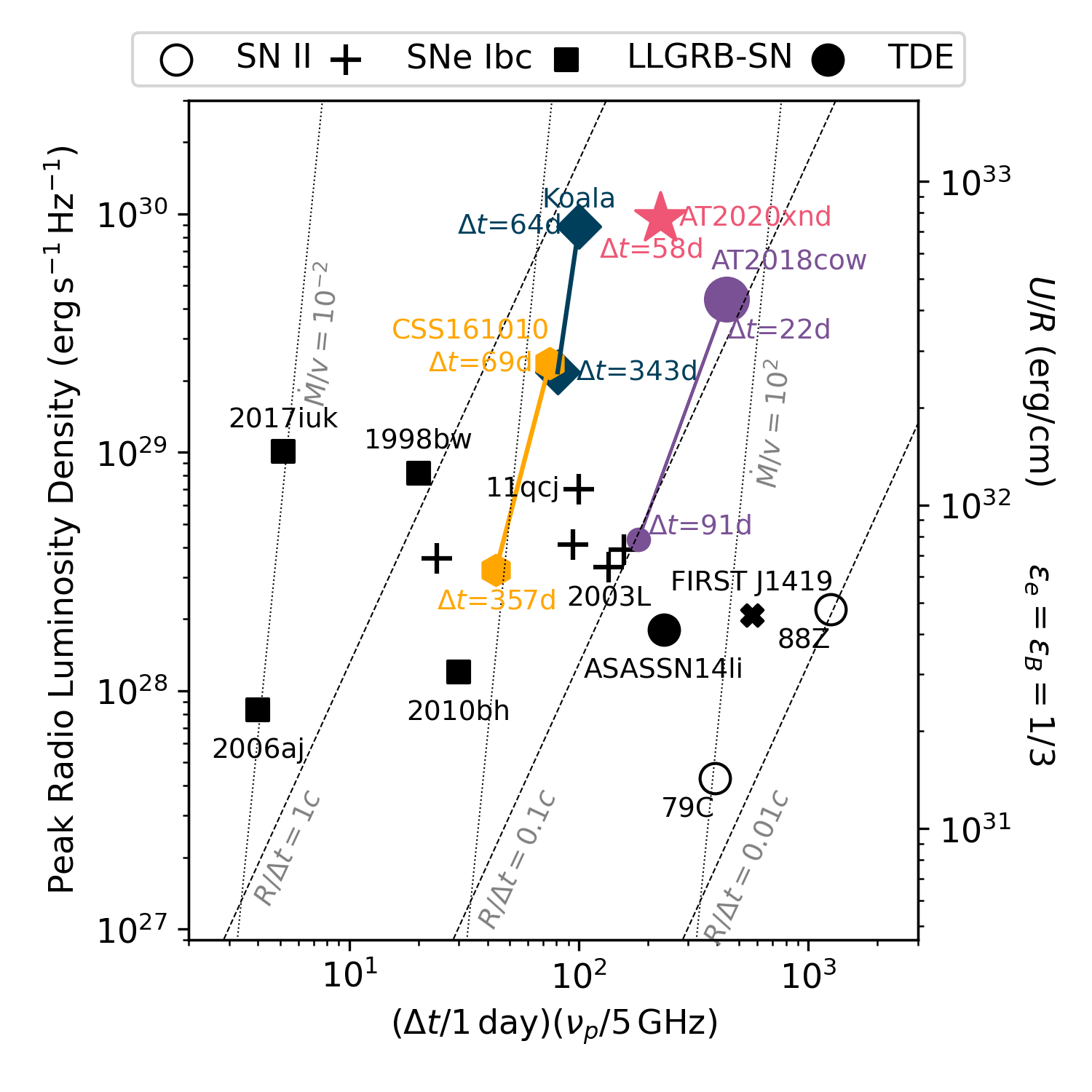}
    \caption{The peak luminosity of AT2020xnd and other AT2018cow-like explosions on two different epochs, compared to classes of energetic transients (cf. \citealt{Chevalier1998,Soderberg2010,Ho2019cow}).
    Lines of constant mass-loss rate (scaled to wind velocity) are shown in units of $10^{-4}\,M_\odot\,\pyr/1000\,\km\,\psec$. Note that the dotted lines assume that the radio peak is due to synchrotron self-absorption.
    Values for AT2020xnd are from this work.
    Other values are from \citet{Ho2019cow,Ho2020b,Coppejans2020,Margutti2019,Corsi2014,Soderberg2010,Kulkarni1998,Soderberg2006b-06aj, Margutti2013, Horesh2013,Krauss2012,Salas2013,Soderberg2005-03L,Soderberg2006c-03bg,vanDyk1993,Weiler1986}.}
    \label{fig:lum-tnu}
\end{figure*}

Assuming a steady wind,
we can convert the ambient density to a mass-loss rate $\dot{M}$, where

\begin{equation}
    \dot{M} = n_e 4 \pi m_p r^2 v_w
\end{equation}

\noindent and $v_w$ is the velocity of the wind. Taking $v_w=1000\,\km\,\psec$ we have $\dot{M}\approx2\times10^{-4}\,\msol\,\pyr$, while for $10\,\km\,\psec$ we have $2\times10^{-6}\,\msol\,\pyr$.
The inferred velocity and $\dot{M}$ are shown in Figure~\ref{fig:lum-tnu} compared to other energetic explosions.

We can use the shock speed to estimate the minimum Lorentz factor of the electrons $\gamma_m$,

\begin{equation}
    \gamma_m = 1+ \frac{1}{2}\left( \frac{p-2}{p-1} \right) \epsilon_e \frac{m_p}{m_e} \frac{v^2}{c^2} \approx 4.
\end{equation}

\noindent From $\gamma_m$ we can estimate the characteristic synchrotron frequency $\nu_m$,
the frequency of electrons whose Lorentz factor is $\gamma_m$:

\begin{equation}
    \nu_m = \gamma_m^2 \nu_g
\end{equation}

\noindent where

\begin{equation}
    \nu_g = \frac{e B}{2 \pi m_e c}
\end{equation}

\noindent We find $\nu_m \approx 0.05\,\mathrm{GHz}$ which is below our observing frequencies. Finally, we can estimate the cooling frequency $\nu_c$, where

\begin{equation}
    \nu_c = \gamma_c^2 \nu_g
\end{equation}

\noindent and 

\begin{equation}
    \gamma_c = \frac{6 \pi m_e c}{\sigma_T B^2 t}.
\end{equation}

\noindent We find $\nu_c = 100\,\ghz$, significantly above the VLA frequencies, and consistent with our assumption that $\nu_a < \nu_c$ at late times. The forward-shock properties at 71\,d in the observer frame are summarized in Table~\ref{tab:day58-mesaurements}.

\begin{table}[!ht]
    \centering
    \begin{tabular}{c|c}
    \hline\hline
       Parameter  & Value \\
       \hline
       $\nu_a=\nu_p$ (GHz) & $22\pm1$ \\
       $F_{\nu,p}$ (mJy) & $0.68\pm0.08$ \\
       $R$ $(10^{16}\,\cm)$ & $2.2\pm0.2$ \\
       $B$ $(\gauss)$ & $0.87\pm0.04$ \\
       $v/c$ & $0.15\pm0.01$ \\
       $U$ $(10^{48}\,\erg$) & $2.1\pm0.5$ \\
       $n_e$ ($10^{3}\,\pcmcub$) & $3.7\pm0.6$ \\
       $\nu_c$ (GHz) & $100\pm5$ \\
       \hline
    \end{tabular}
    \caption{Quantities derived from measurements on Day 71 (observer-frame), under the standard assumption that the electron energy distribution is a power-law and that the SED peak is governed by synchrotron self-absorption. We assume equipartition, $\epsilon_e=\epsilon_B=1/3$. We provide formal errors from the fit, but caution that the uncertainties on these parameters are dominated by systematics and by our assumptions.}
    \label{tab:day58-mesaurements}
\end{table}

\subsection{Early Stage}
\label{sec:model-early-stage}

In \S\ref{sec:model-late-stage} we modeled the late-time low-frequency emission assuming a power-law distribution of electrons and a radio SED governed by SSA,
a standard approach to modeling radio SNe that has been applied to AT2018cow \citep{Ho2019cow,Margutti2019}, AT2018lug \citep{Ho2020b}, and CSS161010 \citep{Coppejans2020}.
For both AT2018cow and AT2020xnd, however, the SSA model derived from low-frequency late-time observations is not consistent with the early millimeter-wave observations. \citet{Margutti2019} suggested that the early millimeter emission might arise from a separate component like a reverse shock, and \citet{Ho2019cow} suggested that during the mm-bright phase the shock was passing through higher-density material that terminated abruptly, resulting in a rapid decay in both flux and frequency.
In this section we consider several possibilities for the origin of the high-frequency emission from AT2020xnd observed with NOEMA at $\Delta t<50\,\days$:
continuous shock-acceleration with a non-thermal (\S\ref{sec:model-ssa-powerlaw}) or thermal (\S\ref{sec:model-thermal}) electron energy distribution,
and non-steady-state particle injection  (\S\ref{sec:model-adiabatic-expansion}).

\subsubsection{Continuous Shock-Acceleration+SSA+Power-Law Electron Energy Distribution}
\label{sec:model-ssa-powerlaw}

\begin{figure}
    \centering
    \includegraphics[width=\columnwidth]{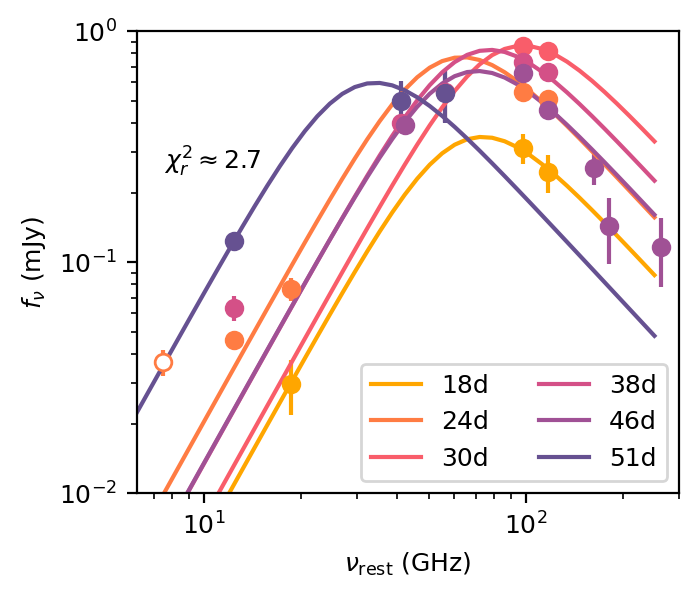}
    \caption{Broken power-law fits to the early-time millimeter and radio data of AT2020xnd. Each epoch is fit independently. We assume an optically thick spectral index of $\beta=5/2$, the expectation for synchrotron self-absorption of a non-thermal electron population, although as discussed in the text it is more likely that a thermal electron population contributes significantly to the emission at these stages.}
    \label{fig:powlaw-fits-early}
\end{figure}

First, we apply the same framework used to model the late-time low-frequency data in \S\ref{sec:model-late-stage}:
continuous shock-acceleration resulting in the acceleration of electrons into a power-law energy distribution,
with the peak in the SED governed by synchrotron self-absorption.
We find that a power-law evolution in the peak flux and frequency does not do a good job of describing the data.
So, we treat each epoch of 79\,\ghz+94\,\ghz\ data independently, fixing $\beta_1=5/2$ and $\beta_2=-1.5$ or $\beta_2=-1$ (depending on whether we find the SED to be in the slow- or fast-cooling regime).
The fits are shown in Figure~\ref{fig:powlaw-fits-early} and the corresponding physical parameters are listed in Table~\ref{tab:early-SSA}.
We caution that only epochs 38\,d and 46\,d have reasonably well-sampled SEDs; the other fits should be regarded as lower limits on the peak frequency. If the optically thick spectral index is shallower (as appears to be the case), the peak would be at a higher frequency and the inferred radius and velocity would be lower. In Table~\ref{tab:early-SSA} we also provide a limit based on $\nu_p\lesssim100\,\ghz$.

\begin{table*}[!ht]
    \centering
    \begin{tabular}{c|c|c|c|c|c}
    \hline\hline
         & 18\,d & 24\,d & 30.3\,d & 38\,d & 46\,d \\
      \hline
      $\nu_a=\nu_p$ (GHz) & 60--100 & 50--100 & 70--100 & $61.7\pm1.2$ & $52.6\pm2.2$  \\
      $F_{\nu,p}$ (mJy) & 0.3--0.6 & 0.6--1.2 & 0.9--1.6 & $1.38\pm0.04$ & $1.05\pm0.04$  \\
      $R$ $(10^{16}\,\cm)$ & 0.3--0.8 & 0.5--1.3 & 0.6--1.0 & $1.11\pm0.03$ & $1.14\pm0.05$ \\
      $B$ $(\gauss)$ & 2.4--4.3 & 1.9--4.0 & 2.6--3.9 & $2.28\pm0.04$ & $2.00\pm0.08$  \\
      $v/c$ & 0.09--0.2 & 0.09--0.3 & 0.09--0.2 & $0.138\pm0.003$ & $0.119\pm0.005$ \\
      $U$ $(10^{48}\,\erg$) & 0.2--0.7 & 0.4--1.8 & 0.7--1.8 & $28.8\pm1.7$ & $1.5\pm0.2$ \\
      $n_e$ ($10^{3}\,\pcmcub$) & 16--270 & 5.5--200 & 30--210 & $19.8\pm0.4$ & $29.7\pm3.7$ \\
      $\nu_c$ (GHz) & 12--71 & 9.5--95 & 6.3--20 & $31.6\pm0.8$ & $20.6\pm0.9$ \\
      \hline
    \end{tabular}
    \caption{Quantities derived from early epochs that have 79\,\ghz+94\,\ghz\ NOEMA observations, under the standard assumption that the SED peak is governed by synchrotron self-absorption and that the electrons are accelerated into a power-law energy distribution.
    We assume equipartition, $\epsilon_e=\epsilon_B=1/3$, and $p=3$. Epochs are listed in the observer-frame. Only epochs 38\,d and 46\,d have well-sampled SEDs. We provide formal uncertainties from our fits, but caution that the true uncertainties are dominated by systematics and our assumptions.}
    \label{tab:early-SSA}
\end{table*}

With our sparsely sampled SEDs, we cannot precisely measure the physical properties of the forward shock from our early observations. However, particularly from the observations at 38\,d and 46\,d, it appears that the density is an order of magnitude higher than at 71\,d, consistent with our inference in \S\ref{sec:model-late-stage} of a steep $n_e \propto r^{-3}$ density profile. The shock speed of $v\approx 0.2$ is similar to our measurement at 71\,d.
We conclude that if the framework presented in this section is correct,
then the shock likely propagated through a particularly high-density region,
and that the density began decreasing abruptly as $n_e \propto r^{-3}$ at $\Delta t\approx50$--60\,\days. This is very similar to the conclusion drawn from the 230\,\ghz\ light curve of AT2018cow, which plateaued for $\approx 50\,\days$ before abruptly declining \citep{Ho2019cow}.
A density profile steeper than a steady wind was inferred for CSS161010 \citep{Coppejans2020}, and---as discussed for that object---implies non-steady mass-loss \citep{Smith2014}.

\subsubsection{Continuous Shock-Acceleration+SSA+Thermal Electron Energy Distribution}
\label{sec:model-thermal}

In \S\ref{sec:model-ssa-powerlaw} we modeled the early high-frequency emission with the same framework used to model the late-time low-frequency emission in \S\ref{sec:model-late-stage}.
We found that at early times the shock likely propagated through a region of high density ($\approx 10^{4}\,\pcmcub$),
and that the density decreased abruptly after 50\,d.
In this section we consider the possibility that instead,
the assumption of all electrons being accelerated into a power-law distribution is incorrect:
that there was a significant population of electrons remaining in a thermal distribution, i.e., a relativistic Maxwellian.

Our primary motivation for considering a thermal population is the steep spectral index observed at 46\,d.
From five NOEMA data points, we measure $\beta=-2.0\pm0.2$, which corresponds to $p=4.0\pm0.5$ in the fast-cooling regime.
In the test particle limit, diffuse shock acceleration predicts $p=2$ for non-relativistic shocks (e.g.,  \citealt{Blandford1987}).
Radio SNe are often inferred to have steeper electron power law indices ($p=3$; e.g., \citealt{Soderberg2005-03L}).
Deviations from $p=2$ may be expected from non-linear effects (departure from the test-particle approximation). For example, \citet{Caprioli2020} recently suggested that self-generated Alfven waves downstream of the shock can enhance particle advective losses and thus steepen the spectrum.

However, to our knowledge, inferred values as steep as $p=$3.5--4 are unusual.
Reviewing the literature, we identified only a handful of events with measured values of $p\geq3.5$.
One is AT2018cow itself: at $\Delta t=10\,$d the spectral index across the SMA observing bands was $\beta=-1.86\pm0.03$, or $p=3.72\pm0.06$ in the fast-cooling regime.
Another is CSS161010: at $\Delta t=99\,\days$, \citet{Coppejans2020} measure $p=3.5^{+0.4}_{-0.1}$.
Finally, 
the ultra-long GRB\,130925A \citep{Horesh2015} had $\beta_2=1.4\pm0.1$, corresponding to $p=3.8\pm0.2$.

\begin{figure*}[!htb]
    \centering
    \includegraphics{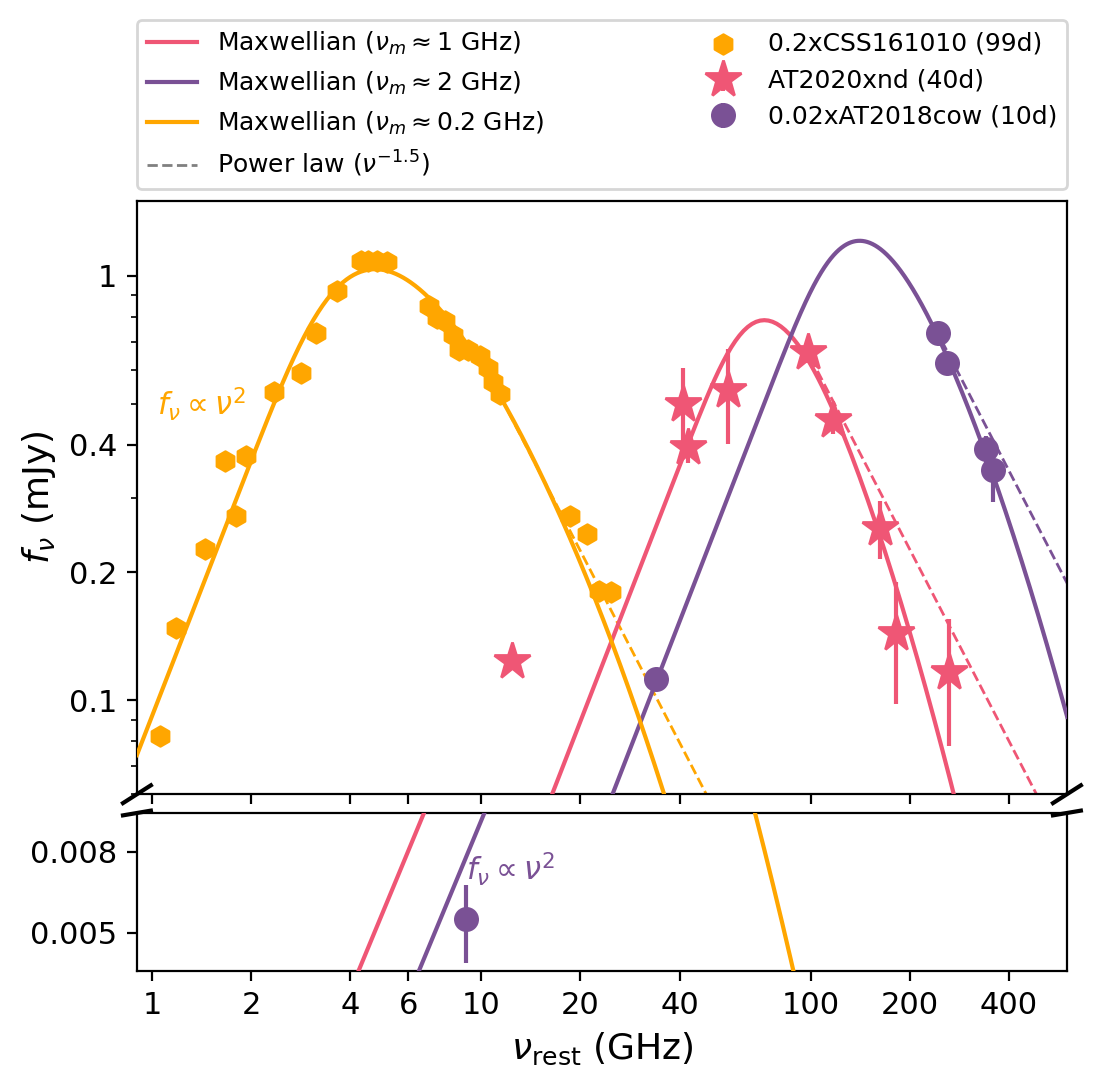}
    \caption{Radio and millimeter SEDs of cosmic explosions in the literature with inferred values of $p\geq3.5$: AT2018cow (circles; \citealt{Ho2019cow}), CSS161010 (hexagons; \citealt{Coppejans2020}), and AT2020xnd (stars; this paper).
    Each SED is well-described by a a self-absorbed relativistic Maxwellian (i.e., $\nu_T<\nu_a$), shown as solid lines. For reference, we show a ``limiting case'' power-law with $\beta=-1.5$ as a dotted line:
    such a steep power law is already difficult to explain in the context of diffuse shock-acceleration theory.
    For clarity, the light curves of AT2018cow and CSS161010 have been scaled in flux, by factors of 0.02 and 0.2 respectively.}
    \label{fig:maxwellian}
\end{figure*}

\citet{Horesh2015} argued that the steep frequency cutoff observed in GRB\,130925A could reflect an underlying steep cutoff in the electron energy distribution, and that a mono-energetic distribution was a better match to the data.
Here we consider whether a Maxwellian energy distribution \citep{Eichler2006} could explain the SED of AT2020xnd, which is physically better motivated than a mono-energetic distribution,
as well as the other events with steep spectra (Figure~\ref{fig:maxwellian}).
Interestingly, we note that CSS161010 had an observed optically thick power-law index of $f_\nu \propto \nu^{2}$, which is an expectation of a thermal rather than non-thermal electron distribution.

The effect of having a thermal electron population in addition to a power-law population has been considered by various authors, primarily in the context of light curves and spectra of GRB afterglows \citep{Giannios2009,Ressler2017,Johannesson2018,Warren2018}.
In general, the Maxwellian adds an excess of flux close to the characteristic synchrotron frequency of the thermal electrons $\nu_T$ (Equation~\ref{eq:nuT}), which results in a steeper spectrum above the peak frequency (an exponential) that eventually reconnects to the power-law.
Over time, the characteristic frequency of this additional component can decrease if the shock decelerates.
At $\nu \gg \nu_T$, following \citet{Mahadevan1996} the spectrum takes the form $f_\nu \propto \nu e^{-1.8899x^{1/3}}$ where $x=\frac{2\nu}{3\nu_T}$.
We use our NOEMA data to analytically estimate $\nu_T$. The local spectral index is $\beta = d\ln{f_\nu}/d\ln{\nu} \approx 1 - (1/3)1.8899x^{1/3}$. So, $\nu_T \approx (2/3)\nu \left[ 3(1-\beta)/1.8899 \right]^{-3}$. Since the spectral index we measure is close to $-2$, we find $\nu_T \approx 0.6\,\ghz$.

Given that $\nu_T\sim1\,\ghz$, and that we observe a steep optically thick spectral index from 10\,\ghz\ to 100\,\ghz\ (instead of the $\nu^{1/3}$ expected for a Maxwellian; \citealt{Mahadevan1996}) we conclude that if the emission is from a thermal population then it must be absorbed: the frequency of the peak of the SED, $\mathbf{\nu_\mathrm{peak}}$, is set by the synchrotron self-absorption frequency, $\mathbf{\nu_\mathrm{peak}=\nu_a}$.
A Maxwellian with synchrotron self-absorption has the form

\begin{equation}
\label{eq:maxwellian}
    f_\nu = f_m \, \left(\frac{\nu}{\nu_T}\right)^2
    \left\{ 1-\exp\left[-\tau_m \left(\frac{\nu}{\nu_T}\right)^{-1} I\left(\frac{2\nu}{3\nu_T}\right)\right]\right\}
\end{equation}

\noindent where the function $I(x)$ is given by \citet{Mahadevan1996},

\begin{equation}
\label{eq:I_of_x}
    I(x) \approx 2.5651 \left(1+\frac{1.92}{x^{1/3}} + \frac{0.9977}{x^{2/3}}\right) e^{-1.8899 x^{1/3}} ,
\end{equation}

\noindent $f_m$ is a scaling constant (it is the flux density at frequency $\nu_T$),
and $\tau_m$ is related to the SSA optical depth at this frequency (up to a factor $I(2/3) \approx 2.2$). 

In Figure~\ref{fig:maxwellian} we show data from AT2020xnd, AT2018cow, and CSS161010, fit with Equation~\ref{eq:maxwellian}, and provide the best-fit parameters in Table~\ref{tab:maxwellian}.
A more general treatment of the self-absorbed Maxwellian model is presented in separate work by \citet{Margalit2021}.
We find that the model describes the data well: in particular, it reproduces the $f_\nu \propto \nu^2$ optically thick spectral index observed in CSS161010 and AT2018cow,
as well as the steep observed optically thin spectral index observed in all three events.
For reference, in Figure~\ref{fig:maxwellian} we also show a power-law with $f_\nu \propto \nu^{-1.5}$. The exponential cutoff does a better job of reproducing the data, and we note that such a steep power law would require $p\geq3$, which is difficult to explain in diffusive shock-acceleration theory.
For AT2020xnd, most likely explanation for the significant flux excess at 10\,GHz is the source geometry: a model of an inhomogeneous medium was successfully used to explain the shallow optically thick index observed in AT2018cow \citep{Nayana2021}.
We do not attempt to fit the spectrum of GRB\,130925A: as an ultra-relativistic event, 
the framework presented here and in \citet{Margalit2021} is not directly applicable.

\begin{table*}[!ht]
    \centering
    \begin{tabular}{c|c|c|c}
    \hline\hline
       Parameter  & AT2020xnd (40d) & AT2018cow (10d) & CSS161010 (99d) \\
       \hline
       $f_m$ (mJy) & 0.0003 & 0.04 & 0.03 \\
       $\tau_m$ & $6\times10^{4}$ & $2\times10^{4}$ & $7\times10^{2}$  \\
       $\nu_T$ (GHz) & 0.7 & 2 & 0.2 \\
       $v/c$ & 0.3 & 0.3 & 0.5 \\
       $B$ (G) & 1 & 4 & 0.04 \\ 
       $n_e$ (cm$^{-3}$) & $4\times10^{3}$ & $9\times10^{3}$ & 40 \\
       \hline
    \end{tabular}
    \caption{Quantities derived from fitting an absorbed relativistic Maxwellian to the SEDs of AT2020xnd, AT2018cow, and CSS161010 shown in Figure~\ref{fig:maxwellian}.}
    \label{tab:maxwellian}
\end{table*}

Our finding that the characteristic frequency of the thermal electrons $\nu_T$ is significantly below the observed peak frequency $\nu_T \ll \nu_\mathrm{peak}=\nu_a$, and that the thermal electrons could dominate the observed emission all the way up to a factor of $\nu \sim 10^2\times\nu_T$, may seem counterintuitive. We defer a detailed discussion of how this can be the case to \citet{Margalit2021}.
In summary, \citet{Margalit2021} show that if most of the electrons are in the thermal distribution, synchrotron emission from the thermal electrons can dominate most of the observed emission. They define a frequency $\nu_j$, the frequency at which the contribution from the thermal electrons equals the contribution from the power-law electrons: the thermal population dominates the emission at $\nu<\nu_j$ and the power-law population dominates the emission at $\nu>\nu_j$. They show that $\nu_j$ can be orders of magnitude larger than $\nu_T$ (in their notation, $\nu_\Theta=\nu_T$). Indeed, later in this section we directly constrain the ratio of power-law electrons to thermal electrons to be $\lesssim0.16$.

We now use our inferred Maxwellian parameters to estimate physical properties of the shock, summarized in Table~3.
The characteristic synchrotron frequency of thermal electrons $\nu_T$ is determined by the electron temperature $\Theta = k_b T_e / m_e c^2$ and the magnetic field strength,
\begin{equation}
\label{eq:nuT}
    \nu_T = \Theta^2 \frac{eB}{2\pi m_e c} .
\end{equation}
Assuming that the electron temperature is set by the post-shock energy density (i.e. that electrons are in equilibrium with the ions), we can relate this quantity to the shock velocity,
\begin{equation}
    \Theta \equiv 
    \frac{k_b T_e}{m_e c^2} \approx \frac{3 m_p v^2}{32 m_e c^2} \approx 15.5 \, \left(\frac{v}{0.3c}\right)^2
    .
\end{equation}

This implies that the thermal synchrotron frequency is
\begin{equation}
    \nu_T \approx 
    0.67\,\mathrm{GHz}\,
    \left( \frac{v}{0.3c} \right)^{4} \left( \frac{B}{1\,\gauss} \right) ,
\end{equation}

\noindent and depends sensitively on the shock velocity.
The flux density at this frequency within the SSA optically-thick regime ($\tau_m \gg 1$; as is applicable in our current situation) is simply given by the Rayleigh–Jeans limit

\begin{align}
    f_m &= \frac{8\pi^2 m_e R^2 \nu_T^2 \Theta}{4\pi D^2} 
    \\ \nonumber
    &\approx 
    7.45 \times 10^{-10} 
    \, {\rm mJy} \, 
    \left( \frac{B}{1\,\gauss} \right)^2 \left( \frac{v}{0.1c} \right)^{12} \left( \frac{t}{50\,\days} \right)^{2} ,
\end{align}
where $D$ is the luminosity distance ($1261 \, {\rm Mpc}$ for AT2020xnd), and $R \approx v t$ is the physical size (radius) of the emitting region. The latter approximation may be incorrect by a factor of a few if the geometry is aspherical, or if the blast-wave has been decelerating as a function of time.

Finally, we can estimate the ambient density from the optical depth parameter $\tau_m$, which is given by
\begin{align}
    \tau_m 
    &= 
    \frac{\pi e}{2 \sqrt{3}} \frac{n_e R}{\Theta^5 B}
    \\ \nonumber
    &\approx 
    3.73 \times 10^7 
    \, \left(\frac{n_e}{100 \, {\rm cm}^{-3}}\right) \left( \frac{B}{1\,\gauss} \right)^{-1} \left( \frac{v}{0.1c} \right)^{-9} \left( \frac{t}{50\,\days} \right).
\end{align}

For AT2020xnd, AT2018cow, and CSS161010, we find values of $v$, $B$, and $n_e$ that are physically realistic (Table~\ref{tab:maxwellian})
and quite similar to the values derived under the assumption of a pure power-law electron distribution.
The fact that the inferred parameters are similar under the thermal and power-law assumptions is not surprising; see Figure~2 of \citet{Margalit2021}.

\begin{figure}[!hbt]
    \centering
    \includegraphics{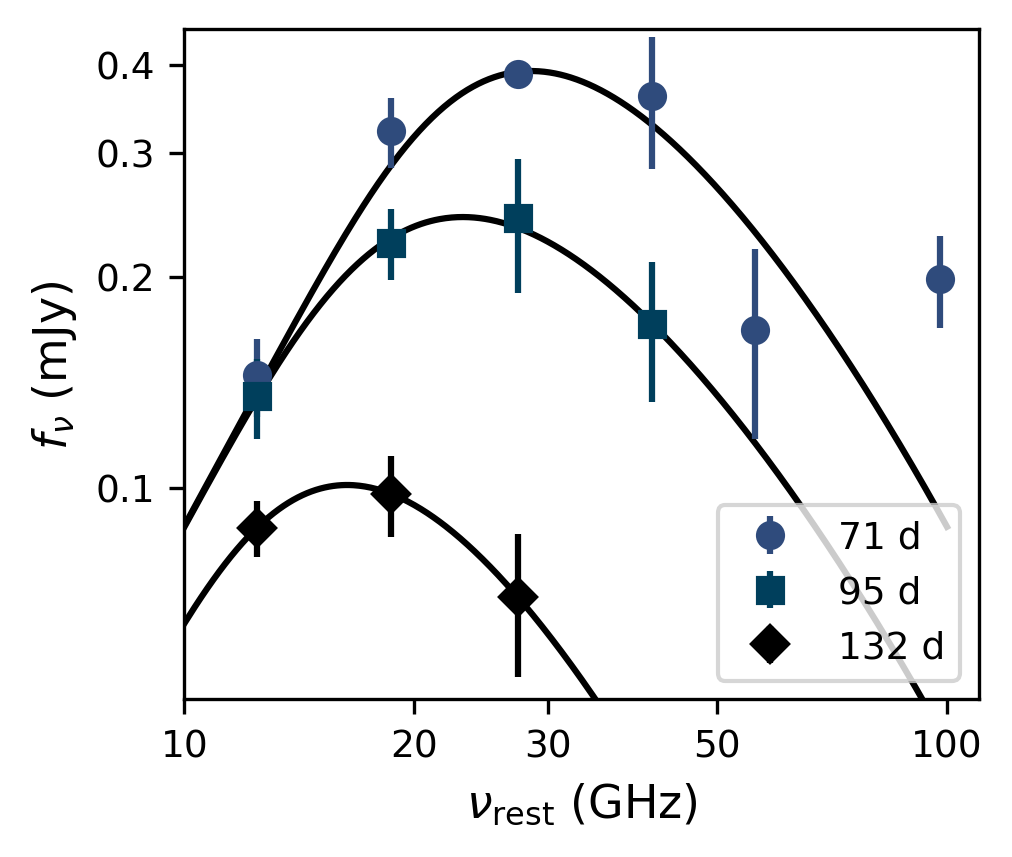}
    \caption{Maxwellian fits to late-time low-frequency VLA observations of AT2020xnd. Each color/symbol combination corresponds to a different observer-frame epoch, the same as in Figure~\ref{fig:radio-sed}. The flux density values have a basic cosmological correction applied and frequency values are reported in the rest-frame.}
    \label{fig:maxwellian-late}
\end{figure}

Finally, we consider whether the Maxwellian could also be used to describe the late-time data of AT2020xnd (from \S\ref{sec:model-late-stage}).
We fit each epoch independently and show the fits in
Figure~\ref{fig:maxwellian-late}, again excluding the 6\,\ghz\ measurement.
Fitting each epoch independently, we find $v\approx0.3c$, $B=0.2\,$G, and $n_e=840\,\pcmcub$ in the first epoch,
$v\approx0.2c$, $B=0.4\,$G, and $n_e=148\,\pcmcub$ in the second epoch,
and $v\approx0.2c$, $B=0.7\,$G, and $n_e=37\,\pcmcub$ in the third epoch.

With this model, we can directly constrain the fraction of electrons that were accelerated into a power-law distribution based on the fact that we do not observe a transition from an exponential to a power-law in the SED, the frequency defined as $\nu_j$ in \citet{Margalit2021}.
The frequency $\nu_j$ is directly related to the relative number of electrons in this power-law distribution to those in the thermal distribution.
Taking the data at 46\,d (rest-frame), we estimate that $\nu_j \gtrsim 200\,\ghz$.
The ratio of this transition frequency to the `thermal' synchrotron frequency is $x_j \equiv 2\nu_j/3\nu_T \gtrsim 190$.

Assuming $\Theta \gtrsim 1$, and that the minimum Lorentz factor of electrons within the putative power-law distribution is $\gamma_m = 3 \Theta$ (the mean Lorentz factor of thermal electrons), the ratio $\delta$ of energy in the power-law distribution versus the thermal distribution
determines the transition frequency $\nu_j$. Using the results of \citet{Margalit2021} we find that our observational constraint $x_j \gtrsim 190$ implies that $\delta \lesssim 0.16$ for any $2.2 \leq p \leq 3.2$ (with very weak $p$ dependence).
If the thermal electron population carries a fraction $\epsilon_T \sim 1$ of the total post-shock energy, then $\delta = \epsilon_e/\epsilon_T$ can be interpreted as $\sim \epsilon_e$, the fraction of post-shock energy that goes into accelerating non-thermal electrons.
Our interpretation above would constitute a novel constraint on this parameter.

If the Maxwellian model is correct, the question is why we are seeing a thermal electron distribution in this group of objects (AT2018cow, AT2020xnd, and CSS161010) and why this has not been inferred from radio observations of SNe.
\citet{Ho2019cow} showed that the luminous millimeter emission observed in AT2018cow implied a high ambient density. It is tempting to think that the high ambient density could lead to
electron collisions, which could in turn produce a thermal distribution.
However, as shown in Figure~\ref{fig:lum-tnu}, similarly high ambient densities have been observed in radio SNe (e.g., \citealt{Soderberg2005-03L,Dong2021}) with no evidence for a sharp high-frequency cutoff or a $f_\nu \propto \nu^{2}$ optically thick spectrum.
So, density cannot be the only important factor.
We defer a detailed discussion of the physical conditions under which a relativistic Maxwellian component is observable to \citet{Margalit2021}.
In summary, the prominence of the thermal population is primarily determined by the shock speed,
with a secondary dependence on the ambient density.
In other words, its prominence in AT2018cow, CSS161010, and AT2020xnd is due to the fact that these events have both faster shock speeds and higher ambient densities than most observed cosmic explosions.

\subsubsection{Non-steady-state Particle Acceleration}
\label{sec:model-adiabatic-expansion}

So far, we have been assuming that the millimeter emission arises from continuous shock-acceleration.
Indeed, the framework typically used to model SNe assumes that particle injection is in a steady state.
Beyond SN studies, however, other classes of radio sources show very steep spectral indices---such as active galactic nuclei (AGN) that are ``switched off'' (e.g., \citealt{Cohen2005,Shulevski2015}).
When particle acceleration is not in a steady state, the optically thin spectral index can be arbitrarily steep, as the highest-energy electrons cool fastest.

In this section we explore the possibility that in AT2020xnd shock-acceleration was also not continuous. We consider a scenario in which shock-acceleration switches off and the electrons cool through inverse Compton scattering, synchrotron emission, or adiabatic expansion.
First we estimate the dynamical time at 46\,d (observer-frame), the epoch when the steep optically thin spectral index was measured,
as a basis of comparison for the cooling processes:

\begin{equation}
    t_\mathrm{dyn} \sim \frac{R}{v} \sim (20\,\days) \left( \frac{R}{10^{16}\,\cm} \right) \left( \frac{v}{0.2c} \right)^{-1}.
\end{equation}

\noindent At 46\,d the shock speed $v\approx0.2c$ and $R\approx 2\times10^{16}\,\cm$, so $t_\mathrm{dyn}\approx 40\,\days$.
Next we estimate the synchrotron-cooling time at 100\,\ghz.
Taking the Lorentz factor of the electrons emitting at $\nu=100\,\ghz$ to be $\gamma = (2\pi m_e c \nu/(eB))^{1/2}$ we have

\begin{equation}
    t_\mathrm{syn} \approx \frac{6\pi m_e c }{\sigma_T B^2 \gamma} \approx (50\,\days) \left( \frac{B}{1\,\gauss} \right)^{-3/2}.
\end{equation}

\noindent At 46\,\days, we have $B\approx1\,\gauss$, so $t_\mathrm{syn}\approx50\,\days$, which is comparable to the dynamical time.

Finally, we estimate the cooling timescale from inverse Compton scattering, by replacing $B^2$ in the expression above with $8\pi u_\mathrm{ph}$, where $u_\mathrm{ph}$ is the photon energy density measured from UVOIR observations,

\begin{align}
    t_\mathrm{IC} &\approx \frac{3 m_e c }{4 \sigma_T u_{\rm ph} \gamma} 
    \\ \nonumber
    &\approx (70\,\days) \left( \frac{R}{10^{16}\,\cm} \right)^{2} \left( \frac{L_\mathrm{UVOIR}}{10^{42}\,\erg\,\psec} \right)^{-1} \left( \frac{B}{1\,\gauss} \right)^{1/2}.
\end{align}

We estimate that the optical luminosity 
at $\Delta t=46\,\days$ (observer-frame) is $10^{42}\,\erg\,\psec$ \citep{Margutti2019,Perley2021}.
Again taking the forward-shock radius $R=2 \times 10^{16}\,\cm$, we have $t_\mathrm{IC}\approx280\,\days$, longer than the dynamical time.
As the cooling timescales from synchrotron radiation and inverse Compton scattering do not appear to be significantly shorter than the dynamical timescale, we conclude that it is unlikely that the ``shut-off'' of shock-acceleration followed by rapid cooling can explain the steep spectrum observed at $\sim 100\,\ghz$.

Because the dynamical time of 40\,\days\ is similar to the synchrotron cooling time of 50\,\days, we next consider whether the expansion of the emitting region (adiabatic cooling) could explain the radio light curves.
In particular, since the picture of shock-interaction with a dense shell has been invoked to explain AT2018cow \citep{Perley2019cow,Margutti2019} as well as other fast and luminous optical transients \citep{Ofek2010,Rest2018,Ho2019gep,Leung2021},
it is interesting to consider whether the same region that produced the optical emission could have expanded and also produced the radio emission.
This is plausible, since the radius of the region responsible for the optical emission is roughly $10^{14}\,\cm$ \citep{Perley2019cow,Margutti2019}, which would be roughly $10^{16}\,\cm$ by 18\,d in the observer-frame assuming $v=0.2c$;
furthermore, because the electron synchrotron cooling timescale is long, there may still be relativistic electrons left to radiate when the shell expands to this radius.

For adiabatic expansion, we have $R \propto t$, $B \propto R^{-2} \propto t^{-2}$ (by flux conservation), and $N_0 \propto R^{-(2+p)} \propto t^{-(2+p)} \propto t^{-5}$ for $p=3$, where the electron energy distribution is $N(E) = N_0 E^{-p}$.
At a given frequency in the optically thin $\nu > \nu_a$ regime, we therefore expect $f_\nu \propto R^3 N_0 B^{(p+1)/2} \propto t^{-6}$ \citep{Chevalier1998}.
At a given frequency in the optically thick regime, we expect $f_\nu \propto R^2 B^{-1/2} \nu^{5/2} \propto t^3$.

Finally, the evolution of $\nu_a$ and $F_a$ can be estimated from Appendix~C. We find that $F_a \propto t^{-3.5}$ and $\nu_a \propto t^{-2.6}$ under the same assumptions above.
So, the adiabatic expansion of a shocked shell---without continuous shock-acceleration---can result in steeply declining light curves, and steeply declining values of $F_a$ and $\nu_a$.
Our observations of AT2020xnd do not quite match the predicted values, however; the observed temporal decline of $F_a$ and $\nu_a$ are more shallow than expected,
and the light curves at optically thick and thin frequencies are slightly shallower.
So, we also consider this picture unlikely.

\subsection{Summary and Model Comparison}

In this section, we considered three possible explanations for the early millimeter emission in AT2020xnd. First we considered the standard framework used in the literature, synchrotron emission from electrons accelerated into a power-law energy distribution.
The challenge for this model is that it implies that the  measured spectral index of $\beta=-2.0\pm0.2$ implies an electron energy power-law index of $p=4.0\pm0.5$, significantly steeper than the predicted $p=2$ from diffuse shock acceleration, and---to our knowledge---steeper than all radio SNe in the literature.
The very steep spectrum led us to consider an alternate model: synchrotron emission from electrons in a thermal distribution. As shown in Figure~\ref{fig:maxwellian},
the thermal model naturally explains the $f_\nu \propto \nu^{2}$ self-absorbed power-law index observed in CSS161010, as well as the steep high-energy spectral index in AT2018cow, AT2020xnd, and CSS161010.
The primary challenge for this model is the fact that it has not been inferred for other cosmic explosions.
However, as discussed in detail in \citet{Margalit2021}, the prominence of thermal-electron emission can naturally be explained by the unusual mildly relativistic shock speeds of these events.
While it may at first seem surprising that the thermal population dominates the emission even at frequencies two orders of magnitude larger than the peak frequency at which the thermal electrons radiate, this is expected when the non-thermal population only has a modest fraction of the total energy of the shock-accelerated electrons, as has been shown in previous work in the context of AGN (e.g., \citealt{Ozel2000}).
Finally, we considered two scenarios in which the emission arises from non-steady-state particle acceleration. However, we concluded that both scenarios were unlikely: the cooling time is not significantly shorter than the dynamical time, and adiabatic expansion predicts different values for the temporal evolution of the radio light curves.

All of the possibilities listed above would be interesting.
For example, if the electron distribution is a power law, then the inferred value of $p>3.5$ is surprising given the wealth of observational and theoretical data favoring $p\sim2$--3.
However, because of the shallow optically thick spectral index, the steep optically thin spectral index, and the natural explanation for the prominent thermal emission presented in \citet{Margalit2021}, we conclude that the theoretically simplest explanation is the presence of a thermal population in addition to a non-thermal tail.
In Section~\ref{sec:discussion} we present observational tests that can rule out or confirm this model in the future.

\section{Origin of the X-ray Emission}
\label{sec:xray}

In the previous section we considered the origin of the radio and millimeter-band emission.
In this section we consider the origin of the X-rays,
which were one of the most peculiar features of AT2018cow.
The X-rays observed from AT2018cow could not be described as an extension of the radio synchrotron spectrum, nor by inverse Compton scattering of UVOIR photons by electrons accelerated in the forward shock; the conclusion was that they must arise from a central compact source \citep{Ho2019cow,Margutti2019}.
We find that similar arguments hold for AT2020xnd, although the X-ray data is more limited in temporal resolution and sensitivity.

\begin{figure}[!htb]
    \centering
    \includegraphics[width=\columnwidth]{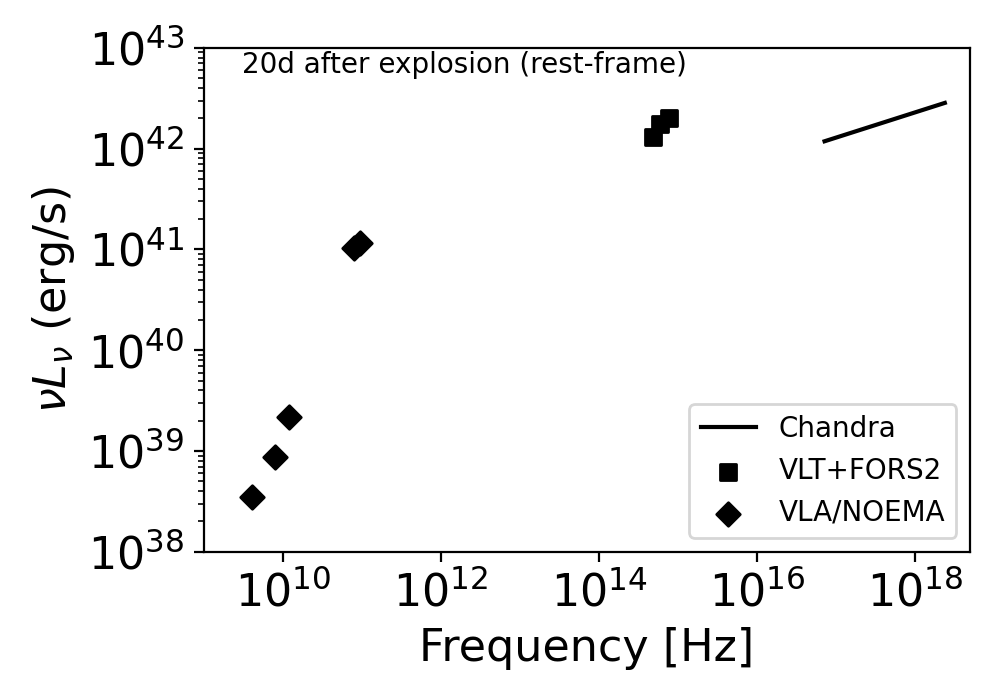}
    \caption{SED of AT2020xnd at 26d after explosion (observer frame) or 21d in the rest-frame. The optical data is taken from \citet{Perley2021}. We plot the \emph{Chandra} data as follows: we take integrated 0.3--10\,keV flux, use the geometric mean of (0.3\,keV, 10\,keV) and the spectral index $f_\nu \propto \nu^{-0.75}$ to solve for the normalization coefficient for the spectrum. We display the spectrum over the full 0.3--10\,keV range.}
    \label{fig:sed}
\end{figure}

In Figure~\ref{fig:sed} we plot the SED from radio to X-ray bands at $\Delta t\sim26\,$d (observer frame; MJD 59158).
The radio to X-ray spectral index at this time is $\beta_{RX} \approx 0.2$ where $f_\nu \propto \nu^{-\beta}$.
Therefore, the value of $\beta_{RX}$ is too shallow for the X-rays to be an extension of the synchrotron spectrum.

Next we consider the possibility that the X-rays arise from inverse Compton scattering.
We begin with the energetics.
The X-ray light curve of AT2020xnd is similar to that of AT2018cow in showing a plateau phase followed by a decline phase (Figure~\ref{fig:xray-lc}).
Assuming that the plateau extends to 30\,d, we estimate that the total X-ray energy emitted in the first month is $10^{49}\,$erg,
similar to the $7 \times 10^{48}\,\erg$ inferred from the AT2018cow X-ray emission,
and similar to the energy we estimated from the AT2020xnd radio observations.
Therefore, if a significant proportion of the X-rays is produced by IC emission, then our assumption of $\epsilon_e=\epsilon_B=1/3$ results in a significant underestimate of the total energy.

The ratio of the X-ray to radio luminosity at $\Delta t\approx26\,\days$ is $L_X/L_\mathrm{radio}\approx40$--80, close to the value of 30 for AT2018cow.
If the X-rays arise from IC scattering off the synchrotron-emitting electrons, we have

\begin{equation}
    \frac{L_\mathrm{X}}{L_\mathrm{radio}}=\frac{L_\mathrm{IC}}{L_\mathrm{syn}}=\frac{u_{ph}}{u_B}
\end{equation}

\noindent where $u_\mathrm{ph}$ is again the photon energy density (measured from UVOIR observations) and $u_B$ is the magnetic energy density (measured from our radio observations; \citealt{RL}).
From the previous section we have $u_\mathrm{ph} \approx 0.03\,\erg\,\pcmcub$.
We require $B\approx$0.03--0.1\,\gauss, where $u_B=B^2/8\pi$. This is smaller than our estimate from modeling the SED using a non-thermal or thermal electron energy distribution.
In addition, the X-ray luminosity does not decline rapidly in keeping with the optical light curve,
and the spectral index is shallower than would be expected.
So, as was the case in AT2018cow, we conclude that the X-rays are unlikely to arise from IC scattering.

\section{Rates in Millimeter Surveys}
\label{sec:rates}

Until recently, there was only one untargeted transient survey specific to the millimeter band \citep{Whitehorn2016}.
Within the past year, the Atacama Cosmology Telescope (ACT; \citealt{Thornton2016}) and the South Pole Telescope (SPT; \citealt{Carlstrom2011}) published blind discoveries of bright (mJy) transients \citep{Guns2021,Naess2021}, including several of extragalactic origin \citep{Guns2021}.
The 100\,\ghz\ light curve of AT2020xnd (Figure~\ref{fig:mm-lc-comparison}) is the most luminous ever obtained for a non-relativistic cosmic explosion,
and in this section we estimate the rate of such events in present and future millimeter transient surveys, summarized in Table~\ref{tab:rates}.

\begin{deluxetable*}{llrrrrr}[htb!]
\tablecaption{The rates of transients similar to AT2018cow in millimeter surveys, compared to other classes of millimeter-bright cosmic explosions. The horizon is set by requiring a 6-$\sigma$ detection. For SPT-3G and CMB-S4 the sensitivity assumes one-week stacks, because these surveys have a cadence of one observation per day or higher.
    \label{tab:rates}
}
\tablehead{
\colhead{Class} & \colhead{Survey} & \colhead{Band} & \colhead{Sensitivity (6-$\sigma$)} & \colhead{Area} & \colhead{Horizon} & Rate \\
&  &  (GHz) & (mJy) & (\% sky) & (Mpc) & ($\pyr$)
}
\startdata
AT2018cow & SPT-3G & 95 & 15 & 4\% & 330 & 0.4 \\
& ACT & 100 & 90 & 40\% & 140 & 0.3 \\
& CMB-S4 Wide & 95 & 18 & 50\% & 300 & 4 \\
& CMB-S4 ultra-deep & 95 & 5 & 3\% & 590 & 2 \\
LGRB & CMB-S4 Wide & 95 & 18 & 50\% & 2200 & 2 \\
& CMB-S4 ultra-deep & 95 & 5 & 3\% & 4200 & 1 \\
LLGRB & CMB-S4 Wide & 95 & 18 & 50\% & 68 & 0.2 \\
& CMB-S4 ultra-deep & 95 & 5 & 3\% & 130 & 0.07 \\
CC SN & CMB-S4 Wide & 95 & 18 & 50\% & 7 & 0.05 \\
& CMB-S4 ultra-deep & 95 & 5 & 3\% & 13 & 0.02 \\
\enddata
\end{deluxetable*}

For the volumetric rate we use the result from \citet{Ho2021},
that events similar to AT2018cow occupy a tight region in optical transient parameter space, with a volumetric rate of 0.001--0.1\% of the CC SN rate \citep{Coppejans2020,Ho2021} or 0.7--70\,\pyr\,\pgpccub.
We take a characteristic 100\,\ghz\ luminosity of $2\times10^{30}\,\erg\,\psec\,\phz$.
We estimate the number of detections per year assuming a 6$\sigma$ threshold (as in \citealt{Whitehorn2016}).

We use the following survey parameters. 
The SPT surveys an area of 1500\,\degsq\ with a 6-$\sigma$ sensitivity of 15\,mJy at 95 and 150\,\ghz\ in one-week stacks \citep{Whitehorn2016}.
The ACT surveys 18,000\,\degsq\ (40\% of the sky) in a raster scan mode, scanning back and forth at constant elevation and allowing sources to pass through the field-of-view.
The cadence has been roughly one week since 2016.
The 1$\sigma$ sensitivity in a single sweep is 30--50\,mJy at 90\,\ghz, but for a source that is steady across the time it takes to traverse the focal plane (approximately 10 minutes),
this is reduced to 10--20\,mJy.
Here we take an RMS sensitivity of 15\,mJy.
CMB-S4 \citep{Abazajian2019}, a next-generation cosmic microwave background (CMB) experiment, will conduct two surveys relevant for the discovery of transients like AT2018cow:
an all-sky wide-area survey (50\% of the sky) and an ultra-deep survey in a smaller region (3\%).
The CMB-S4 $6$-$\sigma$ sensitivity for one-week stacks is 18\,mJy in the wide survey and 5\,mJy in the deep survey \citep{Abazajian2019}.

To estimate the rate of other classes of energetic explosions, in particular long-duration GRBs and low-luminosity GRBs, we take rates from Table~10 of \citet{Ho2020b}.
The GRB luminosity function at 100\,\ghz\ is uncertain.
\citet{deUgartoPostigo2012} found an average peak spectral luminosity of $10^{32.1\pm0.7}\,\erg\,\psec\,\phz$ among detected bursts, although the overall detection rate was only 25\%.
For now we adopt a characteristic luminosity of $10^{32}$\,\erg\,\psec\,\phz.
Slightly off-axis GRBs are expected to have a similar luminosity to those observed directly on-axis \citep{Metzger2015}, so the correction from including off-axis bursts may roughly compensate for the correction for bursts that are mm-faint.
For LLGRBs we adopt a characteristic 100\,GHz luminosity of $10^{29}\,\erg\,\psec\,\phz$ (from Figure~\ref{fig:radio-lc}).
For SNe we adopt a 100\,GHz luminosity of $10^{27}\,\erg\,\psec\,\phz$ (again based on Figure~\ref{fig:radio-lc}).

The number of events detected per year is simply

\begin{equation}
    N_\mathrm{det} = \frac{4\pi}{3} d_\mathrm{lim}^3 \times \cal{R} \times A_\mathrm{survey},
\end{equation}

\noindent where $d_\mathrm{lim}$ is the distance out to which the transient can be detected (second-to-last column in Table~\ref{tab:rates}), $\cal{R}$ is the volumetric rate, and $A_\mathrm{survey}$ is the fraction of the sky observed by the given survey. For the millimeter-band surveys we assume that the duration of the transient is significantly longer than the cadence, and therefore that all transients in that area within the given volume will be detected. 

Table~\ref{tab:rates} shows that for an optimistic estimate of the rate (0.1\% of the CC SN rate), events similar to AT2018cow should be detected routinely by CMB-S4, with a per-year rate higher than what is currently achieved by optical surveys,
and that they may be a dominant population of cataclysmic extragalactic mm-band transients.
However, for a more pessimistic estimate of 0.01\% of the CC SN rate, the number of detected sources would be more similar to that predicted for LLGRBs,
and the expected number would be an order of magnitude less than that of LGRBs.
An interesting scientific question will be whether other classes of cosmic explosions, such as SNe, exhibit millimeter behavior similar to that of AT2018cow, but are not particularly remarkable at optical wavelengths and therefore are not currently followed up at high frequencies.

Our predicted rates for AT2018cow-like events are slightly higher than those predicted using more detailed simulations \citep{Eftekhari2021}. This is primarily because we used the observed 100\,\ghz\ light curve of AT2020xnd, while the simulations used the 230\,GHz light curve of AT2018cow and scaled it to 100\,\ghz\ assuming $F_\nu \propto \nu^{-0.7}$.
Our back-of-the-envelope estimate for LGRBs is consistent with the more detailed prediction for the CMB-S4 wide survey, but significantly lower than the prediction for the deep survey.
The differences may be due to the fact that the simulations incorporate the evolution of the cosmic star-formation rate.
We also point out that the detection rate of extragalactic transients is predicted to be dominated by the reverse shock from LGRBs \citep{Eftekhari2021}, which we have not considered here.

\section{Summary and Discussion}
\label{sec:discussion}

We presented millimeter, radio, and X-ray observations of AT2020xnd, a transient with luminous ($M\approx-21\,$mag) and short-duration ($t_{1/2}\approx3\,\days$) optical emission.
Our early discovery enabled only the second-ever detailed high-frequency ($\nu\gtrsim 100\,\ghz$) observations of such an object.

AT2018cow and AT2020xnd comprise a growing class of objects with millimeter and radio properties that are unusual among cosmic explosions: a steep optically thin spectral index,
and early high-frequency emission that is difficult to reconcile with the late-time low-frequency behavior.
The discrepancy between the early-time and late-time radio emission in AT2020xnd was also noted by \citet{Bright2022}, who suggested that it may arise from a steepening density distribution.

The basic shock properties from a standard analysis, assuming that the peak of the SED is governed by synchrotron self-absorption and that the electrons are in a power-law distribution, are a fast speed ($v\approx0.2c$) and a high ambient density, similar to that inferred for AT2018cow.
Furthermore, the X-ray emission is in excess of what would be predicted from an extrapolation of the synchrotron spectrum.
\citet{Bright2022} independently reached a similar conclusion from their 1--100\,\ghz\ data.
We also found that the X-rays are in excess of that predicted from inverse Compton scattering.

However, based on our 100--200\,\ghz\ NOEMA data,
we conclude that a thermal electron distribution (a relativistic Maxwellian) likely significantly contributes to the synchrotron emission at early times,
and likely also contributed to the emission observed in AT2018cow and CSS161010.
The Maxwellian model predicts an optically thick spectral index of $f_\nu \propto \nu^2$, which was observed in CSS161010 and AT2018cow; the optically thick spectral index of AT2020xnd was even shallower.

The presence of a Maxwellian is not a surprise: it is expected that only a small fraction of electrons should be accelerated into the power-law tail, with the majority accelerated into a thermal distribution \citep[e.g.][]{Park+15}.
The question then arises why it has not been definitively seen in previous SNe.
As presented in more detail in \citet{Margalit2021}, the detectability of the thermal population---its prominence relative to the non-thermal population, and its peak frequency---is highly sensitive to the shock speed (and to a lesser extent to the ambient density).
The mildly relativistic shock speeds of transients like AT2018cow, CSS161010, and AT2020xnd---together with the fact that early high-frequency observations were obtained---explains why the thermal population is more prominent in these events than in most observed SNe.
The fast speed (and high ambient density) is also why the influence of the Maxwellian is best observed at high frequencies ($\gtrsim 100\,$GHz).
Testable predictions of this model are that the transition from the thermal to power-law distribution should be detectable in even higher-frequency observations ($\gtrsim200\,\ghz$), and that explosions with fast shock speeds ($v\gg0.1c$) should have $\nu^{2}$ rather than $\nu^{5/2}$ optically thick spectral indices.

Accounting for the Maxwellian does not dramatically change the inferred physical parameters: the shock speeds remain in the range of 0.1--0.5$c$, and the ambient densities close to $10^{4}\,\pcmcub$.
However, for AT2018cow the Maxwellian model implies a shock speed of $\approx0.3c$ at $10\,\days$ and a decelerating shock, different from the constant shock speed of $0.1c$ inferred in previous work.
In addition, the Maxwellian model enables a novel constraint on the fraction of electrons accelerated by the shock, which we constrain to be $<20\%$ from our observations of AT2020xnd.
We defer a thorough re-analysis of the evolution of AT2018cow, CSS161010, and AT2020xnd in the context of a Maxwellian to future work.
For now we caution that the usual assumption of all electrons being accelerated into a power-law distribution is not well-motivated for fast shock speeds ($v\gtrsim0.2c$) and high ambient densities for the observing frequencies involved here (1--100\,GHz).

It appears that a short light-curve duration and high peak luminosity are predictive of luminous millimeter and X-ray emission.
Indeed, this sets AT2018cow and AT2020xnd apart from optical transients that have similar spectroscopic properties and rapid light-curve evolution: interacting SNe of Type~Ibn.
This suggests that the essential difference between AT2018cow and Type~Ibn SNe is the presence of high-velocity ejecta, perhaps from a central engine like a newly formed black hole \citep{Kashiyama2018,Quataert2019}---analogous to the fact that most stripped-envelope SNe do not exhibit relativistic ejecta, while a small subset (those associated with GRBs) do.

The limitation of our approach---identifying transients via optical surveys, and following them up with millimeter telescopes---is that it prevents us from identifying the subset of other SN classes that may exhibit similar behavior.
Indeed, a handful of transients had similar shock properties to AT2018cow and AT2020xnd, including SN\,2003L and PTF11qcj.
The shock properties, together with the observation of optically thick emission at early times, suggests that they would also have been luminous millimeter transients, perhaps also with a prominent Maxwellian component to the SED.
Future wide-field millimeter cosmology experiments will enable luminous millimeter transients to be detected routinely without relying on an optical discovery.
Based on our NOEMA 100\,\ghz\ light curve of AT2020xnd,
we estimate that events like AT2018cow and AT2020xnd (and likely other types of SNe) will be detected blindly by CMB-S4.
Our work is a direct demonstration of how these discoveries, together with multi-band follow-up observations,
can shed light on how particles are accelerated in astrophysical shocks produced by cosmic explosions.

The code used to produce the figures in this paper can be found in a public Github repository\footnote{\url{ https://github.com/annayqho/ZTF20acigmel}}.

\appendix

\section{Radio Observations \& Reduction}
\label{sec:appendix-radio-data}

\subsection{Very Large Array (VLA)}
\label{sec:vla-obs}

Our VLA observations are summarized
in Table~\ref{tab:radio-log}.
Observations were obtained in standard continuum imaging mode and spanned BnA, A, and D configuration.
We used 3C48 as the flux density and bandpass calibrator and J2218-0335 as the complex gain calibrator.
Data were calibrated using the automated pipeline available in the Common Astronomy Software Applications (CASA; \citealt{McMullin2007}),
with additional flagging performed manually,
and imaged\footnote{Cell size was 1/5 of the synthesized beamwidth, field size was the smallest magic number
($10 \times 2^n$) larger than the number of cells needed to cover the primary beam.} using the CLEAN algorithm \citep{Hogbom1974}.
In each image we verified that the source was a point source using \texttt{imfit} and that the image was free of artifacts, then measured the peak flux density in a region centered on the source using \texttt{imstat}.

To estimate the uncertainty on the source flux density, we measured the RMS pixel value in an area of the image close to the source with no substantial emission.
We added this in quadrature to two additional sources of systematic error.
First, the VLA flux density scale calibration accuracy is 5\% at L- through Ku-bands and 10-15\% for the three higher bands\footnote{\url{https://science.nrao.edu/facilities/vla/docs/manuals/oss/performance/fdscale}}.
Second, the flux density calibrator 3C48 has been undergoing a flare since January 2018. To account for this, we added an additional 10\% systematic error at low frequencies (C-band through Ku-band),
an additional 15\% at K and Ka-band, and 20\% at Q-band.

\subsection{Australia Telescope Compact Array (ATCA)}
\label{sec:obs-atca}

We obtained three observations with the ATCA under project CX472, with two 2048\,MHz bands centered on 33 and 35\,GHz.
Observations were carried out in the 6B, H168 and 1.5A array configurations, with maximum baselines of 6\,km, 192\,m  (after removing antenna 6 to ensure more even sampling of the u--v plane) and 4.5\,km.

The data were reduced using standard \textsc{Miriad} routines \citep{1995ASPC...77..433S}. The first and third observation used the standard continuum correlator setup with 1\,MHz channels, while the second observation was carried out in a hybrid correlator mode with 1\,MHz channels in the 33\,GHz band and 64\,MHz coarse channels in the 35\,GHz band. A single zoom band with 2048$\times$64\,kHz channels was placed at 35\,GHz to allow for initial calibration of the coarse channels delays, although this data was not used further.

For all observations we performed an initial bandpass calibration using observations of 1921-293, and calibrated the gain and polarisation using the secondary calibrator, 2216-038. We used 1934-638 to calibrate the flux density scale, and then improved the bandpass calibration using the standard bootstrapping procedure outlined in the ATCA User Guide\footnote{\url{https://www.narrabri.atnf.csiro.au/observing/users\_guide/html/atug.html\#Calibration2}}. Both bands were combined and imaged with a cell size corresponding to approximately one fifth of the synthesised beam width using CLEAN \citep{Hogbom1974} and robust=0.5 weighting. We used \textsc{IMFIT} to fit a point source, allowing the position to vary in a $20\times20$ pixel box centered on the location of AT2020xnd and in the event of a detection, report the measured flux density and associated uncertainty. In the event of a non-detection we report an upper limit of 3 times the image noise.

We have also independently analysed the ATCA observations reported by \citet{Bright2021a,Bright2021b} under project CX471, as well as a third epoch that was not reported. These observations were carried out with two 2048\,MHz bands centered on 17 and 19\,GHz with the same flux, bandpass and phase calibrators described above. The same overall process was used to reduce the data, but substantially more manual flagging was carried out to remove data irregularities discovered via inspection of the visibilities. In all three observations we found noise spikes near the centre of the 19\,GHz band on some baselines\footnote{Baselines 3--4, 3--5 and 4--5 in the first observation; 1--4, 2--3, 3--4, 3--5 in the second; and 1--2 in the third} and flagged channels 500-1250 to remove them. In the third observation we removed similar spikes at the edges of the 19\,GHz band on baseline 1--2 (flagging channels 1--300 and 1800--2048) and near the centre of the 19\,GHz band on baselines 1--3 and 2--3 (flagging channels 850-1150). In the third observation we also removed antenna 6 due to an irregular bandpass response, and flagged all 19\,GHz data from antenna 4 due to noise spikes in the Stokes YY visibilities across the full band.

\subsection{The Submillimeter Array (SMA)}
\label{sec:obs-sma}

We obtained three observations,
all in the Sub-Compact configuration, using all 8 antennas.  
During the first two observations, the receivers were tuned to local oscillator (LO) frequencies of 225.5\,GHz USB and 232.5\,GHz, which provides continous frequency coverage from 209.5\,GHz to 249.5\,GHz (with 10 GHz overlap) and 48\,GHz bandwidth available for continuum channel generation. The third attempt piggybacked anther science project with one of the receivers tuned to an LO of 225.3\,GHz giving coverage of 209.5\,GHz to 241.5\,GHz while the second was tuned to an LO of 256.5\,GHz giving coverage from 240.5\,GHz to 272\,GHz (both with an 8\,GHz gap in the middle between sidebands) giving a total of 48\,GHz of bandwidth for continuum centered on 241.0\,GHz. The quasars 2232+117 and 2148+069 were used as primary phase and amplitude gain calibrators, with absolute flux calibration performed by comparison to Neptune and Uranus, while the quasar 3C\,84 was used for bandpass calibration. Data were calibrated in IDL using the MIR package then exported for additional analysis and imaging using the \textsc{Miriad} package. On the first night the atmospheric opacity was 0.22 ($\sim$4 mm precipital water vapour) and after 5.0 hours on-source an RMS of 0.38\,mJy was achieved. The second night the opacity was 0.1 and after 5.0 hours on-source an RMS of 0.16\,mJy was reached. On the final night the opacity was better (around 0.06) but with a shorter observation (4.3 hours on-source) an RMS of 0.16\,mJy was again reached.

\subsection{NOrthern Extended Millimeter Array (NOEMA)}
\label{sec:obs-noema}

NOEMA is situated on the Plateau de Bure (France) at an altitude of 2550 m. The number of available 15-m antennas varied between 9 and 11, and the antenna spacings changed between intermediate-extended C and compact D configurations. The PolyFiX backend was configured in low-resolution continuum mode (2 MHz resolution) covering both sidebands of the 2SB receivers in dual polarization, resulting in a spectral coverage of $4\times7.744$~GHz. The spectral bandpass was calibrated on strong quasars and the time-dependent amplitude and phase calibrations done on the quasars 2216-038 and and 2227-088 that are close to AT2020xnd. In the primary flux calibration the radio continuum of the emission line stars MWC349 and LKHA101 was used; based on the observatory-internal flux monitoring we assume that MWC349 was 8\% brighter at the time of the AT2020xnd monitoring than its CLIC internal flux model predicts. This improves the overall consistency of the flux calibration in the 3\,mm band from about 10\% to 5\%. The inherent errors of the 2\,mm and 1.3\,mm bands are higher, we assume them to be at 15\% and 20\%, respectively. The data reduction was done with the CLIC software (GILDAS package\footnote{\url{https://www.iram.fr/IRAMFR/GILDAS/}}). Dual-polarization UV tables were written for each of the receiver sidebands, their central sky frequencies are given in Tab.~\ref{tab:radio-log}. The resulting calibrated UV tables were analyzed in the MAPPING software (also from the GILDAS package) and point-source UV plane fits were performed. We constrained the fit position to the coordinates found in our VLA observations (Section~\ref{sec:vla-obs}), the difference in the derived flux as compared to a free fit is typically a small fraction of one sigma. The advantage of this procedure compared to map deconvolution is the straightforward error propagation in the UV fitting process. Weather conditions were good, with the exception of the 3\,mm and 2\,mm data points taken on December 16th, 2020.

\startlongtable
\begin{deluxetable*}{lrrrrrr}
\tablecaption{
    Observations of AT2020xnd with the SMA, NOEMA, the ATCA, and the VLA. Upper limits are reported as $3\times$ the image RMS (in the case of NOEMA, the rms of the UV plane fits).
    For NOEMA, the absolute flux scale calibration accuracy is 5\% at 3\,mm, 15\% at 2\,mm and 20\% at 1.3\,mm.
    The VLA uncertainty is the quadrature sum of the image RMS, the standard flux density scale calibration accuracy (5\% at L- through Ku-bands, 15\% for K, Ka, and Q-bands), and additional uncertainty from the fact that the flux density calibrator 3C48 is currently undergoing a flare (additional 10\% at C-band through Ku-band, 15\% at K- and Ka-band, 20\% at Q-band).
    All measurements are reported in the observer-frame.
    \label{tab:radio-log}
}
\tablehead{
\colhead{Start Date} & \colhead{$\Delta t$} & \colhead{Facility} & \colhead{$\nu$} & \colhead{Flux Density} & \colhead{Array Configuration} \\
\colhead{(UT)} & \colhead{(days)} &  & \colhead{(GHz)} & \colhead{(mJy)}
}
\startdata
2020 Oct 20.1 & 10.1 & SMA & 230 & $<1.14$ & subcompact \\
2020 Oct 23.0 & 13.0 & VLA & 10 & $0.024 \pm 0.006$ & BnA \\
2020 Oct 27.0 & 17.0 & VLA & 10 & $<0.024$ & BnA \\
2020 Oct 27.8 & 17.8 & NOEMA & 79 & $0.389 \pm 0.059$ & 10C\\
2020 Oct 27.8 & 17.8 & NOEMA & 94 & $0.304 \pm 0.057$ & 10C\\
2020 Oct 28.0 & 18.0 & VLA & 10 & $<0.051$ & BnA \\
2020 Oct 28.0 & 18.0 & VLA & 6 & $<0.030$ & BnA \\
2020 Oct 28.0 & 18.0 & VLA & 15 & $0.037\pm0.010$ & BnA \\
2020 Oct 29.2 & 19.2 & ATCA & 34 & $<0.108$ & H168 \\
2020 Oct 29.2 & 19.2$^{a}$ & ATCA & 18 & $0.135\pm0.040$ & H168 \\
2020 Oct 31.1 & 21.1 & SMA & 230 & $<0.48$ & subcompact \\
2020 Nov 02.8 & 23.8 & NOEMA & 79 & $0.675 \pm 0.047$ & 10C\\
2020 Nov 02.8 & 23.8 & NOEMA & 94 & $0.634 \pm 0.045$ & 10C\\
2020 Nov 04.0 & 25.0 & VLA & 15 & $0.095\pm0.011$ & BnA \\
2020 Nov 04.0 & 25.0 & VLA & 10 & $0.057\pm0.005$ & BnA \\
2020 Nov 04.0 & 25.0 & VLA & 6 & $0.046\pm0.006$ & BnA \\
2020 Nov 07.3 & 28.3 & ATCA & 34 & $0.310\pm0.020$ & H168\\
2020 Nov 10.8 & 31.8 & NOEMA & 79 & $1.076 \pm 0.049$ & 10C\\
2020 Nov 10.8 & 31.8 & NOEMA & 94 & $1.018 \pm 0.044$ & 10C\\
2020 Nov 14.1 & 35.1 & SMA & 230 & $<0.48$ & subcompact \\
2020 Nov 15.9 & 36.9 & VLA & 33 & $0.497\pm0.011$ & BnA to A \\
2020 Nov 15.9 & 36.9 &  VLA & 45 & $0.675\pm0.171$ & BnA to A \\
2020 Nov 15.9 & 36.9 & VLA & 10 & $0.079\pm0.010$ & BnA to A \\
2020 Nov 18.8 & 39.8 & NOEMA & 79 & $0.912\pm0.050$  & 10C\\
2020 Nov 18.8 & 39.8 & NOEMA & 94 & $0.825\pm0.046$  &10C \\
2020 Nov 19.2 & 40.2$^{b}$ & ATCA & 18 & $0.320\pm0.060$ & H168 \\
2020 Nov 24.7 & 45.7 & NOEMA & 79 & $0.822\pm0.048$ & 10D\\
2020 Nov 24.7 & 45.7 & NOEMA & 94 & $0.569\pm0.042$ & 10D \\
2020 Nov 24.8 & 45.8 & NOEMA & 211 & $0.145\pm0.048$ & 10D \\
2020 Nov 24.8 & 45.8 & NOEMA & 227 & $<0.156$  & 10D \\
2020 Nov 25.8 & 46.8 & NOEMA & 131 & $0.317\pm0.049$  & 9D \\
2020 Nov 25.8 & 46.8 & NOEMA & 146 & $0.179\pm0.057$  & 9D \\
2020 Nov 27.2 & 48.2 & ATCA & 34 & $0.490\pm0.040$ & H168 \\
2020 Nov 30.9 & 51.9 & VLA & 10 & $0.154\pm0.005$ & BnA to A \\
2020 Nov 30.9 & 51.9 & VLA & 33 & $0.621\pm0.132$ & BnA to A \\
2020 Nov 30.9 & 51.9 & VLA & 45 & $0.668\pm0.168$ & BnA to A \\
2020 Dec 16.6$^c$ & 67.6 & NOEMA & 79 & $0.247\pm0.037$ & 11D\\
2020 Dec 16.6$^c$ & 67.6 & NOEMA & 94 & $<0.105$ & 11D\\
2020 Dec 16.7$^c$ & 67.7 & NOEMA & 131 & $<0.213$ & 9D \\
2020 Dec 16.7$^c$ & 67.7 & NOEMA & 146 & $<0.279$ & 9D \\
2020 Dec 20.9 & 71.9 & VLA & 10 & $0.180\pm0.023$ & A \\
2020 Dec 20.9 &71.9 & VLA & 15 & $0.401\pm0.046$ & A \\
2020 Dec 20.9 &71.9 & VLA & 22 & $0.484\pm0.010$ & A \\
2020 Dec 20.9 &71.9 & VLA & 33 & $0.450\pm0.096$ & A \\
2020 Dec 20.9 &71.9 & VLA & 45 & $0.209\pm0.063$ & A \\
2020 Dec 26.3 & 77.3 & ATCA & 18 & $0.300\pm0.035$ & H168 \\
2021 Jan 06.6 & 88.6 & NOEMA & 131 & $<0.114$ & 11D\\
2021 Jan 06.6 & 88.6 & NOEMA & 146 & $<0.120$ & 11D\\
2021 Jan 12.7 & 94.7 & VLA & 10 & $0.168\pm0.022$ & A \\
2021 Jan 12.7 & 94.7 & VLA & 15 & $0.278\pm0.032$ & A \\
2021 Jan 12.7 & 94.7 & VLA & 22 & $0.301\pm0.065$ & A \\
2021 Jan 12.7 & 94.7 & VLA & 33 & $0.213\pm0.048$ & A \\
2021 Feb 18.6 & 131.6 & VLA & 6 & $0.117\pm0.016$ & A \\
2021 Feb 18.6 & 131.6 & VLA & 10 & $0.109\pm0.010$ & A \\
2021 Feb 18.6 & 131.6 & VLA & 15 & $0.122\pm0.016$ & A \\
2021 Feb 18.6 & 131.6 & VLA & 22 & $0.087\pm0.020$ & A \\
2021 Feb 18.6 & 131.6 & VLA & 33 & $<0.042$ & A \\
\enddata
\tablenotetext{a}{Re-analysis of the data first reported in ATel\#14148}
\tablenotetext{b}{Re-analysis of the data first reported ATel\#14249}
\tablenotetext{c}{Weather conditions unstable}
\end{deluxetable*}

\onecolumngrid

\section{X-ray Observations and Reduction}
\label{sec:appendix-xray}

We used the \texttt{CIAO} v4.12 \citep{Fruscione2006} tool \texttt{specextract} to extract the spectrum, using a circular region with a radius of 1$^{\prime \prime}$ centered on the apparent position of the source. 
The background was extracted from a neaby source-free region with a radius of 10$^{\prime \prime}$.
We performed spectral fitting on the 0.5--8\,keV specrum with \texttt{xspec} v12.11.0 \citep{Arnaud1996}, using $C$-statistics via \texttt{cstat} \citep{Cash1979}. 
We adopted an absorbed powerlaw model (\texttt{tbabs*powerlaw} in \texttt{xspec}, \citealt{Wilms2000}), and fixed the column density at the Galactic value of $N_{\rm H} =6.33 \times 10^{20} \,\pcmsq$ \citep{Willingale2013}. 
The resulting powerlaw photon index $\Gamma$ and the 0.3--10\,keV flux are listed in Table~\ref{tab:xray}.

In the 3rd--6th observations, AT2020xnd was not clearly detected. 
In order to determine the position of the source, we first ran \texttt{wavdetect} on the observations to obtain lists of positions for all sources in the \chandra\ ACIS-S3 FoV.
We then cross-matched the \chandra\ source lists with the \textit{Gaia} DR2 catalog \citep{GaiaCollaboration2018} to obtain the astrometric shifts. 
For obsID 23549, 
$\delta {\rm RA}= -0.73\pm0.12^{\prime\prime}$ and 
$\delta {\rm Dec}= -0.93 \pm 0.49^{\prime\prime}$; 
For obsID 23550, 
$\delta {\rm RA}= -0.75\pm0.14^{\prime\prime}$ and 
$\delta {\rm Dec}= -0.45 \pm 0.69^{\prime\prime}$; 
For obsID 23551, 
$\delta {\rm RA}= 0.03\pm0.23^{\prime\prime}$ and 
$\delta {\rm Dec}= 0.66 \pm 0.44^{\prime\prime}$; 
For obsID 25008, 
$\delta {\rm RA}= -0.26\pm0.33^{\prime\prime}$ and 
$\delta {\rm Dec}= 1.02 \pm 0.19^{\prime\prime}$. 

We used \texttt{srcflux} to estimate the 0.5--7\,keV count rate and the uncertainty. 
For obsID 23549, 2 counts were detected in a $1.5^{\prime\prime}$ circular region, corresponding to a 2.46-$\sigma$ (Gaussian equivalent) confidence-limit detection. 
For obsID 23550, no count was detected in a $1.7^{\prime\prime}$ circular region. 
For obsID 23551, 1 count was detected in a $1.5^{\prime\prime}$ circular region, corresponding to a 1.75-$\sigma$ (Gaussian equivalent) confidence-limit detection. 
For obsID 25008, no count was detected in a $1.4^{\prime\prime}$ circular region. 
We then converted the count rate to 0.3--10\,keV flux with \texttt{WebPIMMS}\footnote{\url{https://heasarc.gsfc.nasa.gov/cgi-bin/Tools/w3pimms/w3pimms.pl}}, assuming an absorbed powerlaw model with $\Gamma=1.5$ and $N_{\rm H} =6.33 \times 10^{20} \,\pcmsq$. 

\begin{deluxetable*}{ccccccc}[htb!]
\label{tab:xray}
\tablecaption{
    \chandra\ observations of AT2020xnd. 
}
\tablehead{
\colhead{ObsID} & \colhead{Exp. time}  & \colhead{Obs. time} & $\Delta t$ & 0.5--7\,keV count rate  & $\Gamma$ & 0.3--10\,keV flux\\
& (ks)   & (MJD) & (days) & ($10^{-3}\,\rm count\,s^{-1}$) & & ($10^{-14}\,\rm erg\,cm^{-2}\,s^{-1}$)
}
\startdata
23547 & 19.82 &  59157.8 & 20.8 & $1.24^{+0.28}_{-0.25}$ &
$1.23\pm0.48$ & $3.46^{+0.96}_{-1.27}$ \\
23548 & 19.82 &  59163.8 & 25.6 & $1.24^{+0.28}_{-0.25}$ &
$1.75^{+0.56}_{-0.54}$ & $2.79^{+0.75}_{-0.67}$ \\
23549 & 19.82 &  59179.1 & 37.9 & $0.09^{+0.10}_{-0.06}$ &
1.5 (fixed) &
$0.15^{+0.17}_{-0.11}$ \\
23550 & 19.75 &  59207.2 & 60.5 & $<0.13$ &
1.5 (fixed) & $<0.24$ \\
23551 & 16.86 & 59316.6 & 148.5 & $<0.23$ & 1.5 (fixed) & $<0.44$ \\
25008 & 19.82 & 59317.1 & 148.9 & $<0.13$ & 1.5 (fixed) & $<0.24$\\
\enddata
\tablecomments{For obsID 23547--23549, all uncertianties are represented by the 68\% confidence intervals. For the last three obsIDs, the limits are given by the upper bound of the 90\% confidence intervals. $\Delta t$ is is rest-frame days since the reference epoch of 59132\,MJD.}
\end{deluxetable*}

\onecolumngrid

\section{Synchrotron Self-Absorption Model}
\label{sec:appendix-model}

Below, we derive expressions for the source properties (size, magnetic field, density) as a function of observationally accessible properties, specifically the self-absorption frequency $\nu_a$ and the corresponding (peak) flux $F_a \equiv f_\nu(\nu_a)$. Following the notation in the main text, we assume a spherical shock of radius $R$ that propogates into an upstream medium whose density profile is $\rho \propto r^{-k}$. Similar to the standard \cite{Chevalier1998} model, we assume: that a non-thermal population of electrons is accelerated at the shock front, and that magnetic fields are amplified behind it; that the energy density of reltativistic electrons is a factor $\epsilon$ of the magnetic field energy density;\footnote{This is related to $\epsilon_{e}$ ($\epsilon_{B}$) that are often used to express the ratio of electron (magnetic field) energy density to the total kinetic shock power via $\epsilon \equiv \epsilon_e/\epsilon_B$.}
that the non-thermal electron population can be modeled as a power-law in Lorentz factor, $dN/d\gamma \propto \gamma^{-p}$, above $\gamma > \gamma_m$; and that the minimum Lorentz factor is $\gamma_m \approx 1$, and does not evolve with time.
The final assumption follows the standard \cite{Chevalier1998} hypothesis. We note however that an alternative hypothesis---related to the so-called deep-Newtonian regime first discussed by \cite{Sironi&Giannios13}---assumes that $\epsilon_e$ accounts for the energy of all electrons participating in diffusive-shock acceleration (not only those that are relativistic), is akin to an effectively time-dependent $\gamma_m$ \citep{Sironi&Giannios13}. In the present work, we focus on the ``standard'' model \citep{Chevalier1998}, and leave consideration of the ``deep-Newtonian'' ansatz for future work.

The self-absorption frequency $\nu_a$ is defined as the frequency at which asymptotic expressions for the optically thin and optically-thick flux equal one another, $f_\nu^{\rm thin}(\nu_a) = f_\nu^{\rm thick}(\nu_a)$.
In the standard, slow-cooling, case where $\nu_a < \nu_c$ (here $\nu_c$ is the cooling frequency), this leads to the relations
\begin{equation}
    R = \eta_1^{-\frac{p+4}{2(2p+13}} \zeta^{-\frac{p+6}{2p+13}} \epsilon^{-\frac{1}{2p+13}} (F_a D^2)^{\frac{p+6}{2p+13}} \nu_a^{-1} 
    ,
\end{equation}
\begin{equation}
    B = \eta_1^{-\frac{2(p+4)}{2p+13}} \zeta^{\frac{2}{2p+13}} \epsilon^{-\frac{4}{2p+13}} (F_a D^2)^{\frac{-2}{2p+13}} \nu_a
    ,
\end{equation}
between the source size and magnetic field ($R$, $B$) and self-absorbtion frequency $\nu_a$, flux density $F_a$ and angular-diameter distance $D$ (see \citet{Bright2022} for a discussion of the various required cosmological corrections). Note that the measured peak flux is expected to be slightly smaller than $F_a$ (as defined above), because the transition between optically thin/thick limits in reality is smooth (and depends on geometry) rather than a broken power-law.

Above, we have defined the constants
\begin{equation}
    \eta_1(p) \equiv \left[\frac{(p-2) \sigma_T}{12 \pi^2 m_e^2 c}\right]^{\frac{2}{p+4}} \left(\frac{e}{2\pi m_e c}\right)^{\frac{p-2}{p+4}}
    \approx \left[ (p-2) \times 2.26 \times 10^{17} \right]^{\frac{2}{p+4}} \left( 2.80 \times 10^6 \right)^{\frac{p-2}{p+4}}
\end{equation}
and 
\begin{equation}
    \zeta \equiv \frac{1}{3} \left( 2\pi m_e \right)^{3/2} \left(c/e\right)^{1/2}
    \approx 1.14 \times 10^{-30}
    ,
\end{equation}
where the numerical values are given in cgs units.

For the specific case where $p=3$, we find
\begin{equation}
    R \approx (5.1 \times 10^{15}\,\cm) \, \epsilon^{-1/19} \left( \frac{F_a}{\jy} \right)^{9/19} \left(\frac{D}{\mpc}\right)^{18/19} \left( \frac{\nu_a}{5\,\ghz} \right)^{-1} ,
\end{equation}
\begin{equation}
    B \approx (0.30\,\gauss) \, \epsilon^{-4/19} \left( \frac{F_a}{\jy} \right)^{-2/19} \left( \frac{D}{\mpc} \right)^{-4/19} \left( \frac{\nu_a}{5\,\ghz} \right)
    .
\end{equation}
This is consistent with the results of \cite{Chevalier1998}.

We now also consider the novel regime where the cooling frquency is below the self-absorption frequency.
In this limit, we instead find
\begin{equation}
    R = \xi^{\frac{1}{2(2p+7)}} \eta_1(p)^{-\frac{p+4}{2(2p+7)}} \zeta^{-\frac{p+3}{2p+7}} \epsilon^{-\frac{1}{2p+7}} (F_a D^2)^{\frac{p+3}{2p+7}} \nu_a^{-\frac{2p+5}{2p+7}} t^{\frac{1}{2p+7}} 
    ,
\end{equation}
\begin{equation}
    B = \xi^{\frac{2}{2p+7}} \eta_1(p)^{-\frac{2(p+4)}{2p+7}} \zeta^{\frac{2}{2p+7}} \epsilon^{-\frac{4}{2p+7}} (F_a D^2)^{\frac{-2}{2p+7}} \nu_a^{\frac{2p+15}{2p+7}} t^{\frac{4}{2p+7}}
    ,
\end{equation}
where the solution now depends explicitly on the elapsed time $t$, and the constant $\xi$ in cgs units is given by
\begin{equation}
    \xi \equiv \frac{\sigma_T^2}{18\pi m_e c e} \approx 5.97 \times 10^{-25}
    .
\end{equation}

Focusing again on the case where $p=3$, we find that in the fast-cooling regime ($\nu_c < \nu_a$)
\begin{equation}
    R \approx (4.2 \times 10^{15}\,\cm) \, \epsilon^{-1/13} \left( \frac{F_a}{\jy} \right)^{6/13} \left(\frac{D}{\mpc}\right)^{12/13} \left( \frac{\nu_a}{5\,\ghz} \right)^{-11/13} \left(\frac{t}{100\,{\rm day}}\right)^{1/13} ,
\end{equation}
\begin{equation}
    B \approx (0.14\,\gauss) \, \epsilon^{-4/13} \left( \frac{F_a}{\jy} \right)^{-2/13} \left( \frac{D}{\mpc} \right)^{-4/13} \left( \frac{\nu_a}{5\,\ghz} \right)^{21/13} \left(\frac{t}{100\,{\rm day}}\right)^{4/13}
    .
\end{equation}

The number density of relativistic electrons $n_{e,rel}$ (those with Lorentz factor $>\gamma_m$) can then be inferred using the magnetic field expressions above, it is 
\begin{equation}
    n_{e,rel} = \frac{p-2}{p-1} \epsilon \frac{B^2}{m_e c^2}
    \underset{p=3}{\approx}
    \begin{cases}
    (5.5 \times 10^4 \,{\rm cm}^{-3}) \, 
    \epsilon^{11/19} \left( \frac{F_a}{\jy} \right)^{-4/19} \left( \frac{D}{\mpc} \right)^{-8/19} \left( \frac{\nu_a}{5\,\ghz} \right)^2
    &, \nu_a < \nu_c
    \\
    (1.2 \times 10^4 \,{\rm cm}^{-3}) \, 
    \epsilon^{5/13} \left( \frac{F_a}{\jy} \right)^{-4/13} \left( \frac{D}{\mpc} \right)^{-8/13} \left( \frac{\nu_a}{5\,\ghz} \right)^{42/13} \left(\frac{t}{100\,{\rm day}}\right)^{8/13}
    &, \nu_a > \nu_c
    \end{cases}
\end{equation}
This is related to the upstream density $n$ as $n \approx n_{e,rel} \times 2\gamma_m \epsilon_e^{-1} (m_e/m_p) (v/c)^{-2} (p-1)/(p-2)$ where $v$ is the shock velocity.

One can verify the regime of relevance, i.e. whether $\nu_a \lessgtr \nu_c$, by comparing $\nu_a$ from the expressions above, to $\nu_c = 18\pi m_e c e / \sigma_T^2 B^3 t^2$.
This gives
\begin{equation}
    \frac{\nu_a}{\nu_c} \underset{p=3}{\approx}
    \begin{cases}
    6.2 \times 10^{-3} \, \epsilon^{12/19} \left( \frac{F_a}{\jy} \right)^{6/19} \left( \frac{D}{\mpc} \right)^{12/19} \left( \frac{\nu_a}{5\,\ghz} \right)^{-3} \left(\frac{t}{100\,{\rm day}}\right)^{-2}
    &, \nu_a < \nu_c
    \\
    5.9 \times 10^{-4} \, \epsilon^{12/13} \left( \frac{F_a}{\jy} \right)^{6/13} \left( \frac{D}{\mpc} \right)^{12/13} \left( \frac{\nu_a}{5\,\ghz} \right)^{-63/13} \left(\frac{t}{100\,{\rm day}}\right)^{-38/13}
    &, \nu_a > \nu_c
    \end{cases}
\end{equation}

\subsection{Temporal Scaling}

The light-curve evolution can be found from the equations above by introducing the assumption that $R \propto t^{\alpha_r}$, i.e that the position of the shock front scales as a power-law in time. Along with the assumption of a power-law density profile, $n \propto r^{-k}$, we find that $B \sim \sqrt{16\pi \epsilon_B n m_p v^2} \propto t^{-(1-\alpha_r)-\alpha_r k/2}$.
Altogether, this leads to 
\begin{equation}
    \nu_a \propto
    \begin{cases}
    t^{ -\frac{2(p+6)-2\alpha_r(p+8)+\alpha_r k(p+6)}{2(p+4)} }
    \gamma_m(t)^{\frac{2(p-2)}{p+4}}
    &, \nu_a < \nu_c
    \\
    t^{ \alpha_r-1-\alpha_r\frac{k(p+3)}{2(p+5)} }
    \gamma_m(t)^{\frac{2(p-2)[4(p+5)^2+5p]}{(2p+15)(2p+5)(p+5)}}
    &, \nu_a > \nu_c
    \end{cases}
\end{equation}
and 
\begin{equation}
    F_a \propto
    \begin{cases}
    t^{ -\frac{(2p+13)[2-\alpha_r (4-k)]}{2(p+4)} }
    \gamma_m(t)^{\frac{5(p-2)}{p+4}}
    &, \nu_a < \nu_c
    \\
    t^{ 4\alpha_r - 2 - \alpha_r \frac{k(2p+5)}{2(p+5)} }
    \gamma_m(t)^{\frac{5(p-2)}{p+5}}
    &, \nu_a > \nu_c
    \end{cases}
    ,
\end{equation}
where the minimal electron Lorentz factor is typically constant in time, $\gamma_m \approx 1$. For shocks of sufficiently high velocity (or if the fraction of swept up electrons that participates in diffusive shock acceleration is very small), $\gamma_m \propto v^2 \propto t^{2(\alpha_r-1)}$.
In the case of constant shock velocity ($\alpha_r=1$) and a wind density profile ($k=2$), we recover the well known result in the slow-cooling regime ($\nu_a < \nu_c$), that $\nu_a \propto t^{-1}$ and $F_a \propto t^0$.
In the fast-cooling regime, however, one finds that the self-absorption frequency decreases more gradually with time ($\nu_a \propto t^{-\frac{p+3}{p+5}}$) and the peak flux increases (as $F_a \propto t^{\frac{5}{p+5}}$).

The cooling frequency scales as $\nu_c \propto B^{-3} t^{-2} \propto t^{-2 + 3(1-\alpha_r) + 3\alpha_r k/2}$, so that the ratio of self-absorption to cooling frequency evolves as
\begin{equation}
    \frac{\nu_a}{\nu_c} \propto
    \begin{cases}
    t^{ -\frac{(4\alpha_r-2)(p+5)-\alpha_r k(2p+9)}{p+4} }
    &, \nu_a < \nu_c
    \\
    t^{ -\frac{(4\alpha_r-2)(p+5)-\alpha_r k(2p+9)}{p+5} }
    &, \nu_a > \nu_c
    \end{cases}
    .
\end{equation}
This {\it increases} with time if $k < (4\alpha_r-2)(p+5)/\alpha_r(2p+9) \approx 1.1 (2 - 1/\alpha_r)$, that is---if the ambient density profile is relatively shallow and the shock does not decelerate too dramatically ($\alpha_r \approx 1$).

Overall, for the fiducial case where $\gamma_m \approx 1$, the light-curve evolution at a given frequency scales as
\begin{equation}
    f_\nu 
    = F_a
    \begin{cases}
    \left(\frac{\nu_*}{\nu_a}\right)^{-\frac{p-1}{2}} \left(\frac{\nu}{\nu_*}\right)^{-\frac{p}{2}}
    &, \nu > \nu_* \equiv \max( \nu_c, \nu_a)
    \\
    \left(\frac{\nu}{\nu_a}\right)^{-\frac{p-1}{2}}
    &, \nu_a < \nu < \nu_c
    \\
    \left(\frac{\nu}{\nu_a}\right)^{5/2}
    &, \nu < \nu_a
    \end{cases}
    \propto 
    \begin{cases}
    t^{-[2(p+4)+\alpha_r k(p+2) - 2\alpha_r (p+8)]/4}
    &, \nu > \max( \nu_c, \nu_a)
    \\
    t^{-[(2+\alpha_r k)(p+5) - 2\alpha_r (p+11)]/4}
    &, \nu_a < \nu < \nu_c
    \\
    t^{-[2+\alpha_r (k+6)]/4}
    &, \nu < \nu_a
    \end{cases}
\end{equation}

\vspace{5mm}
\facilities{EVLA, VLA, IRAM:NOEMA, SMA}

\software{{\tt CASA} \citep{McMullin2007},
          {\tt astropy} \citep{Astropy2013,Astropy2018},
          {\tt matplotlib} \citep{Hunter2007},
          {\tt scipy} \citep{Virtanen2020},
          {\tt pyne2001} 
}

\acknowledgements

A.Y.Q.H. would like to thank Eliot Quataert, Dan Kasen, Sterl Phinney, Anatoly Spitkovsky, and Shri Kulkarni for useful discussions about steep-spectrum radio sources and relativistic Maxwellians;
Kunal Mooley for assistance with VLA calibration;
Fabian Walter for advice regarding NOEMA data reduction; and Joe Bright, Greg Sivakoff, and Susan Clark for a careful reading of the manuscript.  D.K. and A.O. are supported by NSF grant AST-1816492.
B.M. is supported by NASA through the NASA Hubble Fellowship grant \#HST-HF2-51412.001-A awarded by the Space Telescope Science Institute, which is operated by the Association of Universities for Research in Astronomy, Inc., for NASA, under contract NAS5-26555.
The authors would like to thank the anonymous referee for detailed comments that greatly improved the clarity of the paper.

The Submillimeter Array is a joint project between the Smithsonian Astrophysical Observatory and the Academia Sinica Institute of Astronomy and Astrophysics and is funded by the Smithsonian Institution and the Academia Sinica.

This work is based on observations carried out under project numbers D20AF and D20AG with the IRAM NOEMA Interferometer. IRAM is supported by INSU/CNRS (France), MPG (Germany) and IGN (Spain).

The National Radio Astronomy Observatory is a facility of the National Science Foundation operated under cooperative agreement by Associated Universities, Inc.

The Australia Telescope Compact Array is part of the Australia Telescope National Facility (grid.421683.a) which is funded by the Australian Government for operation as a National Facility managed by CSIRO. We acknowledge the Gomeroi people as the traditional owners of the Observatory site.

The scientific results reported in this article are based in part on observations made by the Chandra X-ray Observatory. This research has made use of software provided by the Chandra X-ray Center (CXC) in the application package CIAO.

\bibliography{refs}

\begin{thebibliography}{}
\expandafter\ifx\csname natexlab\endcsname\relax\def\natexlab#1{#1}\fi
\providecommand{\url}[1]{\href{#1}{#1}}
\providecommand{\dodoi}[1]{doi:~\href{http://doi.org/#1}{\nolinkurl{#1}}}
\providecommand{\doeprint}[1]{\href{http://ascl.net/#1}{\nolinkurl{http://ascl.net/#1}}}
\providecommand{\doarXiv}[1]{\href{https://arxiv.org/abs/#1}{\nolinkurl{https://arxiv.org/abs/#1}}}

\bibitem[{{Abazajian} {et~al.}(2019){Abazajian}, {Addison}, {Adshead}, {Ahmed},
  {Allen}, {Alonso}, {Alvarez}, {Anderson}, {Arnold}, {Baccigalupi}, {Bailey},
  {Barkats}, {Barron}, {Barry}, {Bartlett}, {Basu Thakur}, {Battaglia},
  {Baxter}, {Bean}, {Bebek}, {Bender}, {Benson}, {Berger}, {Bhimani},
  {Bischoff}, {Bleem}, {Bocquet}, {Boddy}, {Bonato}, {Bond}, {Borrill},
  {Bouchet}, {Brown}, {Bryan}, {Burkhart}, {Buza}, {Byrum}, {Calabrese},
  {Calafut}, {Caldwell}, {Carlstrom}, {Carron}, {Cecil}, {Challinor}, {Chang},
  {Chinone}, {Cho}, {Cooray}, {Crawford}, {Crites}, {Cukierman}, {Cyr-Racine},
  {de Haan}, {de Zotti}, {Delabrouille}, {Demarteau}, {Devlin}, {Di Valentino},
  {Dobbs}, {Duff}, {Duivenvoorden}, {Dvorkin}, {Edwards}, {Eimer}, {Errard},
  {Essinger-Hileman}, {Fabbian}, {Feng}, {Ferraro}, {Filippini}, {Flauger},
  {Flaugher}, {Fraisse}, {Frolov}, {Galitzki}, {Galli}, {Ganga}, {Gerbino},
  {Gilchriese}, {Gluscevic}, {Green}, {Grin}, {Grohs}, {Gualtieri}, {Guarino},
  {Gudmundsson}, {Habib}, {Haller}, {Halpern}, {Halverson}, {Hanany},
  {Harrington}, {Hasegawa}, {Hasselfield}, {Hazumi}, {Heitmann}, {Henderson},
  {Henning}, {Hill}, {Hlozek}, {Holder}, {Holzapfel}, {Hubmayr},
  {Huffenberger}, {Huffer}, {Hui}, {Irwin}, {Johnson}, {Johnstone}, {Jones},
  {Karkare}, {Katayama}, {Kerby}, {Kernovsky}, {Keskitalo}, {Kisner}, {Knox},
  {Kosowsky}, {Kovac}, {Kovetz}, {Kuhlmann}, {Kuo}, {Kurita}, {Kusaka},
  {Lahteenmaki}, {Lawrence}, {Lee}, {Lewis}, {Li}, {Linder}, {Loverde},
  {Lowitz}, {Madhavacheril}, {Mantz}, {Matsuda}, {Mauskopf}, {McMahon},
  {McQuinn}, {Meerburg}, {Melin}, {Meyers}, {Millea}, {Mohr}, {Moncelsi},
  {Mroczkowski}, {Mukherjee}, {M{\"u}nchmeyer}, {Nagai}, {Nagy}, {Namikawa},
  {Nati}, {Natoli}, {Negrello}, {Newburgh}, {Niemack}, {Nishino}, {Nordby},
  {Novosad}, {O'Connor}, {Obied}, {Padin}, {Pandey}, {Partridge}, {Pierpaoli},
  {Pogosian}, {Pryke}, {Puglisi}, {Racine}, {Raghunathan}, {Rahlin},
  {Rajagopalan}, {Raveri}, {Reichanadter}, {Reichardt}, {Remazeilles}, {Rocha},
  {Roe}, {Roy}, {Ruhl}, {Salatino}, {Saliwanchik}, {Schaan}, {Schillaci},
  {Schmittfull}, {Scott}, {Sehgal}, {Shandera}, {Sheehy}, {Sherwin},
  {Shirokoff}, {Simon}, {Slosar}, {Somerville}, {Spergel}, {Staggs}, {Stark},
  {Stompor}, {Story}, {Stoughton}, {Suzuki}, {Tajima}, {Teply}, {Thompson},
  {Timbie}, {Tomasi}, {Treu}, {Tristram}, {Tucker}, {Umilt{\`a}}, {van
  Engelen}, {Vieira}, {Vieregg}, {Vogelsberger}, {Wang}, {Watson}, {White},
  {Whitehorn}, {Wollack}, {Kimmy Wu}, {Xu}, {Yasini}, {Yeck}, {Yoon}, {Young},
  \& {Zonca}}]{Abazajian2019}
{Abazajian}, K., {Addison}, G., {Adshead}, P., {et~al.} 2019, arXiv e-prints,
  arXiv:1907.04473.
\newblock \doarXiv{1907.04473}

\bibitem[{{Alexander} {et~al.}(2016){Alexander}, {Berger}, {Guillochon},
  {Zauderer}, \& {Williams}}]{Alexander2016}
{Alexander}, K.~D., {Berger}, E., {Guillochon}, J., {Zauderer}, B.~A., \&
  {Williams}, P.~K.~G. 2016, \apjl, 819, L25,
  \dodoi{10.3847/2041-8205/819/2/L25}

\bibitem[{{Alexander} {et~al.}(2020){Alexander}, {van Velzen}, {Horesh}, \&
  {Zauderer}}]{Alexander2020}
{Alexander}, K.~D., {van Velzen}, S., {Horesh}, A., \& {Zauderer}, B.~A. 2020,
  \ssr, 216, 81, \dodoi{10.1007/s11214-020-00702-w}

\bibitem[{{Arnaud}(1996)}]{Arnaud1996}
{Arnaud}, K.~A. 1996, in Astronomical Society of the Pacific Conference Series,
  Vol. 101, Astronomical Data Analysis Software and Systems V, ed. G.~H.
  {Jacoby} \& J.~{Barnes}, 17

\bibitem[{{Astropy Collaboration} {et~al.}(2013){Astropy Collaboration},
  {Robitaille}, {Tollerud}, {Greenfield}, {Droettboom}, {Bray}, {Aldcroft},
  {Davis}, {Ginsburg}, {Price-Whelan}, {Kerzendorf}, {Conley}, {Crighton},
  {Barbary}, {Muna}, {Ferguson}, {Grollier}, {Parikh}, {Nair}, {Unther},
  {Deil}, {Woillez}, {Conseil}, {Kramer}, {Turner}, {Singer}, {Fox}, {Weaver},
  {Zabalza}, {Edwards}, {Azalee Bostroem}, {Burke}, {Casey}, {Crawford},
  {Dencheva}, {Ely}, {Jenness}, {Labrie}, {Lim}, {Pierfederici}, {Pontzen},
  {Ptak}, {Refsdal}, {Servillat}, \& {Streicher}}]{Astropy2013}
{Astropy Collaboration}, {Robitaille}, T.~P., {Tollerud}, E.~J., {et~al.} 2013,
  \aap, 558, A33, \dodoi{10.1051/0004-6361/201322068}

\bibitem[{{Astropy Collaboration} {et~al.}(2018){Astropy Collaboration},
  {Price-Whelan}, {Sip{\H{o}}cz}, {G{\"u}nther}, {Lim}, {Crawford}, {Conseil},
  {Shupe}, {Craig}, {Dencheva}, {Ginsburg}, {Vand erPlas}, {Bradley},
  {P{\'e}rez-Su{\'a}rez}, {de Val-Borro}, {Aldcroft}, {Cruz}, {Robitaille},
  {Tollerud}, {Ardelean}, {Babej}, {Bach}, {Bachetti}, {Bakanov}, {Bamford},
  {Barentsen}, {Barmby}, {Baumbach}, {Berry}, {Biscani}, {Boquien}, {Bostroem},
  {Bouma}, {Brammer}, {Bray}, {Breytenbach}, {Buddelmeijer}, {Burke},
  {Calderone}, {Cano Rodr{\'\i}guez}, {Cara}, {Cardoso}, {Cheedella}, {Copin},
  {Corrales}, {Crichton}, {D'Avella}, {Deil}, {Depagne}, {Dietrich}, {Donath},
  {Droettboom}, {Earl}, {Erben}, {Fabbro}, {Ferreira}, {Finethy}, {Fox},
  {Garrison}, {Gibbons}, {Goldstein}, {Gommers}, {Greco}, {Greenfield},
  {Groener}, {Grollier}, {Hagen}, {Hirst}, {Homeier}, {Horton}, {Hosseinzadeh},
  {Hu}, {Hunkeler}, {Ivezi{\'c}}, {Jain}, {Jenness}, {Kanarek}, {Kendrew},
  {Kern}, {Kerzendorf}, {Khvalko}, {King}, {Kirkby}, {Kulkarni}, {Kumar},
  {Lee}, {Lenz}, {Littlefair}, {Ma}, {Macleod}, {Mastropietro}, {McCully},
  {Montagnac}, {Morris}, {Mueller}, {Mumford}, {Muna}, {Murphy}, {Nelson},
  {Nguyen}, {Ninan}, {N{\"o}the}, {Ogaz}, {Oh}, {Parejko}, {Parley}, {Pascual},
  {Patil}, {Patil}, {Plunkett}, {Prochaska}, {Rastogi}, {Reddy Janga},
  {Sabater}, {Sakurikar}, {Seifert}, {Sherbert}, {Sherwood-Taylor}, {Shih},
  {Sick}, {Silbiger}, {Singanamalla}, {Singer}, {Sladen}, {Sooley},
  {Sornarajah}, {Streicher}, {Teuben}, {Thomas}, {Tremblay}, {Turner},
  {Terr{\'o}n}, {van Kerkwijk}, {de la Vega}, {Watkins}, {Weaver}, {Whitmore},
  {Woillez}, {Zabalza}, \& {Astropy Contributors}}]{Astropy2018}
{Astropy Collaboration}, {Price-Whelan}, A.~M., {Sip{\H{o}}cz}, B.~M., {et~al.}
  2018, \aj, 156, 123, \dodoi{10.3847/1538-3881/aabc4f}

\bibitem[{{Bellm} {et~al.}(2019){Bellm}, {Kulkarni}, {Graham}, {Dekany},
  {Smith}, {Riddle}, {Masci}, {Helou}, {Prince}, {Adams}, {Barbarino},
  {Barlow}, {Bauer}, {Beck}, {Belicki}, {Biswas}, {Blagorodnova}, {Bodewits},
  {Bolin}, {Brinnel}, {Brooke}, {Bue}, {Bulla}, {Burruss}, {Cenko}, {Chang},
  {Connolly}, {Coughlin}, {Cromer}, {Cunningham}, {De}, {Delacroix}, {Desai},
  {Duev}, {Eadie}, {Farnham}, {Feeney}, {Feindt}, {Flynn}, {Franckowiak},
  {Frederick}, {Fremling}, {Gal-Yam}, {Gezari}, {Giomi}, {Goldstein},
  {Golkhou}, {Goobar}, {Groom}, {Hacopians}, {Hale}, {Henning}, {Ho}, {Hover},
  {Howell}, {Hung}, {Huppenkothen}, {Imel}, {Ip}, {Ivezi{\'c}}, {Jackson},
  {Jones}, {Juric}, {Kasliwal}, {Kaspi}, {Kaye}, {Kelley}, {Kowalski},
  {Kramer}, {Kupfer}, {Landry}, {Laher}, {Lee}, {Lin}, {Lin}, {Lunnan},
  {Giomi}, {Mahabal}, {Mao}, {Miller}, {Monkewitz}, {Murphy}, {Ngeow},
  {Nordin}, {Nugent}, {Ofek}, {Patterson}, {Penprase}, {Porter}, {Rauch},
  {Rebbapragada}, {Reiley}, {Rigault}, {Rodriguez}, {van Roestel}, {Rusholme},
  {van Santen}, {Schulze}, {Shupe}, {Singer}, {Soumagnac}, {Stein}, {Surace},
  {Sollerman}, {Szkody}, {Taddia}, {Terek}, {Van Sistine}, {van Velzen},
  {Vestrand}, {Walters}, {Ward}, {Ye}, {Yu}, {Yan}, \&
  {Zolkower}}]{Bellm2019_ztf}
{Bellm}, E.~C., {Kulkarni}, S.~R., {Graham}, M.~J., {et~al.} 2019, \pasp, 131,
  018002, \dodoi{10.1088/1538-3873/aaecbe}

\bibitem[{{Berger} {et~al.}(2012){Berger}, {Zauderer}, {Pooley}, {Soderberg},
  {Sari}, {Brunthaler}, \& {Bietenholz}}]{Berger2012}
{Berger}, E., {Zauderer}, A., {Pooley}, G.~G., {et~al.} 2012, \apj, 748, 36,
  \dodoi{10.1088/0004-637X/748/1/36}

\bibitem[{{Bietenholz} {et~al.}(2021){Bietenholz}, {Bartel}, {Argo}, {Dua},
  {Ryder}, \& {Soderberg}}]{Bietenholz2021}
{Bietenholz}, M.~F., {Bartel}, N., {Argo}, M., {et~al.} 2021, \apj, 908, 75,
  \dodoi{10.3847/1538-4357/abccd9}

\bibitem[{{Blandford} \& {Eichler}(1987)}]{Blandford1987}
{Blandford}, R., \& {Eichler}, D. 1987, \physrep, 154, 1,
  \dodoi{10.1016/0370-1573(87)90134-7}

\bibitem[{{Bright} {et~al.}(2020{\natexlab{a}}){Bright}, {Wieringa}, {Laskar},
  {Margutti}, {Coppejans}, {Alexander}, {DeMarchi}, \&
  {Matthews}}]{Bright2021a}
{Bright}, J., {Wieringa}, M., {Laskar}, T., {et~al.} 2020{\natexlab{a}}, The
  Astronomer's Telegram, 14148, 1

\bibitem[{{Bright} {et~al.}(2020{\natexlab{b}}){Bright}, {Wieringa},
  {Margutti}, {Laskar}, {Alexander}, {DeMarchi}, \& {Matthews}}]{Bright2021b}
{Bright}, J., {Wieringa}, M., {Margutti}, R., {et~al.} 2020{\natexlab{b}}, The
  Astronomer's Telegram, 14249, 1

\bibitem[{{Bright} {et~al.}(2022){Bright}, {Margutti}, {Matthews}, {Brethauer},
  {Coppejans}, {Wieringa}, {Metzger}, {DeMarchi}, {Laskar}, {Romero},
  {Alexander}, {Horesh}, {Migliori}, {Chornock}, {Berger}, {Bietenholz},
  {Devlin}, {Dicker}, {Jacobson-Gal{\'a}n}, {Mason}, {Milisavljevic}, {Motta},
  {Mroczkowski}, {Ramirez-Ruiz}, {Rhodes}, {Sarazin}, {Sfaradi}, \&
  {Sievers}}]{Bright2022}
{Bright}, J.~S., {Margutti}, R., {Matthews}, D., {et~al.} 2022, \apj, 926, 112,
  \dodoi{10.3847/1538-4357/ac4506}

\bibitem[{{Caprioli} {et~al.}(2020){Caprioli}, {Haggerty}, \&
  {Blasi}}]{Caprioli2020}
{Caprioli}, D., {Haggerty}, C.~C., \& {Blasi}, P. 2020, \apj, 905, 2,
  \dodoi{10.3847/1538-4357/abbe05}

\bibitem[{{Carlstrom} {et~al.}(2011){Carlstrom}, {Ade}, {Aird}, {Benson},
  {Bleem}, {Busetti}, {Chang}, {Chauvin}, {Cho}, {Crawford}, {Crites}, {Dobbs},
  {Halverson}, {Heimsath}, {Holzapfel}, {Hrubes}, {Joy}, {Keisler}, {Lanting},
  {Lee}, {Leitch}, {Leong}, {Lu}, {Lueker}, {Luong-Van}, {McMahon}, {Mehl},
  {Meyer}, {Mohr}, {Montroy}, {Padin}, {Plagge}, {Pryke}, {Ruhl}, {Schaffer},
  {Schwan}, {Shirokoff}, {Spieler}, {Staniszewski}, {Stark}, {Tucker},
  {Vanderlinde}, {Vieira}, \& {Williamson}}]{Carlstrom2011}
{Carlstrom}, J.~E., {Ade}, P.~A.~R., {Aird}, K.~A., {et~al.} 2011, \pasp, 123,
  568, \dodoi{10.1086/659879}

\bibitem[{{Cash}(1979)}]{Cash1979}
{Cash}, W. 1979, \apj, 228, 939, \dodoi{10.1086/156922}

\bibitem[{{Chandra} \& {Frail}(2012)}]{Chandra2012}
{Chandra}, P., \& {Frail}, D.~A. 2012, \apj, 746, 156,
  \dodoi{10.1088/0004-637X/746/2/156}

\bibitem[{{Chevalier}(1998)}]{Chevalier1998}
{Chevalier}, R.~A. 1998, \apj, 499, 810, \dodoi{10.1086/305676}

\bibitem[{{Chevalier} \& {Fransson}(2006)}]{Chevalier2006}
{Chevalier}, R.~A., \& {Fransson}, C. 2006, \apj, 651, 381,
  \dodoi{10.1086/507606}

\bibitem[{{Cohen} {et~al.}(2005){Cohen}, {Clarke}, {Feretti}, \&
  {Kassim}}]{Cohen2005}
{Cohen}, A.~S., {Clarke}, T.~E., {Feretti}, L., \& {Kassim}, N.~E. 2005, \apjl,
  620, L5, \dodoi{10.1086/428572}

\bibitem[{{Coppejans} {et~al.}(2020){Coppejans}, {Margutti}, {Terreran},
  {Nayana}, {Coughlin}, {Laskar}, {Alexander}, {Bietenholz}, {Caprioli},
  {Chandra}, {Drout}, {Frederiks}, {Frohmaier}, {Hurley}, {Kochanek},
  {MacLeod}, {Meisner}, {Nugent}, {Ridnaia}, {Sand}, {Svinkin}, {Ward}, {Yang},
  {Baldeschi}, {Chilingarian}, {Dong}, {Esquivia}, {Fong}, {Guidorzi},
  {Lundqvist}, {Milisavljevic}, {Paterson}, {Reichart}, {Shappee}, {Stroh},
  {Valenti}, {Zauderer}, \& {Zhang}}]{Coppejans2020}
{Coppejans}, D.~L., {Margutti}, R., {Terreran}, G., {et~al.} 2020, \apjl, 895,
  L23, \dodoi{10.3847/2041-8213/ab8cc7}

\bibitem[{{Cordes} \& {Lazio}(2002)}]{Cordes2002}
{Cordes}, J.~M., \& {Lazio}, T.~J.~W. 2002, arXiv e-prints, astro.
\newblock \doarXiv{astro-ph/0207156}

\bibitem[{{Corsi} {et~al.}(2014){Corsi}, {Ofek}, {Gal-Yam}, {Frail},
  {Kulkarni}, {Fox}, {Kasliwal}, {Sullivan}, {Horesh}, {Carpenter}, {Maguire},
  {Arcavi}, {Cenko}, {Cao}, {Mooley}, {Pan}, {Sesar}, {Sternberg}, {Xu},
  {Bersier}, {James}, {Bloom}, \& {Nugent}}]{Corsi2014}
{Corsi}, A., {Ofek}, E.~O., {Gal-Yam}, A., {et~al.} 2014, \apj, 782, 42,
  \dodoi{10.1088/0004-637X/782/1/42}

\bibitem[{{de Ugarte Postigo} {et~al.}(2012){de Ugarte Postigo}, {Lundgren},
  {Mart{\'\i}n}, {Garcia-Appadoo}, {de Gregorio Monsalvo}, {Peck},
  {Micha{\l}owski}, {Th{\"o}ne}, {Campana}, {Gorosabel}, {Tanvir}, {Wiersema},
  {Castro-Tirado}, {Schulze}, {De Breuck}, {Petitpas}, {Hjorth}, {Jakobsson},
  {Covino}, {Fynbo}, {Winters}, {Bremer}, {Levan}, {Llorente},
  {S{\'a}nchez-Ram{\'\i}rez}, {Tello}, \& {Salvaterra}}]{deUgartoPostigo2012}
{de Ugarte Postigo}, A., {Lundgren}, A., {Mart{\'\i}n}, S., {et~al.} 2012,
  \aap, 538, A44, \dodoi{10.1051/0004-6361/201117848}

\bibitem[{{Dong} {et~al.}(2021){Dong}, {Hallinan}, {Nakar}, {Ho}, {Hughes},
  {Hotokezaka}, {Myers}, {De}, {Mooley}, {Ravi}, {Horesh}, {Kasliwal}, \&
  {Kulkarni}}]{Dong2021}
{Dong}, D.~Z., {Hallinan}, G., {Nakar}, E., {et~al.} 2021, Science, 373, 1125,
  \dodoi{10.1126/science.abg6037}

\bibitem[{{Eftekhari} {et~al.}(2018){Eftekhari}, {Berger}, {Zauderer},
  {Margutti}, \& {Alexander}}]{Eftekhari2018}
{Eftekhari}, T., {Berger}, E., {Zauderer}, B.~A., {Margutti}, R., \&
  {Alexander}, K.~D. 2018, \apj, 854, 86, \dodoi{10.3847/1538-4357/aaa8e0}

\bibitem[{{Eftekhari} {et~al.}(2021){Eftekhari}, {Berger}, {Metzger}, {Laskar},
  {Villar}, {Alexander}, {Holder}, {Vieira}, {Whitehorn}, \&
  {Williams}}]{Eftekhari2021}
{Eftekhari}, T., {Berger}, E., {Metzger}, B.~D., {et~al.} 2021, arXiv e-prints,
  arXiv:2110.05494.
\newblock \doarXiv{2110.05494}

\bibitem[{{Eichler} \& {Granot}(2006)}]{Eichler2006}
{Eichler}, D., \& {Granot}, J. 2006, \apjl, 641, L5, \dodoi{10.1086/503667}

\bibitem[{{Frater} {et~al.}(1992){Frater}, {Brooks}, \&
  {Whiteoak}}]{Frater1992}
{Frater}, R.~H., {Brooks}, J.~W., \& {Whiteoak}, J.~B. 1992, Journal of
  Electrical and Electronics Engineering Australia, 12, 103

\bibitem[{{Fruscione} {et~al.}(2006){Fruscione}, {McDowell}, {Allen},
  {Brickhouse}, {Burke}, {Davis}, {Durham}, {Elvis}, {Galle}, {Harris},
  {Huenemoerder}, {Houck}, {Ishibashi}, {Karovska}, {Nicastro}, {Noble},
  {Nowak}, {Primini}, {Siemiginowska}, {Smith}, \& {Wise}}]{Fruscione2006}
{Fruscione}, A., {McDowell}, J.~C., {Allen}, G.~E., {et~al.} 2006, Society of
  Photo-Optical Instrumentation Engineers (SPIE) Conference Series, Vol. 6270,
  {CIAO: Chandra's data analysis system}, 62701V, \dodoi{10.1117/12.671760}

\bibitem[{{Gaia Collaboration} {et~al.}(2018){Gaia Collaboration}, {Brown},
  {Vallenari}, {Prusti}, {de Bruijne}, {Babusiaux}, {Bailer-Jones}, {Biermann},
  {Evans}, {Eyer}, {Jansen}, {Jordi}, {Klioner}, {Lammers}, {Lindegren},
  {Luri}, {Mignard}, {Panem}, {Pourbaix}, {Randich}, {Sartoretti}, {Siddiqui},
  {Soubiran}, {van Leeuwen}, {Walton}, {Arenou}, {Bastian}, {Cropper},
  {Drimmel}, {Katz}, {Lattanzi}, {Bakker}, {Cacciari}, {Casta{\~n}eda},
  {Chaoul}, {Cheek}, {De Angeli}, {Fabricius}, {Guerra}, {Holl}, {Masana},
  {Messineo}, {Mowlavi}, {Nienartowicz}, {Panuzzo}, {Portell}, {Riello},
  {Seabroke}, {Tanga}, {Th{\'e}venin}, {Gracia-Abril}, {Comoretto},
  {Garcia-Reinaldos}, {Teyssier}, {Altmann}, {Andrae}, {Audard},
  {Bellas-Velidis}, {Benson}, {Berthier}, {Blomme}, {Burgess}, {Busso},
  {Carry}, {Cellino}, {Clementini}, {Clotet}, {Creevey}, {Davidson}, {De
  Ridder}, {Delchambre}, {Dell'Oro}, {Ducourant},
  {Fern{\'a}ndez-Hern{\'a}ndez}, {Fouesneau}, {Fr{\'e}mat}, {Galluccio},
  {Garc{\'\i}a-Torres}, {Gonz{\'a}lez-N{\'u}{\~n}ez}, {Gonz{\'a}lez-Vidal},
  {Gosset}, {Guy}, {Halbwachs}, {Hambly}, {Harrison}, {Hern{\'a}ndez},
  {Hestroffer}, {Hodgkin}, {Hutton}, {Jasniewicz}, {Jean-Antoine-Piccolo},
  {Jordan}, {Korn}, {Krone-Martins}, {Lanzafame}, {Lebzelter}, {L{\"o}ffler},
  {Manteiga}, {Marrese}, {Mart{\'\i}n-Fleitas}, {Moitinho}, {Mora}, {Muinonen},
  {Osinde}, {Pancino}, {Pauwels}, {Petit}, {Recio-Blanco}, {Richards},
  {Rimoldini}, {Robin}, {Sarro}, {Siopis}, {Smith}, {Sozzetti}, {S{\"u}veges},
  {Torra}, {van Reeven}, {Abbas}, {Abreu Aramburu}, {Accart}, {Aerts},
  {Altavilla}, {{\'A}lvarez}, {Alvarez}, {Alves}, {Anderson}, {Andrei},
  {Anglada Varela}, {Antiche}, {Antoja}, {Arcay}, {Astraatmadja}, {Bach},
  {Baker}, {Balaguer-N{\'u}{\~n}ez}, {Balm}, {Barache}, {Barata}, {Barbato},
  {Barblan}, {Barklem}, {Barrado}, {Barros}, {Barstow}, {Bartholom{\'e}
  Mu{\~n}oz}, {Bassilana}, {Becciani}, {Bellazzini}, {Berihuete}, {Bertone},
  {Bianchi}, {Bienaym{\'e}}, {Blanco-Cuaresma}, {Boch}, {Boeche}, {Bombrun},
  {Borrachero}, {Bossini}, {Bouquillon}, {Bourda}, {Bragaglia}, {Bramante},
  {Breddels}, {Bressan}, {Brouillet}, {Br{\"u}semeister}, {Brugaletta},
  {Bucciarelli}, {Burlacu}, {Busonero}, {Butkevich}, {Buzzi}, {Caffau},
  {Cancelliere}, {Cannizzaro}, {Cantat-Gaudin}, {Carballo}, {Carlucci},
  {Carrasco}, {Casamiquela}, {Castellani}, {Castro-Ginard}, {Charlot},
  {Chemin}, {Chiavassa}, {Cocozza}, {Costigan}, {Cowell}, {Crifo}, {Crosta},
  {Crowley}, {Cuypers}, {Dafonte}, {Damerdji}, {Dapergolas}, {David}, {David},
  {de Laverny}, {De Luise}, {De March}, {de Martino}, {de Souza}, {de Torres},
  {Debosscher}, {del Pozo}, {Delbo}, {Delgado}, {Delgado}, {Di Matteo},
  {Diakite}, {Diener}, {Distefano}, {Dolding}, {Drazinos}, {Dur{\'a}n},
  {Edvardsson}, {Enke}, {Eriksson}, {Esquej}, {Eynard Bontemps}, {Fabre},
  {Fabrizio}, {Faigler}, {Falc{\~a}o}, {Farr{\`a}s Casas}, {Federici},
  {Fedorets}, {Fernique}, {Figueras}, {Filippi}, {Findeisen}, {Fonti},
  {Fraile}, {Fraser}, {Fr{\'e}zouls}, {Gai}, {Galleti}, {Garabato},
  {Garc{\'\i}a-Sedano}, {Garofalo}, {Garralda}, {Gavel}, {Gavras}, {Gerssen},
  {Geyer}, {Giacobbe}, {Gilmore}, {Girona}, {Giuffrida}, {Glass}, {Gomes},
  {Granvik}, {Gueguen}, {Guerrier}, {Guiraud}, {Guti{\'e}rrez-S{\'a}nchez},
  {Haigron}, {Hatzidimitriou}, {Hauser}, {Haywood}, {Heiter}, {Helmi}, {Heu},
  {Hilger}, {Hobbs}, {Hofmann}, {Holland}, {Huckle}, {Hypki}, {Icardi},
  {Jan{\ss}en}, {Jevardat de Fombelle}, {Jonker}, {Juh{\'a}sz}, {Julbe},
  {Karampelas}, {Kewley}, {Klar}, {Kochoska}, {Kohley}, {Kolenberg},
  {Kontizas}, {Kontizas}, {Koposov}, {Kordopatis}, {Kostrzewa-Rutkowska},
  {Koubsky}, {Lambert}, {Lanza}, {Lasne}, {Lavigne}, {Le Fustec}, {Le
  Poncin-Lafitte}, {Lebreton}, {Leccia}, {Leclerc}, {Lecoeur-Taibi},
  {Lenhardt}, {Leroux}, {Liao}, {Licata}, {Lindstr{\o}m}, {Lister}, {Livanou},
  {Lobel}, {L{\'o}pez}, {Managau}, {Mann}, {Mantelet}, {Marchal}, {Marchant},
  {Marconi}, {Marinoni}, {Marschalk{\'o}}, {Marshall}, {Martino}, {Marton},
  {Mary}, {Massari}, {Matijevi{\v{c}}}, {Mazeh}, {McMillan}, {Messina},
  {Michalik}, {Millar}, {Molina}, {Molinaro}, {Moln{\'a}r}, {Montegriffo},
  {Mor}, {Morbidelli}, {Morel}, {Morris}, {Mulone}, {Muraveva}, {Musella},
  {Nelemans}, {Nicastro}, {Noval}, {O'Mullane}, {Ord{\'e}novic},
  {Ord{\'o}{\~n}ez-Blanco}, {Osborne}, {Pagani}, {Pagano}, {Pailler},
  {Palacin}, {Palaversa}, {Panahi}, {Pawlak}, {Piersimoni}, {Pineau}, {Plachy},
  {Plum}, {Poggio}, {Poujoulet}, {Pr{\v{s}}a}, {Pulone}, {Racero}, {Ragaini},
  {Rambaux}, {Ramos-Lerate}, {Regibo}, {Reyl{\'e}}, {Riclet}, {Ripepi}, {Riva},
  {Rivard}, {Rixon}, {Roegiers}, {Roelens}, {Romero-G{\'o}mez}, {Rowell},
  {Royer}, {Ruiz-Dern}, {Sadowski}, {Sagrist{\`a} Sell{\'e}s}, {Sahlmann},
  {Salgado}, {Salguero}, {Sanna}, {Santana-Ros}, {Sarasso}, {Savietto},
  {Schultheis}, {Sciacca}, {Segol}, {Segovia}, {S{\'e}gransan}, {Shih},
  {Siltala}, {Silva}, {Smart}, {Smith}, {Solano}, {Solitro}, {Sordo}, {Soria
  Nieto}, {Souchay}, {Spagna}, {Spoto}, {Stampa}, {Steele},
  {Steidelm{\"u}ller}, {Stephenson}, {Stoev}, {Suess}, {Surdej}, {Szabados},
  {Szegedi-Elek}, {Tapiador}, {Taris}, {Tauran}, {Taylor}, {Teixeira},
  {Terrett}, {Teyssandier}, {Thuillot}, {Titarenko}, {Torra Clotet}, {Turon},
  {Ulla}, {Utrilla}, {Uzzi}, {Vaillant}, {Valentini}, {Valette}, {van Elteren},
  {Van Hemelryck}, {van Leeuwen}, {Vaschetto}, {Vecchiato}, {Veljanoski},
  {Viala}, {Vicente}, {Vogt}, {von Essen}, {Voss}, {Votruba}, {Voutsinas},
  {Walmsley}, {Weiler}, {Wertz}, {Wevers}, {Wyrzykowski}, {Yoldas},
  {{\v{Z}}erjal}, {Ziaeepour}, {Zorec}, {Zschocke}, {Zucker}, {Zurbach}, \&
  {Zwitter}}]{GaiaCollaboration2018}
{Gaia Collaboration}, {Brown}, A.~G.~A., {Vallenari}, A., {et~al.} 2018, \aap,
  616, A1, \dodoi{10.1051/0004-6361/201833051}

\bibitem[{{Giannios} \& {Spitkovsky}(2009)}]{Giannios2009}
{Giannios}, D., \& {Spitkovsky}, A. 2009, \mnras, 400, 330,
  \dodoi{10.1111/j.1365-2966.2009.15454.x}

\bibitem[{{Graham} {et~al.}(2019){Graham}, {Kulkarni}, {Bellm}, {Adams},
  {Barbarino}, {Blagorodnova}, {Bodewits}, {Bolin}, {Brady}, {Cenko}, {Chang},
  {Coughlin}, {De}, {Eadie}, {Farnham}, {Feindt}, {Franckowiak}, {Fremling},
  {Gezari}, {Ghosh}, {Goldstein}, {Golkhou}, {Goobar}, {Ho}, {Huppenkothen},
  {Ivezi{\'c}}, {Jones}, {Juric}, {Kaplan}, {Kasliwal}, {Kelley}, {Kupfer},
  {Lee}, {Lin}, {Lunnan}, {Mahabal}, {Miller}, {Ngeow}, {Nugent}, {Ofek},
  {Prince}, {Rauch}, {van Roestel}, {Schulze}, {Singer}, {Sollerman}, {Taddia},
  {Yan}, {Ye}, {Yu}, {Barlow}, {Bauer}, {Beck}, {Belicki}, {Biswas}, {Brinnel},
  {Brooke}, {Bue}, {Bulla}, {Burruss}, {Connolly}, {Cromer}, {Cunningham},
  {Dekany}, {Delacroix}, {Desai}, {Duev}, {Feeney}, {Flynn}, {Frederick},
  {Gal-Yam}, {Giomi}, {Groom}, {Hacopians}, {Hale}, {Helou}, {Henning},
  {Hover}, {Hillenbrand}, {Howell}, {Hung}, {Imel}, {Ip}, {Jackson}, {Kaspi},
  {Kaye}, {Kowalski}, {Kramer}, {Kuhn}, {Landry}, {Laher}, {Mao}, {Masci},
  {Monkewitz}, {Murphy}, {Nordin}, {Patterson}, {Penprase}, {Porter},
  {Rebbapragada}, {Reiley}, {Riddle}, {Rigault}, {Rodriguez}, {Rusholme}, {van
  Santen}, {Shupe}, {Smith}, {Soumagnac}, {Stein}, {Surace}, {Szkody}, {Terek},
  {Van Sistine}, {van Velzen}, {Vestrand}, {Walters}, {Ward}, {Zhang}, \&
  {Zolkower}}]{Graham2019}
{Graham}, M.~J., {Kulkarni}, S.~R., {Bellm}, E.~C., {et~al.} 2019, \pasp, 131,
  078001, \dodoi{10.1088/1538-3873/ab006c}

\bibitem[{{Granot} \& {Sari}(2002)}]{Granot2002_shape_spectral_breaks}
{Granot}, J., \& {Sari}, R. 2002, \apj, 568, 820, \dodoi{10.1086/338966}

\bibitem[{{Guns} {et~al.}(2021){Guns}, {Foster}, {Daley}, {Rahlin},
  {Whitehorn}, {Ade}, {Ahmed}, {Anderes}, {Anderson}, {Archipley}, {Avva},
  {Aylor}, {Balkenhol}, {Barry}, {Basu Thakur}, {Benabed}, {Bender}, {Benson},
  {Bianchini}, {Bleem}, {Bouchet}, {Bryant}, {Byrum}, {Carlstrom}, {Carter},
  {Cecil}, {Chang}, {Chaubal}, {Chen}, {Cho}, {Chou}, {Cliche}, {Crawford},
  {Cukierman}, {de Haan}, {Denison}, {Dibert}, {Ding}, {Dobbs}, {Dutcher},
  {Everett}, {Feng}, {Ferguson}, {Fu}, {Galli}, {Gambrel}, {Gardner},
  {Goeckner-Wald}, {Gualtieri}, {Gupta}, {Guyser}, {Halverson},
  {Harke-Hosemann}, {Harrington}, {Henning}, {Hilton}, {Hivon}, {Holder},
  {Holzapfel}, {Hood}, {Howe}, {Huang}, {Irwin}, {Jeong}, {Jonas}, {Jones},
  {Khaire}, {Knox}, {Kofman}, {Korman}, {Kubik}, {Kuhlmann}, {Kuo}, {Lee},
  {Leitch}, {Lowitz}, {Lu}, {Marrone}, {Meyer}, {Michalik}, {Millea},
  {Montgomery}, {Nadolski}, {Natoli}, {Nguyen}, {Noble}, {Novosad}, {Omori},
  {Padin}, {Pan}, {Paschos}, {Pearson}, {Phadke}, {Posada}, {Prabhu}, {Quan},
  {Reichardt}, {Riebel}, {Riedel}, {Rouble}, {Ruhl}, {Sayre}, {Schiappucci},
  {Shirokoff}, {Smecher}, {Sobrin}, {Stark}, {Stephen}, {Story}, {Suzuki},
  {Thompson}, {Thorne}, {Tucker}, {Umilta}, {Vale}, {Vieira}, {Wang}, {Wu},
  {Yefremenko}, {Yoon}, {Young}, \& {Zhang}}]{Guns2021}
{Guns}, S., {Foster}, A., {Daley}, C., {et~al.} 2021, \apj, 916, 98,
  \dodoi{10.3847/1538-4357/ac06a3}

\bibitem[{{Ho} {et~al.}(2020{\natexlab{a}}){Ho}, {Perley}, \&
  {Yao}}]{Ho2020ATel}
{Ho}, A.~Y.~Q., {Perley}, D.~A., \& {Yao}, Y. 2020{\natexlab{a}}, Transient
  Name Server AstroNote, 204, 1

\bibitem[{{Ho} {et~al.}(2019{\natexlab{a}}){Ho}, {Phinney}, {Ravi}, {Kulkarni},
  {Petitpas}, {Emonts}, {Bhalerao}, {Blundell}, {Cenko}, {Dobie}, {Howie},
  {Kamraj}, {Kasliwal}, {Murphy}, {Perley}, {Sridharan}, \& {Yoon}}]{Ho2019cow}
{Ho}, A.~Y.~Q., {Phinney}, E.~S., {Ravi}, V., {et~al.} 2019{\natexlab{a}},
  \apj, 871, 73, \dodoi{10.3847/1538-4357/aaf473}

\bibitem[{{Ho} {et~al.}(2019{\natexlab{b}}){Ho}, {Goldstein}, {Schulze},
  {Khatami}, {Perley}, {Ergon}, {Gal-Yam}, {Corsi}, {Andreoni}, {Barbarino},
  {Bellm}, {Blagorodnova}, {Bright}, {Burns}, {Cenko}, {Cunningham}, {De},
  {Dekany}, {Dugas}, {Fender}, {Fransson}, {Fremling}, {Goldstein}, {Graham},
  {Hale}, {Horesh}, {Hung}, {Kasliwal}, {Kuin}, {Kulkarni}, {Kupfer}, {Lunnan},
  {Masci}, {Ngeow}, {Nugent}, {Ofek}, {Patterson}, {Petitpas}, {Rusholme},
  {Sai}, {Sfaradi}, {Shupe}, {Sollerman}, {Soumagnac}, {Tachibana}, {Taddia},
  {Walters}, {Wang}, {Yao}, \& {Zhang}}]{Ho2019gep}
{Ho}, A.~Y.~Q., {Goldstein}, D.~A., {Schulze}, S., {et~al.} 2019{\natexlab{b}},
  \apj, 887, 169, \dodoi{10.3847/1538-4357/ab55ec}

\bibitem[{{Ho} {et~al.}(2020{\natexlab{b}}){Ho}, {Perley}, {Beniamini},
  {Cenko}, {Kulkarni}, {Andreoni}, {Singer}, {De}, {Kasliwal}, {Fremling},
  {Bellm}, {Dekany}, {Delacroix}, {Duev}, {Goldstein}, {Golkhou}, {Goobar},
  {Graham}, {Hale}, {Kupfer}, {Laher}, {Masci}, {Miller}, {Neill}, {Riddle},
  {Rusholme}, {Shupe}, {Smith}, {Sollerman}, \& {van Roestel}}]{Ho2020d}
{Ho}, A. Y.~Q., {Perley}, D.~A., {Beniamini}, P., {et~al.} 2020{\natexlab{b}},
  \apj, 905, 98, \dodoi{10.3847/1538-4357/abc34d}

\bibitem[{{Ho} {et~al.}(2020{\natexlab{c}}){Ho}, {Perley}, {Kulkarni}, {Dong},
  {De}, {Chandra}, {Andreoni}, {Bellm}, {Burdge}, {Coughlin}, {Dekany},
  {Feeney}, {Frederiks}, {Fremling}, {Golkhou}, {Graham}, {Hale}, {Helou},
  {Horesh}, {Kasliwal}, {Laher}, {Masci}, {Miller}, {Porter}, {Ridnaia},
  {Rusholme}, {Shupe}, {Soumagnac}, \& {Svinkin}}]{Ho2020b}
{Ho}, A. Y.~Q., {Perley}, D.~A., {Kulkarni}, S.~R., {et~al.}
  2020{\natexlab{c}}, \apj, 895, 49, \dodoi{10.3847/1538-4357/ab8bcf}

\bibitem[{{Ho} {et~al.}(2021){Ho}, {Perley}, {Gal-Yam}, {Lunnan}, {Sollerman},
  {Schulze}, {Das}, {Dobie}, {Yao}, {Fremling}, {Adams}, {Anand}, {Andreoni},
  {Bellm}, {Bruch}, {Burdge}, {Castro-Tirado}, {Dahiwale}, {De}, {Dekany},
  {Drake}, {Duev}, {Graham}, {Helou}, {Kaplan}, {Karambelkar}, {Kasliwal},
  {Kool}, {Kulkarni}, {Mahabal}, {Medford}, {Miller}, {Nordin}, {Ofek},
  {Petitpas}, {Riddle}, {Sharma}, {Smith}, {Stewart}, {Taggart}, {Tartaglia},
  {Tzanidakis}, \& {Winters}}]{Ho2021}
{Ho}, A. Y.~Q., {Perley}, D.~A., {Gal-Yam}, A., {et~al.} 2021, arXiv e-prints,
  arXiv:2105.08811.
\newblock \doarXiv{2105.08811}

\bibitem[{{Ho} {et~al.}(2004){Ho}, {Moran}, \& {Lo}}]{Ho2004}
{Ho}, P. T.~P., {Moran}, J.~M., \& {Lo}, K.~Y. 2004, \apjl, 616, L1,
  \dodoi{10.1086/423245}

\bibitem[{{H{\"o}gbom}(1974)}]{Hogbom1974}
{H{\"o}gbom}, J.~A. 1974, \aaps, 15, 417

\bibitem[{{Horesh} {et~al.}(2015){Horesh}, {Cenko}, {Perley}, {Kulkarni},
  {Hallinan}, \& {Bellm}}]{Horesh2015}
{Horesh}, A., {Cenko}, S.~B., {Perley}, D.~A., {et~al.} 2015, \apj, 812, 86,
  \dodoi{10.1088/0004-637X/812/1/86}

\bibitem[{{Horesh} {et~al.}(2013){Horesh}, {Stockdale}, {Fox}, {Frail},
  {Carpenter}, {Kulkarni}, {Ofek}, {Gal-Yam}, {Kasliwal}, {Arcavi}, {Quimby},
  {Cenko}, {Nugent}, {Bloom}, {Law}, {Poznanski}, {Gorbikov}, {Polishook},
  {Yaron}, {Ryder}, {Weiler}, {Bauer}, {Van Dyk}, {Immler}, {Panagia},
  {Pooley}, \& {Kassim}}]{Horesh2013}
{Horesh}, A., {Stockdale}, C., {Fox}, D.~B., {et~al.} 2013, \mnras, 436, 1258,
  \dodoi{10.1093/mnras/stt1645}

\bibitem[{{Hunter}(2007)}]{Hunter2007}
{Hunter}, J.~D. 2007, Computing in Science and Engineering, 9, 90,
  \dodoi{10.1109/MCSE.2007.55}

\bibitem[{{J{\'o}hannesson} \& {Bj{\"o}rnsson}(2018)}]{Johannesson2018}
{J{\'o}hannesson}, G., \& {Bj{\"o}rnsson}, G. 2018, \apjl, 859, L11,
  \dodoi{10.3847/2041-8213/aac380}

\bibitem[{{Kashiyama} {et~al.}(2018){Kashiyama}, {Hotokezaka}, \&
  {Murase}}]{Kashiyama2018}
{Kashiyama}, K., {Hotokezaka}, K., \& {Murase}, K. 2018, \mnras, 478, 2281,
  \dodoi{10.1093/mnras/sty1145}

\bibitem[{{Krauss} {et~al.}(2012){Krauss}, {Soderberg}, {Chomiuk}, {Zauderer},
  {Brunthaler}, {Bietenholz}, {Chevalier}, {Fransson}, \& {Rupen}}]{Krauss2012}
{Krauss}, M.~I., {Soderberg}, A.~M., {Chomiuk}, L., {et~al.} 2012, \apjl, 750,
  L40, \dodoi{10.1088/2041-8205/750/2/L40}

\bibitem[{{Kuin} {et~al.}(2019){Kuin}, {Wu}, {Oates}, {Lien}, {Emery},
  {Kennea}, {de Pasquale}, {Han}, {Brown}, {Tohuvavohu}, {Breeveld}, {Burrows},
  {Cenko}, {Campana}, {Levan}, {Markwardt}, {Osborne}, {Page}, {Page},
  {Sbarufatti}, {Siegel}, \& {Troja}}]{Kuin2019}
{Kuin}, N. P.~M., {Wu}, K., {Oates}, S., {et~al.} 2019, \mnras, 487, 2505,
  \dodoi{10.1093/mnras/stz053}

\bibitem[{{Kulkarni} {et~al.}(1998){Kulkarni}, {Frail}, {Wieringa}, {Ekers},
  {Sadler}, {Wark}, {Higdon}, {Phinney}, \& {Bloom}}]{Kulkarni1998}
{Kulkarni}, S.~R., {Frail}, D.~A., {Wieringa}, M.~H., {et~al.} 1998, Nature,
  395, 663, \dodoi{10.1038/27139}

\bibitem[{{Laskar} {et~al.}(2018){Laskar}, {Alexander}, {Berger}, {Guidorzi},
  {Margutti}, {Fong}, {Kilpatrick}, {Milne}, {Drout}, {Mundell}, {Kobayashi},
  {Lunnan}, {Barniol Duran}, {Menten}, {Ioka}, \&
  {Williams}}]{Laskar2018_GRB161219B}
{Laskar}, T., {Alexander}, K.~D., {Berger}, E., {et~al.} 2018, \apj, 862, 94,
  \dodoi{10.3847/1538-4357/aacbcc}

\bibitem[{{Laskar} {et~al.}(2019){Laskar}, {van Eerten}, {Schady}, {Mundell},
  {Alexander}, {Barniol Duran}, {Berger}, {Bolmer}, {Chornock}, {Coppejans},
  {Fong}, {Gomboc}, {Jordana-Mitjans}, {Kobayashi}, {Margutti}, {Menten},
  {Sari}, {Yamazaki}, {Lipunov}, {Gorbovskoy}, {Kornilov}, {Tyurina},
  {Zimnukhov}, {Podesta}, {Levato}, {Buckley}, {Tlatov}, {Rebolo}, \&
  {Serra-Ricart}}]{Laskar2019}
{Laskar}, T., {van Eerten}, H., {Schady}, P., {et~al.} 2019, \apj, 884, 121,
  \dodoi{10.3847/1538-4357/ab40ce}

\bibitem[{{Law} {et~al.}(2018){Law}, {Gaensler}, {Metzger}, {Ofek}, \&
  {Sironi}}]{Law2018}
{Law}, C.~J., {Gaensler}, B.~M., {Metzger}, B.~D., {Ofek}, E.~O., \& {Sironi},
  L. 2018, \apjl, 866, L22, \dodoi{10.3847/2041-8213/aae5f3}

\bibitem[{{Leung} {et~al.}(2021){Leung}, {Fuller}, \& {Nomoto}}]{Leung2021}
{Leung}, S.-C., {Fuller}, J., \& {Nomoto}, K. 2021, \apj, 915, 80,
  \dodoi{10.3847/1538-4357/abfcbe}

\bibitem[{{Maeda} {et~al.}(2021){Maeda}, {Chandra}, {Matsuoka}, {Ryder},
  {Moriya}, {Kuncarayakti}, {Lee}, {Kundu}, {Patnaude}, {Saito}, \&
  {Folatelli}}]{Maeda2021}
{Maeda}, K., {Chandra}, P., {Matsuoka}, T., {et~al.} 2021, \apj, 918, 34,
  \dodoi{10.3847/1538-4357/ac0dbc}

\bibitem[{{Mahadevan} {et~al.}(1996){Mahadevan}, {Narayan}, \&
  {Yi}}]{Mahadevan1996}
{Mahadevan}, R., {Narayan}, R., \& {Yi}, I. 1996, \apj, 465, 327,
  \dodoi{10.1086/177422}

\bibitem[{{Margalit} \& {Quataert}(2021)}]{Margalit2021}
{Margalit}, B., \& {Quataert}, E. 2021, \apjl, 923, L14,
  \dodoi{10.3847/2041-8213/ac3d97}

\bibitem[{{Margutti} {et~al.}(2013){Margutti}, {Soderberg}, {Wieringa},
  {Edwards}, {Chevalier}, {Morsony}, {Barniol Duran}, {Sironi}, {Zauderer},
  {Milisavljevic}, {Kamble}, \& {Pian}}]{Margutti2013}
{Margutti}, R., {Soderberg}, A.~M., {Wieringa}, M.~H., {et~al.} 2013, \apj,
  778, 18, \dodoi{10.1088/0004-637X/778/1/18}

\bibitem[{{Margutti} {et~al.}(2019){Margutti}, {Metzger}, {Chornock}, {Vurm},
  {Roth}, {Grefenstette}, {Savchenko}, {Cartier}, {Steiner}, {Terreran},
  {Margalit}, {Migliori}, {Milisavljevic}, {Alexand er}, {Bietenholz},
  {Blanchard}, {Bozzo}, {Brethauer}, {Chilingarian}, {Coppejans}, {Ducci},
  {Ferrigno}, {Fong}, {G{\"o}tz}, {Guidorzi}, {Hajela}, {Hurley}, {Kuulkers},
  {Laurent}, {Mereghetti}, {Nicholl}, {Patnaude}, {Ubertini}, {Banovetz},
  {Bartel}, {Berger}, {Coughlin}, {Eftekhari}, {Frederiks}, {Kozlova},
  {Laskar}, {Svinkin}, {Drout}, {MacFadyen}, \& {Paterson}}]{Margutti2019}
{Margutti}, R., {Metzger}, B.~D., {Chornock}, R., {et~al.} 2019, \apj, 872, 18,
  \dodoi{10.3847/1538-4357/aafa01}

\bibitem[{{Matthews} {et~al.}(2020){Matthews}, {Margutti}, {Brethauer},
  {Bright}, {Coppejans}, {Alexander}, {Laskar}, {DeMarchi}, {Jacobson-Galan},
  {Migliori}, {Baldeschi}, {Stroh}, \& {Auchettl}}]{Matthews2020}
{Matthews}, D., {Margutti}, R., {Brethauer}, D., {et~al.} 2020, The
  Astronomer's Telegram, 14154, 1

\bibitem[{{McMullin} {et~al.}(2007){McMullin}, {Waters}, {Schiebel}, {Young},
  \& {Golap}}]{McMullin2007}
{McMullin}, J.~P., {Waters}, B., {Schiebel}, D., {Young}, W., \& {Golap}, K.
  2007, Astronomical Society of the Pacific Conference Series, Vol. 376, {CASA
  Architecture and Applications}, ed. R.~A. {Shaw}, F.~{Hill}, \& D.~J. {Bell},
  127

\bibitem[{{Metzger} {et~al.}(2015){Metzger}, {Williams}, \&
  {Berger}}]{Metzger2015}
{Metzger}, B.~D., {Williams}, P.~K.~G., \& {Berger}, E. 2015, \apj, 806, 224,
  \dodoi{10.1088/0004-637X/806/2/224}

\bibitem[{{Mooley} {et~al.}(2022){Mooley}, {Margalit}, {Law}, {Perley},
  {Deller}, {Lazio}, {Bietenholz}, {Shimwell}, {Intema}, {Gaensler}, {Metzger},
  {Dong}, {Hallinan}, {Ofek}, \& {Sironi}}]{Mooley2022}
{Mooley}, K.~P., {Margalit}, B., {Law}, C.~J., {et~al.} 2022, \apj, 924, 16,
  \dodoi{10.3847/1538-4357/ac3330}

\bibitem[{{Naess} {et~al.}(2021){Naess}, {Battaglia}, {Richard Bond},
  {Calabrese}, {Choi}, {Cothard}, {Devlin}, {Duell}, {Duivenvoorden},
  {Dunkley}, {D{\"u}nner}, {Gallardo}, {Gralla}, {Guan}, {Halpern}, {Colin
  Hill}, {Hilton}, {Huffenberger}, {Koopman}, {Kosowsky}, {Madhavacheril},
  {McMahon}, {Nati}, {Niemack}, {Page}, {Partridge}, {Salatino}, {Sehgal},
  {Spergel}, {Staggs}, {Wollack}, \& {Xu}}]{Naess2021}
{Naess}, S., {Battaglia}, N., {Richard Bond}, J., {et~al.} 2021, \apj, 915, 14,
  \dodoi{10.3847/1538-4357/abfe6d}

\bibitem[{{Narayan}(1992)}]{Narayan1992}
{Narayan}, R. 1992, Philosophical Transactions of the Royal Society of London
  Series A, 341, 151, \dodoi{10.1098/rsta.1992.0090}

\bibitem[{{Nayana} \& {Chandra}(2021)}]{Nayana2021}
{Nayana}, A.~J., \& {Chandra}, P. 2021, \apjl, 912, L9,
  \dodoi{10.3847/2041-8213/abed55}

\bibitem[{{Ofek} {et~al.}(2010){Ofek}, {Rabinak}, {Neill}, {Arcavi}, {Cenko},
  {Waxman}, {Kulkarni}, {Gal-Yam}, {Nugent}, {Bildsten}, {Bloom}, {Filippenko},
  {Forster}, {Howell}, {Jacobsen}, {Kasliwal}, {Law}, {Martin}, {Poznanski},
  {Quimby}, {Shen}, {Sullivan}, {Dekany}, {Rahmer}, {Hale}, {Smith},
  {Zolkower}, {Velur}, {Walters}, {Henning}, {Bui}, \& {McKenna}}]{Ofek2010}
{Ofek}, E.~O., {Rabinak}, I., {Neill}, J.~D., {et~al.} 2010, \apj, 724, 1396,
  \dodoi{10.1088/0004-637X/724/2/1396}

\bibitem[{{{\"O}zel} {et~al.}(2000){{\"O}zel}, {Psaltis}, \&
  {Narayan}}]{Ozel2000}
{{\"O}zel}, F., {Psaltis}, D., \& {Narayan}, R. 2000, \apj, 541, 234,
  \dodoi{10.1086/309396}

\bibitem[{{Park} {et~al.}(2015){Park}, {Caprioli}, \& {Spitkovsky}}]{Park+15}
{Park}, J., {Caprioli}, D., \& {Spitkovsky}, A. 2015, \prl, 114, 085003,
  \dodoi{10.1103/PhysRevLett.114.085003}

\bibitem[{{Perley} {et~al.}(2020{\natexlab{a}}){Perley}, {Schulze}, \&
  {Bruch}}]{Perley2020}
{Perley}, D., {Schulze}, S., \& {Bruch}, R. 2020{\natexlab{a}}, Transient Name
  Server AstroNote, 37, 1

\bibitem[{{Perley} {et~al.}(2020{\natexlab{b}}){Perley}, {Ho}, {Fremling}, \&
  {Yao}}]{Perley2020ATel}
{Perley}, D.~A., {Ho}, A.~Y.~Q., {Fremling}, C., \& {Yao}, Y.
  2020{\natexlab{b}}, The Astronomer's Telegram, 14105, 1

\bibitem[{{Perley} {et~al.}(2017){Perley}, {Schulze}, \& {de Ugarte
  Postigo}}]{Perley2017-alma}
{Perley}, D.~A., {Schulze}, S., \& {de Ugarte Postigo}, A. 2017, GRB
  Coordinates Network, 22252, 1

\bibitem[{{Perley} {et~al.}(2014){Perley}, {Cenko}, {Corsi}, {Tanvir}, {Levan},
  {Kann}, {Sonbas}, {Wiersema}, {Zheng}, {Zhao}, {Bai}, {Bremer},
  {Castro-Tirado}, {Chang}, {Clubb}, {Frail}, {Fruchter},
  {G{\"o}{\u{g}}{\"u}{\textcommabelow s}}, {Greiner}, {G{\"u}ver}, {Horesh},
  {Filippenko}, {Klose}, {Mao}, {Morgan}, {Pozanenko}, {Schmidl}, {Stecklum},
  {Tanga}, {Volnova}, {Volvach}, {Wang}, {Winters}, \&
  {Xin}}]{Perley2014_130427A}
{Perley}, D.~A., {Cenko}, S.~B., {Corsi}, A., {et~al.} 2014, \apj, 781, 37,
  \dodoi{10.1088/0004-637X/781/1/37}

\bibitem[{{Perley} {et~al.}(2019){Perley}, {Mazzali}, {Yan}, {Cenko}, {Gezari},
  {Taggart}, {Blagorodnova}, {Fremling}, {Mockler}, {Singh}, {Tominaga},
  {Tanaka}, {Watson}, {Ahumada}, {Anupama}, {Ashall}, {Becerra}, {Bersier},
  {Bhalerao}, {Bloom}, {Butler}, {Copperwheat}, {Coughlin}, {De}, {Drake},
  {Duev}, {Frederick}, {Gonz{\'a}lez}, {Goobar}, {Heida}, {Ho}, {Horst},
  {Hung}, {Itoh}, {Jencson}, {Kasliwal}, {Kawai}, {Khanam}, {Kulkarni},
  {Kumar}, {Kumar}, {Kutyrev}, {Lee}, {Maeda}, {Mahabal}, {Murata}, {Neill},
  {Ngeow}, {Penprase}, {Pian}, {Quimby}, {Ramirez-Ruiz}, {Richer},
  {Rom{\'a}n-Z{\'u}{\~n}iga}, {Sahu}, {Srivastav}, {Socia}, {Sollerman},
  {Tachibana}, {Taddia}, {Tinyanont}, {Troja}, {Ward}, {Wee}, \&
  {Yu}}]{Perley2019cow}
{Perley}, D.~A., {Mazzali}, P.~A., {Yan}, L., {et~al.} 2019, \mnras, 484, 1031,
  \dodoi{10.1093/mnras/sty3420}

\bibitem[{{Perley} {et~al.}(2021){Perley}, {Ho}, {Yao}, {Fremling}, {Anderson},
  {Schulze}, {Kumar}, {Anupama}, {Barway}, {Bellm}, {Bhalerao}, {Chen}, {Duev},
  {Galbany}, {Graham}, {Gromadzki}, {Guti{\'e}rrez}, {Ihanec}, {Inserra},
  {Kasliwal}, {Kool}, {Kulkarni}, {Laher}, {Masci}, {Neill}, {Nicholl},
  {Pursiainen}, {van Roestel}, {Sharma}, {Sollerman}, {Walters}, \&
  {Wiseman}}]{Perley2021}
{Perley}, D.~A., {Ho}, A. Y.~Q., {Yao}, Y., {et~al.} 2021, \mnras, 508, 5138,
  \dodoi{10.1093/mnras/stab2785}

\bibitem[{{Perley} {et~al.}(2011){Perley}, {Chandler}, {Butler}, \&
  {Wrobel}}]{Perley2011}
{Perley}, R.~A., {Chandler}, C.~J., {Butler}, B.~J., \& {Wrobel}, J.~M. 2011,
  \apjl, 739, L1, \dodoi{10.1088/2041-8205/739/1/L1}

\bibitem[{{Planck Collaboration} {et~al.}(2016){Planck Collaboration}, {Ade},
  {Aghanim}, {Arnaud}, {Ashdown}, {Aumont}, {Baccigalupi}, {Banday},
  {Barreiro}, {Bartlett}, {Bartolo}, {Battaner}, {Battye}, {Benabed},
  {Beno{\^\i}t}, {Benoit-L{\'e}vy}, {Bernard}, {Bersanelli}, {Bielewicz},
  {Bock}, {Bonaldi}, {Bonavera}, {Bond}, {Borrill}, {Bouchet}, {Boulanger},
  {Bucher}, {Burigana}, {Butler}, {Calabrese}, {Cardoso}, {Catalano},
  {Challinor}, {Chamballu}, {Chary}, {Chiang}, {Chluba}, {Christensen},
  {Church}, {Clements}, {Colombi}, {Colombo}, {Combet}, {Coulais}, {Crill},
  {Curto}, {Cuttaia}, {Danese}, {Davies}, {Davis}, {de Bernardis}, {de Rosa},
  {de Zotti}, {Delabrouille}, {D{\'e}sert}, {Di Valentino}, {Dickinson},
  {Diego}, {Dolag}, {Dole}, {Donzelli}, {Dor{\'e}}, {Douspis}, {Ducout},
  {Dunkley}, {Dupac}, {Efstathiou}, {Elsner}, {En{\ss}lin}, {Eriksen},
  {Farhang}, {Fergusson}, {Finelli}, {Forni}, {Frailis}, {Fraisse},
  {Franceschi}, {Frejsel}, {Galeotta}, {Galli}, {Ganga}, {Gauthier}, {Gerbino},
  {Ghosh}, {Giard}, {Giraud-H{\'e}raud}, {Giusarma}, {Gjerl{\o}w},
  {Gonz{\'a}lez-Nuevo}, {G{\'o}rski}, {Gratton}, {Gregorio}, {Gruppuso},
  {Gudmundsson}, {Hamann}, {Hansen}, {Hanson}, {Harrison}, {Helou},
  {Henrot-Versill{\'e}}, {Hern{\'a}ndez-Monteagudo}, {Herranz}, {Hildebrand t},
  {Hivon}, {Hobson}, {Holmes}, {Hornstrup}, {Hovest}, {Huang}, {Huffenberger},
  {Hurier}, {Jaffe}, {Jaffe}, {Jones}, {Juvela}, {Keih{\"a}nen}, {Keskitalo},
  {Kisner}, {Kneissl}, {Knoche}, {Knox}, {Kunz}, {Kurki-Suonio}, {Lagache},
  {L{\"a}hteenm{\"a}ki}, {Lamarre}, {Lasenby}, {Lattanzi}, {Lawrence}, {Leahy},
  {Leonardi}, {Lesgourgues}, {Levrier}, {Lewis}, {Liguori}, {Lilje},
  {Linden-V{\o}rnle}, {L{\'o}pez-Caniego}, {Lubin}, {Mac{\'\i}as-P{\'e}rez},
  {Maggio}, {Maino}, {Mandolesi}, {Mangilli}, {Marchini}, {Maris}, {Martin},
  {Martinelli}, {Mart{\'\i}nez-Gonz{\'a}lez}, {Masi}, {Matarrese}, {McGehee},
  {Meinhold}, {Melchiorri}, {Melin}, {Mendes}, {Mennella}, {Migliaccio},
  {Millea}, {Mitra}, {Miville-Desch{\^e}nes}, {Moneti}, {Montier}, {Morgante},
  {Mortlock}, {Moss}, {Munshi}, {Murphy}, {Naselsky}, {Nati}, {Natoli},
  {Netterfield}, {N{\o}rgaard-Nielsen}, {Noviello}, {Novikov}, {Novikov},
  {Oxborrow}, {Paci}, {Pagano}, {Pajot}, {Paladini}, {Paoletti}, {Partridge},
  {Pasian}, {Patanchon}, {Pearson}, {Perdereau}, {Perotto}, {Perrotta},
  {Pettorino}, {Piacentini}, {Piat}, {Pierpaoli}, {Pietrobon}, {Plaszczynski},
  {Pointecouteau}, {Polenta}, {Popa}, {Pratt}, {Pr{\'e}zeau}, {Prunet},
  {Puget}, {Rachen}, {Reach}, {Rebolo}, {Reinecke}, {Remazeilles}, {Renault},
  {Renzi}, {Ristorcelli}, {Rocha}, {Rosset}, {Rossetti}, {Roudier},
  {Rouill{\'e} d'Orfeuil}, {Rowan-Robinson}, {Rubi{\~n}o-Mart{\'\i}n},
  {Rusholme}, {Said}, {Salvatelli}, {Salvati}, {Sandri}, {Santos},
  {Savelainen}, {Savini}, {Scott}, {Seiffert}, {Serra}, {Shellard}, {Spencer},
  {Spinelli}, {Stolyarov}, {Stompor}, {Sudiwala}, {Sunyaev}, {Sutton},
  {Suur-Uski}, {Sygnet}, {Tauber}, {Terenzi}, {Toffolatti}, {Tomasi},
  {Tristram}, {Trombetti}, {Tucci}, {Tuovinen}, {T{\"u}rler}, {Umana},
  {Valenziano}, {Valiviita}, {Van Tent}, {Vielva}, {Villa}, {Wade}, {Wandelt},
  {Wehus}, {White}, {White}, {Wilkinson}, {Yvon}, {Zacchei}, \&
  {Zonca}}]{Planck2016}
{Planck Collaboration}, {Ade}, P.~A.~R., {Aghanim}, N., {et~al.} 2016, \aap,
  594, A13, \dodoi{10.1051/0004-6361/201525830}

\bibitem[{{Prentice} {et~al.}(2018){Prentice}, {Maguire}, {Smartt}, {Magee},
  {Schady}, {Sim}, {Chen}, {Clark}, {Colin}, {Fulton}, {McBrien}, {O'Neill},
  {Smith}, {Ashall}, {Chambers}, {Denneau}, {Flewelling}, {Heinze}, {Holoien},
  {Huber}, {Kochanek}, {Mazzali}, {Prieto}, {Rest}, {Shappee}, {Stalder},
  {Stanek}, {Stritzinger}, {Thompson}, \& {Tonry}}]{Prentice2018}
{Prentice}, S.~J., {Maguire}, K., {Smartt}, S.~J., {et~al.} 2018, \apjl, 865,
  L3, \dodoi{10.3847/2041-8213/aadd90}

\bibitem[{{Quataert} {et~al.}(2019){Quataert}, {Lecoanet}, \&
  {Coughlin}}]{Quataert2019}
{Quataert}, E., {Lecoanet}, D., \& {Coughlin}, E.~R. 2019, \mnras, 485, L83,
  \dodoi{10.1093/mnrasl/slz031}

\bibitem[{{Ressler} \& {Laskar}(2017)}]{Ressler2017}
{Ressler}, S.~M., \& {Laskar}, T. 2017, \apj, 845, 150,
  \dodoi{10.3847/1538-4357/aa8268}

\bibitem[{{Rest} {et~al.}(2018){Rest}, {Garnavich}, {Khatami}, {Kasen},
  {Tucker}, {Shaya}, {Olling}, {Mushotzky}, {Zenteno}, {Margheim},
  {Strampelli}, {James}, {Smith}, {F{\"o}rster}, \& {Villar}}]{Rest2018}
{Rest}, A., {Garnavich}, P.~M., {Khatami}, D., {et~al.} 2018, Nature Astronomy,
  2, 307, \dodoi{10.1038/s41550-018-0423-2}

\bibitem[{{Rickett}(1990)}]{Rickett1990}
{Rickett}, B.~J. 1990, \araa, 28, 561,
  \dodoi{10.1146/annurev.aa.28.090190.003021}

\bibitem[{{Rivera Sandoval} {et~al.}(2018){Rivera Sandoval}, {Maccarone},
  {Corsi}, {Brown}, {Pooley}, \& {Wheeler}}]{RiveraSandoval2018}
{Rivera Sandoval}, L.~E., {Maccarone}, T.~J., {Corsi}, A., {et~al.} 2018,
  \mnras, 480, L146, \dodoi{10.1093/mnrasl/sly145}

\bibitem[{{Rybicki} \& {Lightman}(1986)}]{RL}
{Rybicki}, G.~B., \& {Lightman}, A.~P. 1986, {Radiative Processes in
  Astrophysics}

\bibitem[{{Salas} {et~al.}(2013){Salas}, {Bauer}, {Stockdale}, \&
  {Prieto}}]{Salas2013}
{Salas}, P., {Bauer}, F.~E., {Stockdale}, C., \& {Prieto}, J.~L. 2013, \mnras,
  428, 1207, \dodoi{10.1093/mnras/sts104}

\bibitem[{{Sault} {et~al.}(1995){Sault}, {Teuben}, \&
  {Wright}}]{1995ASPC...77..433S}
{Sault}, R.~J., {Teuben}, P.~J., \& {Wright}, M.~C.~H. 1995, in Astronomical
  Society of the Pacific Conference Series, Vol.~77, Astronomical Data Analysis
  Software and Systems IV, ed. R.~A. {Shaw}, H.~E. {Payne}, \& J.~J.~E.
  {Hayes}, 433.
\newblock \doarXiv{astro-ph/0612759}

\bibitem[{{Sheth} {et~al.}(2003){Sheth}, {Frail}, {White}, {Das}, {Bertoldi},
  {Walter}, {Kulkarni}, \& {Berger}}]{Sheth2003}
{Sheth}, K., {Frail}, D.~A., {White}, S., {et~al.} 2003, \apjl, 595, L33,
  \dodoi{10.1086/378933}

\bibitem[{{Shulevski} {et~al.}(2015){Shulevski}, {Morganti}, {Barthel},
  {Harwood}, {Brunetti}, {van Weeren}, {R{\"o}ttgering}, {White}, {Horellou},
  {Kunert-Bajraszewska}, {Jamrozy}, {Chyzy}, {Mahony}, {Miley}, {Brienza},
  {B{\^\i}rzan}, {Rafferty}, {Br{\"u}ggen}, {Wise}, {Conway}, {de Gasperin}, \&
  {Vilchez}}]{Shulevski2015}
{Shulevski}, A., {Morganti}, R., {Barthel}, P.~D., {et~al.} 2015, \aap, 583,
  A89, \dodoi{10.1051/0004-6361/201525632}

\bibitem[{{Sironi} \& {Giannios}(2013)}]{Sironi&Giannios13}
{Sironi}, L., \& {Giannios}, D. 2013, \apj, 778, 107,
  \dodoi{10.1088/0004-637X/778/2/107}

\bibitem[{{Smith}(2014)}]{Smith2014}
{Smith}, N. 2014, \araa, 52, 487, \dodoi{10.1146/annurev-astro-081913-040025}

\bibitem[{{Soderberg} {et~al.}(2006{\natexlab{a}}){Soderberg}, {Chevalier},
  {Kulkarni}, \& {Frail}}]{Soderberg2006c-03bg}
{Soderberg}, A.~M., {Chevalier}, R.~A., {Kulkarni}, S.~R., \& {Frail}, D.~A.
  2006{\natexlab{a}}, \apj, 651, 1005, \dodoi{10.1086/507571}

\bibitem[{{Soderberg} {et~al.}(2005){Soderberg}, {Kulkarni}, {Berger},
  {Chevalier}, {Frail}, {Fox}, \& {Walker}}]{Soderberg2005-03L}
{Soderberg}, A.~M., {Kulkarni}, S.~R., {Berger}, E., {et~al.} 2005, \apj, 621,
  908, \dodoi{10.1086/427649}

\bibitem[{{Soderberg} {et~al.}(2006{\natexlab{b}}){Soderberg}, {Nakar},
  {Berger}, \& {Kulkarni}}]{Soderberg2006a-radio}
{Soderberg}, A.~M., {Nakar}, E., {Berger}, E., \& {Kulkarni}, S.~R.
  2006{\natexlab{b}}, \apj, 638, 930, \dodoi{10.1086/499121}

\bibitem[{{Soderberg} {et~al.}(2006{\natexlab{c}}){Soderberg}, {Kulkarni},
  {Nakar}, {Berger}, {Cameron}, {Fox}, {Frail}, {Gal-Yam}, {Sari}, {Cenko},
  {Kasliwal}, {Chevalier}, {Piran}, {Price}, {Schmidt}, {Pooley}, {Moon},
  {Penprase}, {Ofek}, {Rau}, {Gehrels}, {Nousek}, {Burrows}, {Persson}, \&
  {McCarthy}}]{Soderberg2006b-06aj}
{Soderberg}, A.~M., {Kulkarni}, S.~R., {Nakar}, E., {et~al.}
  2006{\natexlab{c}}, \nat, 442, 1014, \dodoi{10.1038/nature05087}

\bibitem[{{Soderberg} {et~al.}(2010){Soderberg}, {Chakraborti}, {Pignata},
  {Chevalier}, {Chandra}, {Ray}, {Wieringa}, {Copete}, {Chaplin},
  {Connaughton}, {Barthelmy}, {Bietenholz}, {Chugai}, {Stritzinger}, {Hamuy},
  {Fransson}, {Fox}, {Levesque}, {Grindlay}, {Challis}, {Foley}, {Kirshner},
  {Milne}, \& {Torres}}]{Soderberg2010}
{Soderberg}, A.~M., {Chakraborti}, S., {Pignata}, G., {et~al.} 2010, \nat, 463,
  513, \dodoi{10.1038/nature08714}

\bibitem[{{Thornton} {et~al.}(2016){Thornton}, {Ade}, {Aiola}, {Angil{\`e}},
  {Amiri}, {Beall}, {Becker}, {Cho}, {Choi}, {Corlies}, {Coughlin}, {Datta},
  {Devlin}, {Dicker}, {D{\"u}nner}, {Fowler}, {Fox}, {Gallardo}, {Gao},
  {Grace}, {Halpern}, {Hasselfield}, {Henderson}, {Hilton}, {Hincks}, {Ho},
  {Hubmayr}, {Irwin}, {Klein}, {Koopman}, {Li}, {Louis}, {Lungu}, {Maurin},
  {McMahon}, {Munson}, {Naess}, {Nati}, {Newburgh}, {Nibarger}, {Niemack},
  {Niraula}, {Nolta}, {Page}, {Pappas}, {Schillaci}, {Schmitt}, {Sehgal},
  {Sievers}, {Simon}, {Staggs}, {Tucker}, {Uehara}, {van Lanen}, {Ward}, \&
  {Wollack}}]{Thornton2016}
{Thornton}, R.~J., {Ade}, P.~A.~R., {Aiola}, S., {et~al.} 2016, \apjs, 227, 21,
  \dodoi{10.3847/1538-4365/227/2/21}

\bibitem[{{van Dyk} {et~al.}(1993){van Dyk}, {Weiler}, {Sramek}, \&
  {Panagia}}]{vanDyk1993}
{van Dyk}, S.~D., {Weiler}, K.~W., {Sramek}, R.~A., \& {Panagia}, N. 1993,
  \apjl, 419, L69, \dodoi{10.1086/187139}

\bibitem[{{Virtanen} {et~al.}(2020){Virtanen}, {Gommers}, {Oliphant},
  {Haberland}, {Reddy}, {Cournapeau}, {Burovski}, {Peterson}, {Weckesser},
  {Bright}, {van der Walt}, {Brett}, {Wilson}, {Jarrod Millman}, {Mayorov},
  {Nelson}, {Jones}, {Kern}, {Larson}, {Carey}, {Polat}, {Feng}, {Moore}, {Vand
  erPlas}, {Laxalde}, {Perktold}, {Cimrman}, {Henriksen}, {Quintero}, {Harris},
  {Archibald}, {Ribeiro}, {Pedregosa}, {van Mulbregt}, \&
  {Contributors}}]{Virtanen2020}
{Virtanen}, P., {Gommers}, R., {Oliphant}, T.~E., {et~al.} 2020, Nature
  Methods, 17, 261, \dodoi{https://doi.org/10.1038/s41592-019-0686-2}

\bibitem[{{Walker}(1998)}]{Walker1998}
{Walker}, M.~A. 1998, \mnras, 294, 307,
  \dodoi{10.1046/j.1365-8711.1998.01238.x}

\bibitem[{{Warren} {et~al.}(2018){Warren}, {Barkov}, {Ito}, {Nagataki}, \&
  {Laskar}}]{Warren2018}
{Warren}, D.~C., {Barkov}, M.~V., {Ito}, H., {Nagataki}, S., \& {Laskar}, T.
  2018, \mnras, 480, 4060, \dodoi{10.1093/mnras/sty2138}

\bibitem[{{Weiler} {et~al.}(1986){Weiler}, {Sramek}, {Panagia}, {van der
  Hulst}, \& {Salvati}}]{Weiler1986}
{Weiler}, K.~W., {Sramek}, R.~A., {Panagia}, N., {van der Hulst}, J.~M., \&
  {Salvati}, M. 1986, \apj, 301, 790, \dodoi{10.1086/163944}

\bibitem[{{Weiler} {et~al.}(1991){Weiler}, {van Dyk}, {Panagia}, {Sramek}, \&
  {Discenna}}]{Weiler1991}
{Weiler}, K.~W., {van Dyk}, S.~D., {Panagia}, N., {Sramek}, R.~A., \&
  {Discenna}, J.~L. 1991, \apj, 380, 161, \dodoi{10.1086/170571}

\bibitem[{{Weiler} {et~al.}(2007){Weiler}, {Williams}, {Panagia}, {Stockdale},
  {Kelley}, {Sramek}, {Van Dyk}, \& {Marcaide}}]{Weiler2007}
{Weiler}, K.~W., {Williams}, C.~L., {Panagia}, N., {et~al.} 2007, \apj, 671,
  1959, \dodoi{10.1086/523258}

\bibitem[{{Whitehorn} {et~al.}(2016){Whitehorn}, {Natoli}, {Ade}, {Austermann},
  {Beall}, {Bender}, {Benson}, {Bleem}, {Carlstrom}, {Chang}, {Chiang}, {Cho},
  {Citron}, {Crawford}, {Crites}, {de Haan}, {Dobbs}, {Everett}, {Gallicchio},
  {George}, {Gilbert}, {Halverson}, {Harrington}, {Henning}, {Hilton},
  {Holder}, {Holzapfel}, {Hoover}, {Hou}, {Hrubes}, {Huang}, {Hubmayr},
  {Irwin}, {Keisler}, {Knox}, {Lee}, {Leitch}, {Li}, {McMahon}, {Meyer},
  {Mocanu}, {Nibarger}, {Novosad}, {Padin}, {Pryke}, {Reichardt}, {Ruhl},
  {Saliwanchik}, {Sayre}, {Schaffer}, {Smecher}, {Stark}, {Story}, {Tucker},
  {Vand erlinde}, {Vieira}, {Wang}, \& {Yefremenko}}]{Whitehorn2016}
{Whitehorn}, N., {Natoli}, T., {Ade}, P.~A.~R., {et~al.} 2016, \apj, 830, 143,
  \dodoi{10.3847/0004-637X/830/2/143}

\bibitem[{{Willingale} {et~al.}(2013){Willingale}, {Starling}, {Beardmore},
  {Tanvir}, \& {O'Brien}}]{Willingale2013}
{Willingale}, R., {Starling}, R.~L.~C., {Beardmore}, A.~P., {Tanvir}, N.~R., \&
  {O'Brien}, P.~T. 2013, \mnras, 431, 394, \dodoi{10.1093/mnras/stt175}

\bibitem[{{Wilms} {et~al.}(2000){Wilms}, {Allen}, \& {McCray}}]{Wilms2000}
{Wilms}, J., {Allen}, A., \& {McCray}, R. 2000, \apj, 542, 914,
  \dodoi{10.1086/317016}

\bibitem[{{Yuan} {et~al.}(2016){Yuan}, {Wang}, {Lei}, {Gao}, \&
  {Zhang}}]{Yuan2016}
{Yuan}, Q., {Wang}, Q.~D., {Lei}, W.-H., {Gao}, H., \& {Zhang}, B. 2016,
  \mnras, 461, 3375, \dodoi{10.1093/mnras/stw1543}

\bibitem[{{Zauderer} {et~al.}(2011){Zauderer}, {Berger}, {Soderberg}, {Loeb},
  {Narayan}, {Frail}, {Petitpas}, {Brunthaler}, {Chornock}, {Carpenter},
  {Pooley}, {Mooley}, {Kulkarni}, {Margutti}, {Fox}, {Nakar}, {Patel},
  {Volgenau}, {Culverhouse}, {Bietenholz}, {Rupen}, {Max-Moerbeck}, {Readhead},
  {Richards}, {Shepherd}, {Storm}, \& {Hull}}]{Zauderer2011}
{Zauderer}, B.~A., {Berger}, E., {Soderberg}, A.~M., {et~al.} 2011, \nat, 476,
  425, \dodoi{10.1038/nature10366}

\end{thebibliography}
\bibliographystyle{aasjournal}

\end{document}